\documentclass[a4paper,fleqn]{cas-sc}

\usepackage[authoryear]{natbib}

\usepackage[makeroom]{cancel}
\usepackage{scrextend} 
\usepackage{bold-extra} 
\usepackage{color}
\usepackage[T1]{fontenc} 
\usepackage{fancyhdr}
\usepackage{ulem}
\usepackage{xcolor}
\usepackage{relsize}
\usepackage{textcomp}
\usepackage{textgreek}
\usepackage{lipsum}
\usepackage{setspace}
\usepackage{empheq}

\usepackage{float} 
\usepackage{epstopdf, epsfig}
\usepackage{graphicx} 
\usepackage{graphics} 
\usepackage{tabu} 
\usepackage{tabulary} 
\usepackage{tabularx} 
\usepackage{booktabs}

\usepackage{framed} 
\usepackage{listings} 

\usepackage{amsmath} 
\usepackage{amssymb} 
\usepackage{mathrsfs}

\usepackage{algorithm} 
\usepackage{algpseudocode} 

\usepackage{tensor}
\usepackage{upgreek}
\usepackage{physics}

\usepackage[toc, xindy, acronym]{glossaries}
\usepackage{natbib}

\usepackage[title]{appendix}

\usepackage[shortlabels]{enumitem}
\usepackage{zref-savepos}
\usepackage{cleveref}
\crefformat{section}{\S#2#1#3} 
\crefformat{subsection}{\S#2#1#3}
\crefformat{subsubsection}{\S#2#1#3}
\crefformat{figure}{figure~#2#1#3}
\crefformat{appendix}{appendix~#2#1#3}
\crefformat{equation}{(#2#1#3)}

\usepackage{tikz}
\usepackage{circuitikz}
\usetikzlibrary{calc}
\usetikzlibrary{patterns}
\usetikzlibrary{arrows}
\usetikzlibrary{shapes}
\usetikzlibrary{plotmarks}
\usetikzlibrary{decorations.markings}

\usepackage{pgfplots}
\usepgfplotslibrary{polar}
\usepgflibrary{shapes.geometric}
\usepackage[]{pstricks}

\tikzset{
	set arrow inside/.code={\pgfqkeys{/tikz/arrow inside}{#1}},
	set arrow inside={end/.initial=>, opt/.initial=},
	/pgf/decoration/Mark/.style={
		mark/.expanded=at position #1 with
		{
			\noexpand\arrow[\pgfkeysvalueof{/tikz/arrow inside/opt}]{\pgfkeysvalueof{/tikz/arrow inside/end}}
		}
	},
	arrow inside/.style 2 args={
		set arrow inside={#1},
		postaction={
			decorate,decoration={
				markings,Mark/.list={#2}
			}
		}
	},
}

\newcommand{\closesymbol}{\!}
\newcommand{\fourier}[1]{\tilde{#1}}  % Fourier-transformed functions
\renewcommand{\fourier}[1]{#1}  % remove the tilde from Fourier-space quantities

\definecolor{googleblue}{HTML}{4285F4}
\definecolor{googlered}{HTML}{DB4437}
\definecolor{googleyellow}{HTML}{F4B400}
\definecolor{googlegreen}{HTML}{0F9D58}

\newcommand{\rmd}{\text{d}}
\newcommand{\rmi}{\text{i}}

\ifdefined\rmJf

\else

\fi

\ifdefined\rmI
\renewcommand{\rmI}{\text{I}}
\else
\newcommand{\rmI}{\text{I}}
\fi

\newcommand{\besselargint}[1][\s]{b_{#1}}

\newcommand{\px}{\partial_x}

\ifdefined\partd
\renewcommand{\partd}[3][]{\frac{{\partial^{#1} #2}}{{\partial #3}^{#1}}}
\else
\newcommand{\partd}[3][]{\frac{{\partial^{#1} #2}}{{\partial #3}^{#1}}}
\fi

\renewcommand{\vec}[1]{\boldsymbol{#1}}
\ifdefined\vec  % Make sure we got bold vectors
\renewcommand{\vec}[1]{\boldsymbol{#1}}
\else
\newcommand{\vec}[1]{\boldsymbol{#1}}
\fi

\ifdefined\div
\renewcommand{\div}{{\vec{\nabla} \cdot}}
\else
\newcommand{\div}{{\vec{\nabla} \cdot}}
\fi

\newcommand{\gpar}{\vec{\nabla}_\parallel}
\newcommand{\gperp}{\vec{\nabla}_\perp}

\ifdefined\vr
\renewcommand{\vr}{{\vec{r}}}
\else
\newcommand{\vr}{{\vec{r}}}
\fi

\newcommand{\vk}{{\vec{k}}}

\newcommand{\kperp}{k_\perp}
\newcommand{\kpar}{k_\parallel}

\newcommand{\s}{s}

\newcommand{\vths}[1][\s]{v_{\text{th}{#1}}}
\newcommand{\vthi}{\vths[i]}
\newcommand{\vthe}{\vths[e]}

\newcommand{\ms}[1][\s]{m_{#1}}
\newcommand{\me}{\ms[e]}
\newcommand{\mi}{\ms[i]}
\newcommand{\rhoi}{\rho_i}
\newcommand{\rhoe}{\rho_e}
\newcommand{\rhos}{\rho_\s}

\newcommand{\taubar}{\bar{\tau}}
\newcommand{\lambdae}{\lambda_{ei}}

\newcommand{\dens}[1][\s]{n_{#1}}

\renewcommand{\ne}{\dens[e]}

\renewcommand{\ni}{\dens[i]}

\newcommand{\Ts}[1][\s]{T_{#1}}
\newcommand{\Te}{\Ts[e]}
\newcommand{\Ti}{\Ts[i]}

\newcommand{\dTs}[1][\s]{\delta T_{#1}}
\newcommand{\dTe}{\dTs[e]}
\newcommand{\dTi}{\dTs[i]}

\newcommand{\pbra}[2]{\left\lbrace #1, #2 \right\rbrace}

\newcommand{\phipot}{\phi}  % electrostatic potential

\newcommand{\dphipot}{\delta \phipot}  % electrostatic potential
\newcommand{\dphipotk}{\fourier{\dphipot}_\vk}

\newcommand{\dApar}{\delta \closesymbol  A_\parallel}
\newcommand{\dApark}{\delta \closesymbol \fourier{A}_{\parallel\vk}}

\newcommand{\dBperp}{{\delta \closesymbol {B}_{\closesymbol\perp}}}

\newcommand{\vE}{\vec{E}}
\newcommand{\vB}{\vec{B}}
\newcommand{\ub}{\vec{b}}

\newcommand{\exb}{\vE\times\vB}

\newcommand{\Lpar}{L_\parallel}
\newcommand{\Lperp}{L_\perp}
\newcommand{\rhoperp}{\rho_\perp}

\newcommand{\LB}{L_B}

\newcommand{\LT}[1][]{L_{T_{#1}}}

\usepackage{etoolbox}
\usepackage{xspace}

\newcommand{\figscale}{0.91}

\newcommand{\flucs}{{\texttt{FLUCS}}\xspace}
\newcommand{\fouriersolver}{{\texttt{FourierSolver}}\xspace}

\newcommand{\tindex}{n}

\newcommand{\dt}{\Delta t}
\newcommand{\dtn}[1][]{\Delta t_{n\ifblank{#1}{}{#1}}}

\newcommand{\fieldssymbol}{X}

\newcommand{\vfields}[1][]{\vec{\fieldssymbol}_{\vec{k}\ifblank{#1}{}{, #1}}}

\newcommand{\oplinear}{L}
\newcommand{\opnonlinear}{N}
\newcommand{\opforcing}{F}

\newcommand{\rindex}{a}
\newcommand{\rindexalt}{b}
\newcommand{\arrayindex}[1][\rindex]{p_{#1}}
\newcommand{\dftindex}[1][\rindex]{q_{#1}}
\newcommand{\varrayindex}{{\vec{p}}}
\newcommand{\vdftindex}{{\vec{q}}}

\newcommand{\domain}[1][\rindex]{L_{#1}}
\newcommand{\resolution}[1][\rindex]{n_{#1}}

\newcommand{\totalmodes}{N}
\newcommand{\solvedmodes}{N_\mathrm{solved}}
\newcommand{\effectivemodes}{N_\mathrm{effective}}

\newcommand{\dimension}{D}

\newcommand{\linearmatrix}{\mathbf{L}_{\vec{k}}}
\newcommand{\propagator}[1][]{\mathbf{P}_{\vec{k}\ifblank{#1}{}{, #1}}}
\newcommand{\rhsexplicit}[1][]{\mathbf{R}_{\vec{k}\ifblank{#1}{}{, #1}}}

\newcommand{\padeorder}{p}
\newcommand{\padeordersum}{q}

\newcommand{\timestepperorder}{n}

\newcommand{\abcoeffs}{c}

\newcommand{\cfl}{\mathrm{CFL}}
\newcommand{\cflmax}{{\texttt{max\_cfl}}\xspace}
\newcommand{\cflomega}[1][\tindex]{\omega_{\cfl, #1}}

\newcommand{\ega}{u}
\newcommand{\egb}{v}

\newcommand{\rtoc}{{\texttt{r2c}}\xspace}
\newcommand{\ctor}{{\texttt{c2r}}\xspace}

\newcommand{\alias}{s}
\newcommand{\valias}{{\vec{\alias}}}

\newcommand{\duperp}{\delta \vec{u}_\perp}

\newcommand{\thetapmk}{\Theta_{\vec{k}}^\pm}
\newcommand{\thetamk}{\Theta_{\vec{k}}^-}
\newcommand{\thetapk}{\Theta_{\vec{k}}^+}
\newcommand{\thetam}{\Theta^-}
\newcommand{\thetap}{\Theta^+}

\newcommand{\vphase}{v_{\text{ph}}}
\newcommand{\fenergy}{W}
\newcommand{\helicity}{H}

\newcommand{\fluxw}{\varepsilon_\fenergy}
\newcommand{\fluxh}{\varepsilon_\helicity}
\newcommand{\fluxpm}{\varepsilon^\pm}
\newcommand{\fluxp}{\varepsilon^+}
\newcommand{\fluxm}{\varepsilon^-}

\renewcommand{\besselargint}[1][\s]{\alpha_{#1}}

\newcommand{\betacrit}{\beta_e^\text{crit}}

\newcommand{\avgzonal}[1]{\overline{#1}}
\newcommand{\zf}{\avgzonal{v}_{E,y}}
\newcommand{\zonaldTi}{\avgzonal{\delta T}_i}
\newcommand{\Piphi}{\Pi_\phi}
\newcommand{\PiT}{\Pi_T}
\newcommand{\Pid}{\Pi_d}

\crefformat{figure}{figure~#2#1#3}
\Crefformat{figure}{Figure~#2#1#3}

\crefrangeformat{figure}{figures #3#1#4--#5#2#6}
\crefmultiformat{figure}{figures~#2#1#3}{ and~#2#1#3}{, #2#1#3}{, and~#2#1#3}

\Crefrangeformat{figure}{Figures #3#1#4--#5#2#6}
\Crefmultiformat{figure}{Figures~#2#1#3}{ and~#2#1#3}{, #2#1#3}{, and~#2#1#3}

\crefformat{equation}{(#2#1#3)}
\Crefformat{equation}{Equation~(#2#1#3)}

\crefrangeformat{equation}{(#3#1#4)--(#5#2#6)}
\crefmultiformat{equation}{(#2#1#3)}{ and~(#2#1#3)}{, (#2#1#3)}{, and~(#2#1#3)}

\Crefrangeformat{equation}{Equations~(#3#1#4)--(#5#2#6)}
\Crefmultiformat{equation}{Equations~(#2#1#3)}{ and~(#2#1#3)}{, (#2#1#3)}{, and~(#2#1#3)}

\begin{document}
\let\WriteBookmarks\relax
\def\floatpagepagefraction{1}
\def\textpagefraction{.001}
\shorttitle{The \flucs code}
\shortauthors{T.~Adkins \& P.~G.~Ivanov}
%\begin{frontmatter}

\title [mode = title]{\flucs: a modular GPU-native framework for fluid and plasma turbulence}                      
%\tnotetext[1]{stuff}

% Authors
\author[1,2,3]{T.~Adkins}[
%		 type=editor,
%        auid=000,bioid=1,
%        prefix=Dr,
%        role=Researcher,
        orcid=0000-0001-5028-8047
]
\cormark[1]
\fnmark[1]
\ead{tadkins@pppl.gov}
%\ead[url]{https://physycola.github.io/}
\credit{Conceptualization, Code development, Data Curation, Writing}

\author[4]{P.~G.~Ivanov}[
%		 type=editor,
%        auid=000,bioid=1,
%        prefix=Dr,
%        role=Researcher,
        orcid=0000-0001-9720-4357
]
\cormark[1]
\fnmark[1]
\ead{plamen.ivanov@ukaea.uk}
\credit{Conceptualization, Code development, Data Curation, Writing}

\affiliation[1]{organization={Princeton Plasma Physics Laboratory},
                addressline={100 Stellarator Road}, 
                city={Princeton},
%               citysep={}, % Uncomment if no comma needed between city and postcode
                state={New Jersey},
                postcode={08542}, 
                country={USA}
            }
            
\affiliation[2]{organization={School of Physics \& Astronomy, The University of Edinburgh},
	city={Edinburgh},
	postcode={EH9 3FD},
	country={UK}
}

\affiliation[3]{organization={School of Mathematics, The University of Edinburgh},
	city={Edinburgh},
	postcode={EH9 3FD},
	country={UK}
}

\affiliation[4]{organization={UKAEA (United Kingdom Atomic Energy Authority)},
	addressline={Culham Campus}, 
	city={Abingdon},
	%               citysep={}, % Uncomment if no comma needed between city and postcode
	state={Oxfordshire},
	postcode={OX14 3DB}, 
	country={UK}
}

\cortext[cor1]{Corresponding authors
%	: tadkins@pppl.gov; plamen.ivanov@ukaea.uk
}
\fntext[fn1]{
	These authors contributed equally to all aspects of this work.
}

\begin{abstract}
We present \flucs, an open-source framework for solving systems of partial differential equations efficiently on graphics processing units (GPUs), targeting models of fluid and plasma turbulence. It comprises a set of numerical \textit{solvers} specialised for a well-defined class of problems and independently developed \textit{systems} that supply specific systems of equations. This allows different models to share numerical infrastructure while retaining control over their individual implementation and optimisation. The present release provides a Fourier pseudospectral solver for one-, two-, and three-dimensional periodic domains solved on a single NVIDIA GPU, with integrating-factor time integration and several time-stepping and dealiasing methods. We verify its convergence, quantify its computational costs, and demonstrate its application to three reduced models of fusion and astrophysical plasma turbulence.
\end{abstract}

\begin{keywords}
Turbulence
\sep
Magnetic confinement fusion
\sep 
GPU acceleration
\sep
Spectral methods 
\end{keywords}

\maketitle

\section{Introduction}
\label{sec:introduction}
Numerical simulations play a central role in the study of turbulence in both neutral fluids and plasmas. Advances in numerical methods and computational hardware have progressively expanded the range of accessible problems, as well as the level of detail with which their dynamics can be resolved. These advances enable two complementary approaches to scientific investigation: developing more comprehensive physical models and studying comparatively simple ones in greater depth. The first permits, for example, additional physical processes and interactions to be incorporated into existing numerical simulations, with the aim of improving the quantitative fidelity of their predictions. The second pursues improvements such as greater spatial resolution, wider scale separation, longer integration times, and more extensive parameter scans without introducing new physical effects. Insights gained from simpler models can guide the development and interpretation of more comprehensive ones, which in turn can reveal where additional physics is essential.

In the domain of fusion-focused plasma physics, the development of modern gyrokinetic codes illustrates the increasing physical fidelity and scope of numerical simulations. For example, \texttt{stella} accommodates an implicit treatment of kinetic electrons and multiple ion species in both axisymmetric and non-axisymmetric magnetic geometries \citep{barnes19STELLA}, while \texttt{ORB5} provides global electromagnetic particle-in-cell simulations in toroidal geometry, including collisions and strong flows \citep{lanti20}. The full-$f$ continuum code \texttt{GENE-X} extends gyrokinetic calculations to the plasma edge and scrape-off layer, including magnetic X-points \citep{michels21}, while \texttt{XGC} can simulate a domain extending from the magnetic axis to the material wall, incorporating neutral particles and atomic interactions \citep{xgc,ku18,hager22}. Computationally ambitious models are also found within fluid descriptions: \texttt{GBS} performs global, three-dimensional fluid turbulence simulations coupled to kinetic neutral dynamics \citep{giacomin22}. In astrophysical, relativistic, and high-energy-density plasma research, particle-in-cell frameworks such as \texttt{OSIRIS} \citep{fonseca02,fonseca08}, \texttt{TRISTAN-MP} \citep{sironi14}, and \texttt{Smilei} \citep{derouillat18} enable investigations of shocks, magnetic reconnection, particle acceleration, radiation, and their interactions. These developments illustrate the substantial numerical and software effort required to combine physical and geometrical complexity within a single computational framework that can make effective use of high-performance computational resources.

Alongside these developments, there has been sustained interest in using reduced fluid models to isolate important physical ingredients and investigate comparatively simple systems of equations. Despite their relative simplicity, such models can reproduce key features of more comprehensive descriptions and provide a means of understanding the underlying physical mechanisms. In fusion research, examples include studies of zonal-flow-regulated transport, intermittent turbulent dynamics, and the Dimits transition \citep{rogers00,ivanov20,zhu20PRL}. Fluid models have similarly supported extensive investigations of space and astrophysical plasma turbulence, including critical balance, alignment, and intermittency in Alfv\'enic turbulence \citep{chandran15,mallet15,mallet19}, as well as the constraints on turbulent cascades associated with the helicity barrier \citep{meyrand21,adkins24,adkins25}. The extent of agreement with more complete descriptions depends on the model and parameter regime, but even qualitative agreement can provide useful physical insight. Obtaining such insight can nevertheless require substantial numerical work: resolving a broad range of spatial scales, sampling intermittent dynamics over sufficiently long times, and performing parameter scans across transitions can all be computationally demanding, even when working with otherwise simple models. Efficient numerical implementations are therefore important not only for comprehensive plasma models, but also for reduced models and fluid systems more generally.

Modern graphics processing units (GPUs) offer an opportunity to extend the scale of these calculations. Their high arithmetic throughput and memory bandwidth are well suited to many of the operations required by fluid solvers, while specialised numerical libraries provide efficient implementations of common operations such as Fourier transforms. GPU acceleration is already used extensively in plasma physics, including in gyrokinetic codes such as \texttt{GX} \citep{mandell24}, \texttt{CGYRO} \citep{candy16}, and \texttt{GENE} \citep{jenko00GENE}. Making effective use of this hardware, however, requires more than simply transferring an existing algorithm or calculation to the GPU: the arrangement of the data, the operations performed on it, and the movement of data between the host and device must all be considered in order to take full advantage of the GPU architecture. Programming environments such as CUDA \citep{cuda26} provide direct control over these operations, while libraries such as CuPy \citep{okuta17} make GPU arrays and numerical routines accessible from high-level languages like Python. These tools make efficient implementations possible, but developing and maintaining one for each new physical model can still entail substantial effort.

Much of this effort concerns numerical machinery that is shared by otherwise different physical models, including time integration, spatial discretisation, and the management of simulation data. Reusable computational frameworks allow these components to be developed and tested once, rather than reconstructed for each new set of equations. For example, the Dedalus Project \citep{burns20} provides a flexible spectral framework in which equations can be entered symbolically, while \texttt{FourierFlows.jl} and its associated \texttt{GeophysicalFlows.jl} package support Fourier pseudospectral fluid calculations on CPUs and GPUs \citep{constantinou21}. The form of the interface between a physical model and its numerical solver is important, however, because it determines which operations can be specialised for that model. Systems that share the same spatial representation and time-integration methods may still require different intermediate quantities, storage arrangements, or combinations of numerical operations. A useful framework should therefore permit the reuse of common numerical machinery while retaining sufficient control over the equation-specific calculations to implement them efficiently.

The \flucs code is designed for this purpose: it provides a common interface and framework within which simple systems of partial differential equations can be implemented and optimised for efficient execution on GPU hardware. This flexibility is not achieved by requiring a single general-purpose solver to accommodate arbitrary equations. Instead, \flucs is structured around a distinction between `solvers', which provide numerical machinery for a chosen spatial representation and class of boundary conditions, and `systems', which supply equation-specific operations through a prescribed interface. Each solver can therefore be specialised for a well-defined class of numerical problems without being restricted to a particular physical model. Individual systems, in turn, reuse tools for standard operations such as time integration and dealiasing while retaining control over the construction and evaluation of their equation-specific terms. Implementing a new system requires these operations to be specified explicitly, but allows them to be tailored to that system's computational requirements. Systems are developed and distributed independently of the core package. Python provides the interface, performs input/output operations, and coordinates the calculations and memory management via CuPy, while CUDA C++ kernels and GPU numerical libraries perform the computationally intensive operations. The aim is thus to retain the opportunities for specialisation associated with a dedicated implementation, while reducing the numerical and software infrastructure that must be developed anew for each physical model. This paper describes the present release of \flucs and its Fourier pseudospectral solver, \fouriersolver, appropriate for systems in one-, two-, or three-dimensional periodic domains solved on a single NVIDIA GPU. The implementations of three models --- isothermal KREHM, cold-ion ITG, and collisional ETG --- provide examples of the interface between the systems and solvers. 

The remainder of this paper is organised as follows. The code structure is described in \cref{sec:code_structure}, the solver algorithm in \cref{sec:fourier_solver_algorithm}, and its dealiasing methods in \cref{sec:dealiasing}. The convergence studies and timing and memory comparisons in \cref{sec:numerics} assess the implementation and the practical tradeoffs between its different options. Details of the time-stepping schemes and benchmarks of the example systems are provided in \cref{app:timestepping_algorithms,app:systems}. The architecture also permits additional solvers and systems, with future developments discussed in \cref{sec:conclusion}.

\section{Code structure}
\label{sec:code_structure}
\begin{figure}[pos=h]
	\centering
	\vspace*{0.5cm}
	\scalebox{1.0}{% Requires:
%   \usepackage{xcolor}
%   \usepackage{tikz}
%   \usetikzlibrary{calc}

\definecolor{googleblue}{HTML}{4285F4}
\definecolor{googlered}{HTML}{DB4437}
\definecolor{googlegreen}{HTML}{0F9D58}

\begin{tikzpicture}[
	scale=1.0,
	every node/.style={
		scale=1.0,
		text=black,
		font=\fontfamily{cmr}\selectfont
	}
	]
	
	% ---------------------------------------------------------------------------
	% Parameters
	% ---------------------------------------------------------------------------
	
	% Box sizes
	\def\rootwidth{3.0cm}
	\def\solverwidth{3.2cm}
	\def\systemwidth{3.7cm}
	\def\itemwidth{2.5cm}
	\def\boxheight{0.7cm}
	
	% Separations
	\def\solverseparation{4.0cm}
	\def\systemseparation{2.8cm}
	\def\othersystemseparation{2.2cm}
	\def\itemseparation{1.0cm}
	\def\branchlength{0.2cm}
	\def\fourieritemshift{0.35cm}
	
	% System layout
	\def\systemhorizontalspacing{1.6cm} % Gap from FourierSolver to system boxes
	\def\systemverticalspacing{1.1cm}   % Vertical separation between systems
	\def\systembranchposition{0.5}     % 0 = solver, 1 = first system
	
	% Line properties
	\def\outlinewidth{0.7pt}
	\def\connectionwidth{0.7pt}
	\def\cornerradius{1.5pt}
	
	% Colours
	\colorlet{rootcolour}{googleblue!20}
	\colorlet{solvercolour}{googlegreen!18}
	\colorlet{systemcolour}{googlered!18}
	\colorlet{itemcolour}{gray!12}
	
	\colorlet{rootoutlinecolour}{googleblue!65!black}
	\colorlet{solveroutlinecolour}{googlegreen!65!black}
	\colorlet{systemoutlinecolour}{googlered!65!black}
	\colorlet{outlinecolour}{gray!70}
	\colorlet{connectioncolour}{gray!70}
	
	% Styles
	\tikzset{
		flucs box/.style={
			draw=outlinecolour,
			line width=\outlinewidth,
			rounded corners=\cornerradius,
			minimum height=\boxheight,
			align=center,
			text=black,
			font=\fontfamily{cmr}\selectfont
		},
		root box/.style={
			flucs box,
			minimum width=\rootwidth,
			fill=rootcolour,
			draw=rootoutlinecolour,
			font=\fontfamily{cmr}\selectfont\bfseries
		},
		solver box/.style={
			flucs box,
			minimum width=\solverwidth,
			fill=solvercolour,
			draw=solveroutlinecolour,
			font=\fontfamily{cmr}\selectfont\bfseries
		},
		system box/.style={
			flucs box,
			minimum width=\systemwidth,
			fill=systemcolour,
			draw=systemoutlinecolour
		},
		item box/.style={
			flucs box,
			minimum width=\itemwidth,
			fill=itemcolour
		},
		connection/.style={
			draw=connectioncolour,
			line width=\connectionwidth,
			rounded corners=\cornerradius
		}
	}
	
	% ---------------------------------------------------------------------------
	% FLUCS and its general components
	% ---------------------------------------------------------------------------
	
	\node[root box] (flucs) at (0,0) {\texttt{flucs}};
	
	\node[item box] (input) at
	($(flucs)+(0,-\itemseparation)$) {Input};
	\node[item box] (output) at
	($(input)+(0,-\itemseparation)$) {Output};
	\node[item box] (restart) at
	($(output)+(0,-\itemseparation)$) {Restart};
	\node[item box] (diagnostics) at
	($(restart)+(0,-\itemseparation)$) {Diagnostics};
	\node[item box] (postprocessing) at
	($(diagnostics)+(0,-\itemseparation)$) {Postprocessing};
	
	\coordinate (flucsbranch) at ($(flucs.west)+(-\branchlength,0)$);
	
	\draw[connection] (flucs.west) -- (flucsbranch) -- (flucsbranch |- postprocessing.west);
	\draw[connection] (flucsbranch |- input.west) -- (input.west);
	\draw[connection] (flucsbranch |- output.west) -- (output.west);
	\draw[connection] (flucsbranch |- diagnostics.west) -- (diagnostics.west);
	\draw[connection] (flucsbranch |- restart.west) -- (restart.west);
	\draw[connection] (flucsbranch |- postprocessing.west) -- (postprocessing.west);
	
	% ---------------------------------------------------------------------------
	% Solvers
	% ---------------------------------------------------------------------------
	
	\node[solver box] (fouriersolver) at
	($(flucs.east)+(\solverseparation,0)$) {\texttt{FourierSolver}};
	
	\draw[connection] (flucs.east) -- (fouriersolver.west);
	
	\node[item box] (timesteppers) at
	($(fouriersolver)+(\fourieritemshift,-\itemseparation)$) {Timesteppers};
	\node[item box] (forcing) at
	($(timesteppers)+(0,-\itemseparation)$) {Forcing};
	\node[item box] (solverdiagnostics) at
	($(forcing)+(0,-\itemseparation)$) {Diagnostics};
%	\node[item box] (solverdots) at
%	($(solverdiagnostics)+(0,-\itemseparation)$) {$\dots$};
	
	\coordinate (fourierbranch) at
	($(fouriersolver.south west)+(\branchlength,0)$);
	
	\draw[connection] (fourierbranch) -- (fourierbranch |- solverdiagnostics.west);
	\draw[connection] (fourierbranch |- timesteppers.west) -- (timesteppers.west);
	\draw[connection] (fourierbranch |- forcing.west) -- (forcing.west);
	\draw[connection] (fourierbranch |- solverdiagnostics.west) -- (solverdiagnostics.west);
%	\draw[connection] (fourierbranch |- solverdots.west) -- (solverdots.west);
	
	\node[solver box] (anothersolver) at
	($(fouriersolver)+(0,-4.2cm)$) {$\dots$};
	
	\coordinate (solverjunction) at ($(flucs)!0.5!(fouriersolver)$);
	\coordinate (solvertrunkend) at
	($(solverjunction |- anothersolver.center)+(0,-1.0cm)$);
	
	\draw[connection] (solverjunction) -- (solvertrunkend);
	\draw[connection] (solverjunction |- anothersolver.west) -- (anothersolver.west);
	\node[inner sep=0pt] at
	($(solvertrunkend)+(0,-0.25cm)$) {$\dots$};
	
	% ---------------------------------------------------------------------------
	% Systems associated with the FourierSolver
	% ---------------------------------------------------------------------------
	
	% anchor=west means \systemhorizontalspacing measures the gap between
	% the edges of FourierSolver and FourierSystem1.
	\node[system box, anchor=west] (systemone) at
	($(fouriersolver.east)+(\systemhorizontalspacing,0)$)
	{\texttt{KREHMFourier}};
	
	\node[system box, anchor=west] (systemtwo) at
	($(systemone.west)+(0,-\systemverticalspacing)$)
	{\texttt{ColdITG2DFourier}};
	
	\node[system box, anchor=west] (systemthree) at
	($(systemtwo.west)+(0,-\systemverticalspacing)$)
	{\texttt{CollisionalETGFourier}};
	
	\node[system box, anchor=west] (systemdots) at
	($(systemthree.west)+(0,-\systemverticalspacing)$)
	{$\dots$};
	
	% Position along the line from FourierSolver to FourierSystem1:
	%   0.0 places it at FourierSolver;
	%   1.0 places it at FourierSystem1.
	\coordinate (systemjunction) at
	($(fouriersolver.east)!\systembranchposition!(systemone.west)$);
	
	% Main connection
	\draw[connection]
	(fouriersolver.east) -- (systemone.west);
	
	% Vertical branch
	\draw[connection]
	(systemjunction) -- (systemjunction |- systemdots.west);
	
	% Horizontal branches
	\draw[connection]
	(systemjunction |- systemtwo.west) -- (systemtwo.west);
	
	\draw[connection]
	(systemjunction |- systemthree.west) -- (systemthree.west);
	
	\draw[connection]
	(systemjunction |- systemdots.west) -- (systemdots.west);
	
\end{tikzpicture}}
	\vspace*{0.5cm}
	\caption{\normalfont A schematic representation of the \flucs code structure. The coloured blocks indicate different objects --- the main function \texttt{flucs} (blue), solvers (green), and systems (red) --- while the grey panels denote functionality handled by the object from which they branch. Ellipses indicate that additional solvers and systems may be added without changing the existing framework.}
	\label{fig:code_structure}
\end{figure}
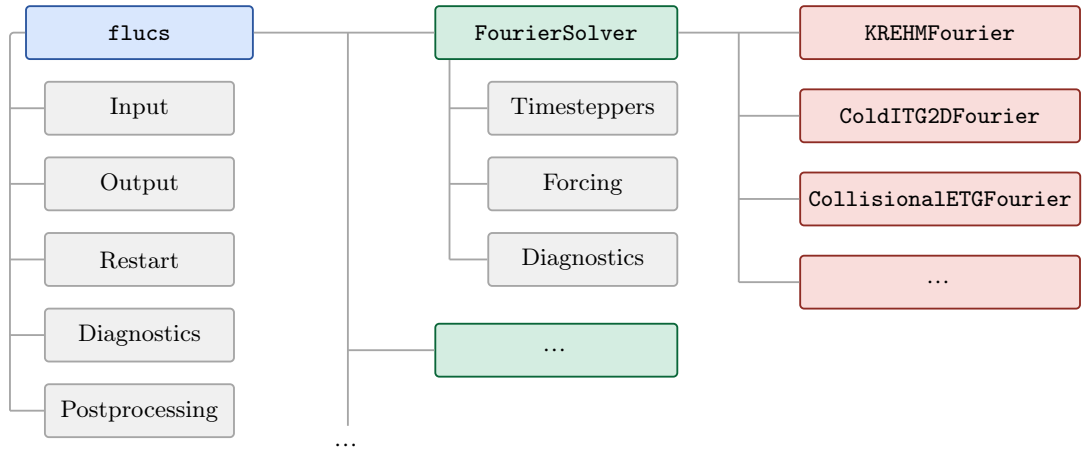

The \flucs code enables users to solve various systems of first-order-in-time partial differential equations (PDEs) within a common computational framework. To allow for such flexibility, it is structured around two principal objects: `solvers' and `systems'. The solvers implement the numerical machinery required to solve these PDEs within a particular spatial representation and corresponding class of boundary conditions (e.g., periodic), while remaining otherwise agnostic to the {particular} set of equations being solved. The systems are then responsible for providing the equation-specific information that is required by the solvers in order to time-advance, including, e.g., information about the linear and nonlinear terms, forcing methods, and runtime diagnostics. Currently, \flucs provides a single solver, \fouriersolver, for systems posed in one-, two-, or three-dimensional homogeneous periodic domains; its algorithm is described in detail in \cref{sec:fourier_solver_algorithm}. The code structure, shown schematically in \cref{fig:code_structure}, is specifically designed to permit the addition of solvers for other spatial representations and boundary conditions. See \cref{sec:conclusion} for further discussion of planned future additions. Further details of the currently implemented systems, shown in red in \cref{fig:code_structure}, can be found in \cref{app:systems}.

The systems act as `plug-ins' associated with a given solver that add to it the functionality to solve a new set of equations. Systems may be developed, distributed, and installed separately from the main \flucs package. At run time, \flucs identifies the solver and system selected in the input file and constructs the corresponding objects. Adding a new system therefore requires the user to express the governing equations in the form expected by the chosen solver and to implement the required equation-specific operations through its prescribed interface (see \cref{sec:timestep_loop}), for which existing systems provide practical templates. Such an approach differs from that of frameworks such as the Dedalus Project \citep{burns20}, in which equations and boundary conditions may be entered symbolically as plain-text expressions. Although our approach makes implementing a new set of equations more involved, it allows each solver to remain highly specialised for a well-defined class of numerical problems. The algorithms and data representation can consequently be optimised for that problem class, rather than being constrained by the need to provide a single general-purpose treatment of arbitrary PDEs. The \flucs code thus lies in a middle ground between ease of adding new systems and computational performance; systems provide flexibility in the physical models that may be implemented, while solver-level specialisation enables excellent performance despite that flexibility.

This balance of flexibility and performance is reflected in the division of labour between Python \citep{python24} and CUDA C++ \citep{cuda26}. Python comprises the majority of the codebase and provides the interface through which users run simulations and process their outputs, with installed systems and solvers discovered through Python package entry points. Adopting a high-level, widely used language for these tasks makes the code straightforward to use, inspect, and modify. The computationally intensive numerical operations are then implemented as CUDA C++ kernels, which are compiled at run time for the selected solver/system combination and launched through the Python package CuPy \citep{okuta17}. CuPy is also used for managing GPU memory, allowing the evolved variables and temporary work arrays to remain device-side during the solver execution. Python coordinates kernel launches and handles only higher-level simulation logic, thereby limiting the contribution of any Python overhead to the overall run time. In this way, \flucs retains the accessibility of Python while leveraging the computational efficiency associated with specialised, compiled execution on GPU hardware. 

\section{\fouriersolver algorithm}
\label{sec:fourier_solver_algorithm}
With the overall code structure described, we now detail the algorithm underlying \fouriersolver, the solver used by all systems currently implemented in \flucs. Given an arbitrary vector of real-space fields $\vec{\fieldssymbol}(\vr, t) = (\fieldssymbol_1, \fieldssymbol_2, \dots)$ in a periodic domain, \fouriersolver evolves the Fourier components $\vfields(t)$ of the fields,
\begin{align}
	\vec{\fieldssymbol} (\vr, t) = \sum_{\vec{k}}  e^{\rmi \vec{k} \cdot \vec{r}} \vfields(t), \quad \vr = (x, y, z),
	\label{eq:fourier_transform}
\end{align}
assuming that the time-evolution of these Fourier components can be written as:
\begin{align}
	\frac{\rmd \vfields}{\rmd t} + \oplinear [\vfields] + \opnonlinear[\vfields] = \opforcing[\vfields].
	\label{eq:continuous_equation}
\end{align}
In \cref{eq:continuous_equation}, the linear operator $\oplinear[\vfields]$ may depend on both time and the wavenumber $\vec{k}$ but is assumed to be local in the latter such that it does not couple distinct Fourier modes; this is generally true in homogeneous systems. The nonlinear operator $\opnonlinear[\vfields]$ may couple different Fourier modes. \flucs does not impose any particular restrictions on the form of the nonlinear operator as the implementation thereof is left to the user. However, \flucs does provide tools for computing and dealiasing polynomial-type nonlinear terms (see \cref{sec:nonlinear_term,sec:dealiasing}). The forcing operator $\opforcing[\vfields]$ represents any externally imposed contribution to the field evolution and is typically integrated together with the nonlinear terms. Note that any part of the forcing that is deemed part of the linear evolution can instead be absorbed into the linear operator $\oplinear[\vfields]$.

\subsection{Spatial discretisation}
\label{sec:spatial_discretisation}
To represent \cref{eq:continuous_equation} on a finite computational grid, we adopt the usual Fourier pseudo-spectral discretisation in each periodic direction, using $\resolution$ uniformly spaced collocation points over a domain of length $\domain$ for $\rindex \in \{x, y, z\}$. The corresponding real-space collocation points are
\begin{align}
	r_\rindex = \frac{\arrayindex \domain}{\resolution}, \quad \arrayindex = 0, \dots , \resolution -1.
	\label{eq:realspace_points}
\end{align}
We define the dimensionless discrete-Fourier-transform (DFT) frequencies $\dftindex$ to take values
\begin{align}
	\dftindex = -\frac{\lfloor\resolution/2\rfloor}{\resolution}, \dots, \frac{\lceil\resolution/2\rceil-1}{\resolution},
	\label{eq:dftindices}
\end{align}
with the corresponding physical wavenumbers given by 
\begin{align}
	k_\rindex = \frac{2\pi \resolution}{\domain} \dftindex.
	\label{eq:wavenumbers}
\end{align}
The continuous Fourier series in \cref{eq:fourier_transform} is then truncated to the $N=\resolution[z]\resolution[x]\resolution[y]$ modes represented on this grid. The associated discrete Fourier transform and its inverse are, respectively,
\begin{align}
	\vfields(t) = \frac{1}{\totalmodes} \sum_\varrayindex \vec{\fieldssymbol}(\vr, t) \exp \left[- 2\pi \rmi \sum_\rindex \arrayindex \dftindex \right],
	\label{eq:discrete_fourier_transform}
\end{align}
and
\begin{align}
	\vec{\fieldssymbol}(\vr, t) = \sum_\vdftindex  \vfields(t) \exp \left[2\pi \rmi \sum_\rindex \arrayindex \dftindex \right].
	\label{eq:discrete_inverse_fourier_transform}
\end{align}
Since the reality of the physical fields $\vec{\fieldssymbol}$ implies that their Fourier transform satisfies $\vfields^\ast = \vec{\fieldssymbol}_{-\vec{k}}$, where `$\ast$' denotes the complex conjugate, it is only necessary to evolve \cref{eq:continuous_equation} for $\dftindex[y] \geqslant 0$, with negative values obtained by complex conjugation. Each Fourier-space field is therefore stored as an array of size $\resolution[z]\resolution[x](\lfloor \resolution[y]/2 \rfloor +1 )$, while its real-space representation remains of size $\totalmodes$, as above. This defines the full computational grid used throughout \fouriersolver. The choice of dealiasing method (see \cref{sec:dealiasing}) determines which Fourier modes on this grid are time-advanced and how the nonlinear terms appearing in \cref{eq:continuous_equation} are evaluated; it does not change the underlying field representation. 

\subsection{Temporal discretisation}
\label{sec:time_integration}
Within this spectral discretisation scheme, the linear operator at each wavenumber can be represented by a matrix $\linearmatrix = \linearmatrix(t)$ acting on the vector of fields, viz.,
\begin{align}
	\oplinear[\vfields] = \linearmatrix \vfields.
	\label{eq:linear_matrix}
\end{align}
Then, our original equation \cref{eq:continuous_equation} can be written as
\begin{align}
	\frac{\rmd \vfields}{\rmd t} + \linearmatrix \vfields = \rhsexplicit\left[\vfields\right], \quad \rhsexplicit\left[\vfields\right] \equiv \opforcing\left[\vfields\right] - \opnonlinear\left[\vfields\right]
	\label{eq:continuous_equation_rhs}.
\end{align}

We now consider solving \cref{eq:continuous_equation_rhs} over the finite time interval $t\in[t_{\tindex},t_{\tindex+1}]$ with the time step size defined by $\dtn=t_{\tindex+1}-t_{\tindex}$. We define the linear propagator $\propagator$ from $t_\tindex$ to $t$ as the linear operator that evolves according to
\begin{align}
	\frac{\rmd \propagator(t_\tindex, t)}{\rmd t} + \linearmatrix(t) \propagator(t_\tindex, t) = 0, \quad \propagator(t_\tindex, t_\tindex) = \mathbf{I}, 
	\label{eq:linear_propagator}
\end{align}
where $\mathbf{I}$ is the identity matrix. The linear propagator satisfies, for any times $t_1$, $t_2$, and $t_3$, the following identities:
\begin{align}
	\propagator^{-1}(t_1, t_2) = \propagator(t_2, t_1), \quad \propagator(t_2, t_3) \propagator(t_1, t_2) = \propagator(t_1, t_3).
	\label{eq:propagator_identities}
\end{align}
We can then rewrite \cref{eq:continuous_equation_rhs} as an equation for the Lawson-style integrating-factor field \citep{lawson67}, viz., 
\begin{align}
	\frac{\rmd}{\rmd t}\left[\propagator^{-1}(t_\tindex,t)\vfields(t) \right]= \propagator^{-1}(t_\tindex,t) \rhsexplicit[]\left[\vfields(t)\right].
	\label{eq:integrating_factor_equation}
\end{align}
Exact integration of \cref{eq:integrating_factor_equation} over this time interval and multiplication by $\propagator(t_\tindex, t_{\tindex+1})$ then yields
\begin{align}
	\vfields(t_{\tindex+1}) = \propagator(t_\tindex, t_{\tindex+1}) \vfields(t_{\tindex}) +\int_{t_{\tindex}}^{t_{\tindex+1}}\rmd t\: \propagator(t, t_{\tindex+1})\:\rhsexplicit\left[\vfields(t)\right],
	\label{eq:discrete_equation_exact}
\end{align}
where we have made use of the identities in \cref{eq:propagator_identities}. The integral appearing on the right-hand side of \cref{eq:discrete_equation_exact} is then approximated by one of the explicit time-integration schemes (time-steppers) provided by \fouriersolver: third-order Adams-Bashforth \citep[henceforth AB3;][]{marciniak20,tavakkol21}, third-order strong-stability-preserving Runge-Kutta \citep[henceforth SSPRK3;][]{shu88}, and classical fourth-order Runge-Kutta \citep[henceforth RK4;][]{runge95,kutta01,hairer93}. Further details about their implementation can be found in \cref{app:timestepping_algorithms}. 

These explicit time-integration schemes require linear propagators over both complete time steps and intermediate stage intervals. To propagate the fields from some time $t$ to another time $t'$, we require a single matrix $\linearmatrix(t,t')$ representing the linear evolution over that same interval, from which the propagator is constructed as
\begin{align}
	\propagator(t, t') \approx \exp\left[-(t'-t)\linearmatrix(t,t')\right].
	\label{eq:interval_linear_propagator}
\end{align}
For a time-independent linear operator, $\linearmatrix(t,t')=\linearmatrix$, and \cref{eq:interval_linear_propagator} is exact; this is the case for all systems currently implemented using \fouriersolver and is the only type of a linear operator currently supported without additional system-specific implementation. For a time-dependent operator, however, the construction of $\linearmatrix(t,t')$ forms part of the system-specific implementation and determines the temporal accuracy with which the linear evolution is represented. It may, for example, be obtained by freezing the operator at a selected time within the interval or through a higher-order approximation to the time-ordered evolution, such as a Magnus expansion \citep{magnus54,blanes09}. The error associated with this interval representation is distinct from that introduced when approximating the resulting matrix exponential, which is done here using a diagonal Pad\'e approximant \citep{pade92,baker96}:
\begin{align}
	\propagator(t,t') \approx \propagator^{(\padeorder)}(t,t') = \left(\sum_{\padeordersum=0}^{\padeorder}c_\padeordersum^{(\padeorder)}\left[(t'-t)\linearmatrix(t,t')\right]^\padeordersum\right)^{-1}\left(\sum_{\padeordersum=0}^{\padeorder}c_\padeordersum^{(\padeorder)}\left[-(t'-t)\linearmatrix(t,t')\right]^\padeordersum\right),
	\label{eq:pade_propagator}
\end{align}
where
\begin{align}
	c_\padeordersum^{(\padeorder)} = \frac{(2\padeorder-\padeordersum)!\,\padeorder!}{(2\padeorder)!\,\padeordersum!\,(\padeorder-\padeordersum)!}.
	\label{eq:pade_coefficients}
\end{align}
The $[\padeorder, \padeorder]$ approximant agrees with the Taylor expansion of the exponential through order $2\padeorder$. The inverse appearing in \cref{eq:pade_propagator} is not calculated explicitly; instead, the propagator is obtained by solving the linear system
\begin{align}
	\left(\sum_{\padeordersum=0}^{\padeorder}c_\padeordersum^{(\padeorder)}\left[(t'-t)\linearmatrix(t,t')\right]^\padeordersum\right)\propagator^{(\padeorder)}(t,t') = \sum_{\padeordersum=0}^{\padeorder}c_\padeordersum^{(\padeorder)}\left[-(t'-t)\linearmatrix(t,t')\right]^\padeordersum
	\label{eq:pade_linear_system}
\end{align}
using Gaussian elimination. Given that the linear operator does not couple distinct wavenumbers, this operation can be performed independently for every wavenumber, and so no global inversion or nonlinear solve is required to obtain \cref{eq:pade_propagator}. The dimension of each linear system is equal to the number of evolved fields. Note that when both the linear matrix and time-step size are fixed, the required propagators can be calculated once at initialisation and reused throughout the simulation, if desired.

The Pad\'e approximation to the propagators therefore provides an implicit treatment of the linear evolution, while the nonlinear term and any explicitly treated forcing are advanced explicitly in the integrating-factor variable. For example, the $[1, 1]$ approximant is
\begin{align}
	\propagator^{(1)}(t,t') = \left[\mathbf{I}+\frac{(t'-t)}{2}\linearmatrix(t,t')\right]^{-1}\left[\mathbf{I}-\frac{(t'-t)}{2}\linearmatrix(t,t')\right],
	\label{eq:crank_nicolson_propagator}
\end{align}
which, for a time-independent linear operator, gives the Crank-Nicolson \citep{crank47} discretisation of the homogeneous linear part of \cref{eq:continuous_equation}.

\subsection{Treatment of the nonlinear term}
\label{sec:nonlinear_term}
As mentioned above, the implementation of the nonlinear terms and forcing comprising \(\rhsexplicit\left[\vfields\right]\) is left to the user. Nevertheless, \flucs provides tools for dealing with nonlinear terms that consist of a number of real-space products. These can, in general, be written in terms of convolutions over the Fourier components of the quantities comprising the products, e.g., for two continuous real-space quantities $\ega$ and $\egb$, we have
\begin{align}
	(\ega \egb)_{\vec{k}} = \ega_\vk \ast \egb_\vk = \sum_{\vk'} \ega_{\vk'} \egb_{\vk - \vk'},
	\label{eq:convolution}
\end{align}
where `$\ast$' denotes the convolution. Rather than evaluating this sum directly --- which would require $\order{\totalmodes^2}$ operations --- one can evaluate the nonlinear products using standard pseudospectral methods \citep{cooley65,orszag72}. Mathematically, these can be expressed as 
\begin{align}
	(\ega \egb)_{\vec{k}} = \totalmodes^{-1} \rtoc\left[\ctor[\ega_\vk] \ctor[\egb_\vk] \right],
	\label{eq:pseudospectral_evaluation}
\end{align}
where $\rtoc$ and $\ctor$ are, respectively, the three-dimensional real-to-complex and complex-to-real fast Fourier transform routines of the NVIDIA cuFFT library, here accessed through CuPy. This combination of inverse transform, pointwise real-space multiplication, and then forward transform requires $\order{\totalmodes\log \totalmodes}$ operations, reducing the asymptotic cost relative to evaluating the convolution. However, the pointwise product in \cref{eq:pseudospectral_evaluation} can contain wavenumbers that fall outside the finite computational grid. These contributions are aliased onto lower wavenumbers by the discrete Fourier transform and must therefore be removed to prevent aliasing errors \citep{orszag71}, requiring dealiasing methods that generally increase the computational overhead. We defer discussion of the various dealiasing methods provided by \fouriersolver to \cref{sec:dealiasing}.

The implementation of \cref{eq:pseudospectral_evaluation} is designed to minimise both the number of CUDA kernel launches and the number of global-memory accesses. All required Fourier-space intermediates, such as spatial derivatives of the fields, are first constructed by a system-specific CUDA kernel. These are then transformed to real space using a batched \ctor cuFFT plan, after which a second system-specific CUDA kernel forms all the required nonlinear real-space intermediates (often products) and, when required, calculates the local real-space contribution to the CFL rate. These real-space intermediates are then transformed back to Fourier space using a batched \rtoc cuFFT plan and combined into the nonlinear term within the relevant time-integration kernel. All intermediate arrays remain in GPU memory throughout this procedure, and the storage for the Fourier- and real-space intermediates may be reused for the resultant terms where they are not required simultaneously. Thus, for a fixed number of both Fourier- and real-space intermediate quantities, each evaluation of the nonlinear operator retains the $\order{\totalmodes\log \totalmodes}$ scaling of the underlying FFTs (see \cref{sec:numerics_performance}).

\subsection{Time step responsibilities}
\label{sec:timestep_loop}
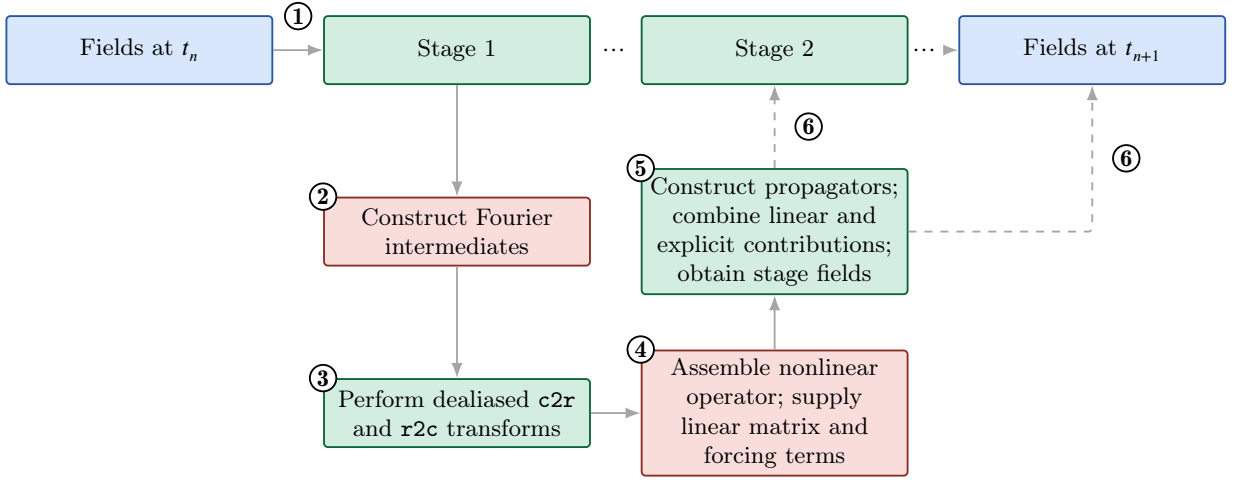
\begin{figure}[pos=h]
	\centering
	\vspace*{0.5cm}
	\scalebox{1.0}{% Requires:
%   \usepackage{xcolor}
%   \usepackage{tikz}
%   \usetikzlibrary{arrows.meta,calc}

\begin{tikzpicture}[
	scale=1.0,
	every node/.style={
		scale=1.0,
		text=black,
		font=\fontfamily{cmr}\selectfont
	}
	]
	
	% ---------------------------------------------------------------------------
	% Parameters
	% ---------------------------------------------------------------------------
	
	% Box sizes
	\def\boxwidth{3.0cm}
	\def\boxtextwidth{3.3cm}
	\def\boxheight{0.9cm}
	
	% Separations
	\def\tophorizontalspacing{4.2cm}
	\def\lowerhorizontalspacing{4.2cm}
	\def\lowerverticalspacing{2.4cm}
	
	% Line properties
	\def\outlinewidth{0.7pt}
	\def\connectionwidth{0.7pt}
	\def\cornerradius{1.5pt}
	\def\arrowlength{2.0mm}
	
	% Colours
	\colorlet{boxcolour}{gray!12}
	\colorlet{outlinecolour}{gray!70}
	\colorlet{connectioncolour}{gray!70}
	
	\definecolor{googleblue}{HTML}{4285F4}
	\definecolor{googlered}{HTML}{DB4437}
	\definecolor{googlegreen}{HTML}{0F9D58}
	
	\colorlet{fieldcolour}{googleblue!20}
	\colorlet{solvercolour}{googlegreen!18}
	\colorlet{systemcolour}{googlered!18}
	
	\colorlet{fieldoutlinecolour}{googleblue!65!black}
	\colorlet{solveroutlinecolour}{googlegreen!65!black}
	\colorlet{systemoutlinecolour}{googlered!65!black}
	
	\newif\ifshownumbers
	\shownumberstrue       % Show numbers
	% \shownumbersfalse    % Hide numbers
	
	% Styles
	\tikzset{
		flow box/.style={
			draw=outlinecolour,
			fill=boxcolour,
			line width=\outlinewidth,
			rounded corners=\cornerradius,
			minimum width=\boxwidth,
			minimum height=\boxheight,
			text width=\boxtextwidth,
			align=center,
			text=black,
			font=\fontfamily{cmr}\selectfont
		},
		field box/.style={
			flow box,
			fill=fieldcolour,
			draw=fieldoutlinecolour
		},
		solver box/.style={
			flow box,
			fill=solvercolour,
			draw=solveroutlinecolour
		},
		system box/.style={
			flow box,
			fill=systemcolour,
			draw=systemoutlinecolour
		},
		timestepper box/.style={
			flow box,
			fill=solvercolour,
			draw=solveroutlinecolour,
			text width={1.0*\boxtextwidth}
		},
		connection/.style={
			draw=connectioncolour,
			line width=\connectionwidth,
			rounded corners=\cornerradius,
			-{Latex[length=\arrowlength]}
		},
		dashed connection/.style={
			connection,
			dashed
		},
		number label/.style={
			circle,
			draw=black,
			fill=white,
			line width=\outlinewidth,
			minimum size=0.35cm,
			inner sep=0pt,
			text=black,
			font=\fontfamily{cmr}\selectfont\bfseries
		}
	}
	
	% ---------------------------------------------------------------------------
	% Boxes
	% ---------------------------------------------------------------------------
	
	% Top row
	\node[field box] (boxone) at (0,0) {Fields at $t_{\tindex}$};
	\node[timestepper box] (boxtwo) at
	($(boxone)+(\tophorizontalspacing,0)$) {Stage 1};
	\node[timestepper box] (boxthree) at
	($(boxtwo)+(\tophorizontalspacing,0)$) {Stage 2};
	\node[field box] (boxfour) at
	($(boxthree)+(\tophorizontalspacing,0)$) {Fields at $t_{\tindex+1}$};
	
	% Lower loop
	\node[system box] (boxfive) at
	($(boxtwo)+(0,-\lowerverticalspacing)$) {Construct Fourier intermediates};
	\node[solver box] (boxsix) at
	($(boxfive)+(0,-\lowerverticalspacing)$) {Perform dealiased \ctor and \rtoc transforms};
	\node[system box] (boxseven) at
	($(boxsix)+(\lowerhorizontalspacing,0)$) {Assemble nonlinear operator; supply linear matrix and forcing terms};
	\node[solver box] (boxeight) at
	($(boxseven)+(0,\lowerverticalspacing)$) {Construct propagators; combine linear and explicit contributions; obtain stage fields};
	
	% ---------------------------------------------------------------------------
	% Main path
	% ---------------------------------------------------------------------------
	
	\draw[connection] (boxone.east) -- (boxtwo.west);
	
	\node[inner sep=0pt] at
	($(boxtwo)!0.5!(boxthree)$) {$\dots$};
	
	\node[inner sep=0pt] (topdots) at
	($(boxthree.east)!0.35!(boxfour.west)$) {$\dots$};
	
	\coordinate (finaljunction) at
	($(boxthree.east)!0.70!(boxfour.west)$);
	
	\draw[connection] (finaljunction) -- (boxfour.west);
	
	% ---------------------------------------------------------------------------
	% Lower loop
	% ---------------------------------------------------------------------------
	
	\draw[connection] (boxtwo.south) -- (boxfive.north);
	\draw[connection] (boxfive.south) -- (boxsix.north);
	\draw[connection] (boxsix.east) -- (boxseven.west);
	\draw[connection] (boxseven.north) -- (boxeight.south);
	
	\draw[dashed connection] (boxeight.north) -- (boxthree.south);
	
	\coordinate (finalreturncorner) at
	(boxfour.south |- boxeight.east);
	
	\draw[dashed connection]
	(boxeight.east)
	-- (finalreturncorner)
	-- (boxfour.south);
	
	% ---------------------------------------------------------------------------
	% Additional numbers
	% ---------------------------------------------------------------------------
	
	\ifshownumbers
	
	% Number above the first connection
	\node[number label] at
	($(boxone.east)!0.5!(boxtwo.west)+(0,0.45cm)$) {1};
	
	% Numbers centred over the top-left corners of boxes
	\node[number label] at (boxfive.north west) {2};
	\node[number label] at (boxsix.north west) {3};
	\node[number label] at (boxseven.north west) {4};
	\node[number label] at (boxeight.north west) {5};
	
	% Number beside the dashed route towards the final box
	\node[number label] at
	($(finalreturncorner)!0.5!(boxfour.south)+(0.45cm,0)$) {6};
	
	% Number beside the dashed connection to box three
	\node[number label] at
	($(boxeight.north)!0.5!(boxthree.south)+(0.45cm,0)$) {6};
	
	\fi
	
\end{tikzpicture}}
	\vspace*{0.5cm}
	\caption{\normalfont A schematic representation of the time step sequence of \fouriersolver. Blue blocks denote the field states, while green and red ones indicate operations coordinated by the solver and supplied by the selected system, respectively. The numbered labels correspond to the steps detailed in the text. The dashed arrows indicate the potential uses of a completed stage field: it is either passed to the next time step stage or, following the final stage, used to update the fields to their end-of-time-step values. Ellipses denote omitted potential intermediate stages.}
	\label{fig:timestep_loop}
\end{figure}

The preceding subsections described the spatial and temporal discretisations employed by \fouriersolver, as well as the numerical operations used to evaluate the nonlinear terms. To clarify the functionality that a system must implement and show how equation-specific operations are incorporated into the general solver machinery, we now describe how these operations are coordinated during a time step, focusing in particular on where responsibility passes between the solver and the selected system. The sequence for advancing from \(t=t_\tindex\) to \(t=t_{\tindex + 1}\) for a system that uses the integrated tools for dealiasing polynomial nonlinearities, illustrated schematically in \cref{fig:timestep_loop}, is as follows:
\begin{enumerate}
	\item \fouriersolver passes the current fields $\vfields(t_\tindex)$ to the selected time-stepper.
	
	\item For each required evaluation of the explicit terms, the time-stepper passes the appropriate stage field and stage time to the system. The system then constructs the required Fourier-space intermediate quantities, such as spatial derivatives of the fields.
	
	\item \fouriersolver applies the selected dealiasing procedure. This comprises the required \ctor transforms, calls to the system kernels that form the real-space nonlinear intermediates, and the corresponding \rtoc transforms.
	
	\item The time-stepper launches a Fourier-space update kernel in which the system assembles the nonlinear operator from the outputs of the preceding operations and supplies any system-specific forcing terms. The system also provides the linear matrix required to construct the propagators. 
	
	\item Using this provided information, \fouriersolver obtains the required linear propagators and combines the linear and explicit contributions according to the selected time-integration scheme.
	
	\item If the time-stepping scheme requires another stage, the resulting stage field is returned to step 2. Otherwise, it is stored as $\vfields(t_{\tindex+1})$, completing the time step.
\end{enumerate}
Thus, \fouriersolver controls the sequence of operations, the Fourier transforms and dealiasing, the construction of the linear propagators, and the time-integration scheme. The system supplies the equation-specific linear matrix, constructs the Fourier- and real-space intermediate quantities required for the nonlinear operator, and assembles these contributions into the explicitly treated right-hand side. It may perform additional operations through optional beginning-of-step, end-of-stage, and end-of-step hooks, allowing for, e.g., projections or constraint enforcement to be applied after every intermediate or final stage. The number and ordering of stage evaluations depend on the selected time-stepper, but the division of responsibility between the system and solver remains unchanged. Consequently, a new system can use the existing numerical machinery without modifying the solver itself.

\section{Dealiasing methods}
\label{sec:dealiasing}
The standard pseudospectral operation provided by \fouriersolver evaluates products of two real-space quantities. We therefore restrict the following discussion to such quadratic products. Consider two represented Fourier modes with DFT frequencies $\vdftindex'$ and $\vdftindex''$ that couple to produce an output at $\vdftindex$. The product returned by the discrete transform may then be written as [cf. \cref{eq:convolution}]
\begin{align}
	(\ega \egb)_\vdftindex = \sum_\valias \sum_{\vdftindex'} \ega_{\vdftindex '} \egb_{\vdftindex - \vdftindex' + \valias}
	\label{eq:convolution_discrete}
\end{align}
where the sum over $\vdftindex'$ now includes only modes represented on the finite grid. Here, $\valias = (s_x, s_y, s_z)$ is the integer-valued aliasing vector whose components satisfy
\begin{align}
	\alias_\rindex = \dftindex' + \dftindex'' - \dftindex,
	\label{eq:alias_vector}
\end{align}
Each $\alias_\rindex$ corresponds to the number of complete DFT periods by which the true product frequency lies outside the represented interval in a given direction $\rindex$. Given that all of the DFT frequencies lie within the principal interval $|\dftindex| \leqslant 1/2$, \cref{eq:alias_vector} implies that $\alias_\rindex \in \{-1, 0, 1\}$. At any collocation point indexed by $\arrayindex$, an integer shift of the DFT frequency changes the Fourier phase by $\exp(2\pi\rmi\arrayindex\alias_\rindex)=1$ (since both $\alias_\rindex$ and $\arrayindex$ are integers). The discrete grid therefore cannot distinguish a mode labelled by $\dftindex$ from one labelled by $\dftindex+\alias_\rindex$  [see \cref{eq:fourier_transform}]. The contributions with $\valias \neq 0$ are therefore aliases and must be removed or cancelled before the nonlinear term is used to time-advance the fields.

The remainder of this section details the various dealiasing methods that are natively supported by \fouriersolver: standard padding (\cref{sec:twothirds_padding}) and phase-shift dealiasing (\cref{sec:phase_shift_methods}), with two possible truncations supported in the latter case. Note that all quantities are here represented using three spatial components, consistent with the implementation of \fouriersolver. For a two-dimensional system, the DFT frequency in the `unused' direction is identically zero, and the dealiasing conditions below apply within the remaining two active directions. For all methods, the dimensions of the Fourier- and real-space arrays remain fixed throughout the \fouriersolver execution. Some array entries may therefore be retained in memory even though the corresponding Fourier modes are excluded from the time advance. Although this increases storage requirements, our testing found the fixed-size representation to be marginally faster because it avoids the cost of repeatedly copying data between differently sized arrays. It also simplifies the implementation by localising the method-specific masking and phase-shift operations within the common dealiasing routines of \fouriersolver, without changing the interface through which systems supply their Fourier- and real-space intermediate quantities. Readers already familiar with the dealiasing methods detailed in \cref{sec:twothirds_padding} and \cref{sec:phase_shift_methods} may wish to skip ahead to \cref{sec:numerics}, working backwards where further clarification is required. 

\subsection{Standard two-thirds dealiasing}
\label{sec:twothirds_padding}
The standard padding method introduced by \citet{orszag71} suppresses aliasing by evolving only a restricted subset of the available Fourier modes while evaluating nonlinear products on a larger transform grid. We refer to the retained modes as the `solved' modes and to the remaining modes, whose Fourier coefficients are fixed to zero, as the `padded' modes. Suppose that, in a given direction $\rindex$, the solved modes satisfy $\abs{\dftindex}\leqslant\dftindex^{\max}$. A quadratic product of two solved modes can then generate a true output frequency with magnitude no greater than
\begin{align}
	\abs{\dftindex'+\dftindex''}\leqslant2\dftindex^{\max}.
\end{align}
Conversely, an aliased contribution into the solved range has $\alias_\rindex\neq0$, so \cref{eq:alias_vector} gives
\begin{align}
	\abs{\dftindex'+\dftindex''}=\abs{\dftindex+\alias_\rindex}\geqslant1-\dftindex^{\max},
\end{align}
where we have used the fact that $\alias_\rindex$ is a nonzero integer and therefore $\abs{\alias_\rindex}\geqslant1$. The true product frequencies and their aliased images cannot overlap within the solved range provided that
\begin{align}
	2\dftindex^{\max}<1-\dftindex^{\max}\quad\Rightarrow\quad\dftindex^{\max}<\frac{1}{3}.
\end{align}
Thus, in each direction, the solved modes occupy two thirds of the complete DFT-frequency interval, giving the usual `two-thirds' dealiasing rule. Equivalently, the nonlinear transforms are evaluated on a (padded) grid whose resolution approaches three halves of the solved resolution.

\begin{figure}
	\centering
%	\vspace*{0.5cm}
	\includegraphics[width=1.0\textwidth]{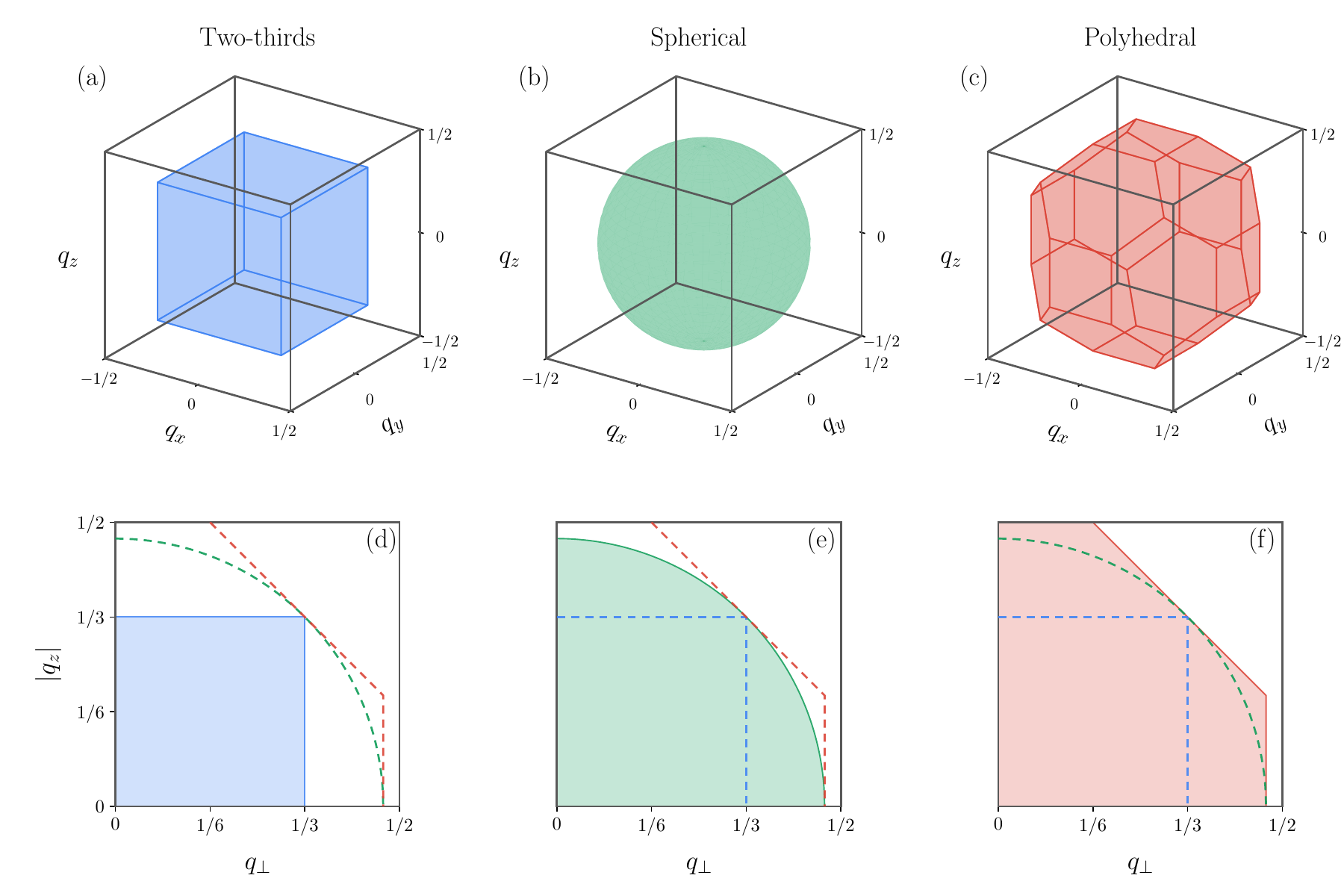}
%	\vspace*{0.5cm}
	\caption{\normalfont A comparison of the solved regions in DFT-frequency space resulting from the various dealiasing methods supported by \fouriersolver: standard two-thirds padding (blue, left column), phase-shift with spherical truncation (green, middle column), and phase-shift with polyhedral truncation (red, right column). The top row shows the solved regions in the full $\vdftindex = (\dftindex[x], \dftindex[y], \dftindex[z])$ space [see \cref{eq:dftindices}] in the different colours, while the bottom row shows the solved regions in the $(\dftindex[\perp], \dftindex[z])$ plane, where $\dftindex[\perp]^2 = {\dftindex[x]^2 + \dftindex[y]^2}$. Note that, in the latter, only the largest axisymmetric subset of the solved modes is shown. The dashed lines are the boundaries of the various truncations to aid visual comparison. It is clear that the phase-shift method with polyhedral truncation retains the largest region of solved modes.}
	\label{fig:dealiasing_regions}
\end{figure}

Within \fouriersolver, the user specifies the resolution $\resolution$ of the full computational grid in each active spatial direction. The solver then determines the corresponding solved-mode cutoff in $\dftindex$, producing a rectangular region of solved modes [see \cref{fig:dealiasing_regions}(a, d)]. All fields and nonlinear intermediates are stored on the full grid: the solved modes are time-advanced, while the padded modes are set to zero during each stage update. The Fourier transforms used to evaluate nonlinear products therefore operate directly on the full, fixed-size arrays.

\subsection{Phase-shift methods}
\label{sec:phase_shift_methods}
Phase-shift dealiasing \citep{patterson71} exploits the difference between the response of the physical and aliased contributions to a translation of the real-space collocation grid \cref{eq:realspace_points}. Consider a set of grids indexed by $j$, with collocation points given by
\begin{align}
	r_\rindex^{(j)} = \frac{\left(\arrayindex + \delta \arrayindex^{(j)}\right)\domain}{\resolution},
	\label{eq:realspace_points_shifted}
\end{align}
where $\delta \arrayindex^{(j)}$ is the shift of the grid relative to \cref{eq:realspace_points} in the direction $\rindex$, measured in units of the corresponding grid spacing. Under this translation, a mode with frequency $\dftindex$ acquires a phase $\exp(2 \pi \rmi \dftindex \delta \arrayindex^{(j)} )$. Evaluating the quadratic product on the set of shifted grids, phase-shifting the results back to the original grid, and performing a weighted average over the results gives
\begin{align}
	(\ega \egb)_\vdftindex = \sum_\valias \left[\sum_j w_j \exp\left(2\pi\rmi\sum_{\rindex}\alias_\rindex\delta \arrayindex^{(j)}\right)\right]  \sum_{\vdftindex'} \ega_{\vdftindex '} \egb_{\vdftindex - \vdftindex' + \valias},
	\label{eq:convolution_phase_shift}
\end{align}
where $w_j$ are arbitrary weights satisfying $\sum_j w_j = 1$. It is evident from \cref{eq:convolution_phase_shift} that the physical contributions, for which $\valias=0$, are unaffected by the shifts, whereas each aliased contribution acquires a phase determined by $\alias_\rindex$. The shifts and weights may then be chosen such that the expression in the square brackets in \cref{eq:convolution_phase_shift} gives zero for selected aliasing vectors, cancelling the corresponding aliased contributions. 

The implementation in \fouriersolver uses two equally weighted evaluations: one on the original grid and one shifted by half a grid cell in every active direction, such that \cref{eq:convolution_phase_shift} becomes:
\begin{align}
	(\ega \egb)_\vdftindex = \frac{1}{2}\sum_\valias \left[1 + \exp\left(\pi\rmi\sum_{\rindex}\alias_\rindex\right)\right]  \sum_{\vdftindex'} \ega_{\vdftindex '} \egb_{\vdftindex - \vdftindex' + \valias},
	\label{eq:convolution_phase_shift_equal}
\end{align}
All aliases for which $\sum_\rindex\alias_\rindex$ is odd therefore vanish. Because each nonzero component of $\valias$ is either $-1$ or $1$, this occurs whenever $\valias$ has an odd number of nonzero components. Thus, all aliases are cancelled in one dimension. In two dimensions, aliases with one nonzero component are cancelled, while in three dimensions those with either one or three nonzero components are cancelled. The remaining aliases have exactly two nonzero components and take the values
\begin{align}
	(\pm1,\pm1,0),\quad(\pm1,0,\pm1),\quad\text{and}\quad(0,\pm1,\pm1).
	\label{eq:phase_shift_aliases}
\end{align}
These remaining contributions must be removed by an additional truncation of the solved modes. The two truncations provided by \fouriersolver are detailed in the following subsections.

\subsubsection{Spherical truncation}
\label{sec:spherical_truncation}
Suppose that all solved modes are chosen to satisfy $|\vdftindex| < \dftindex[]^{\max}$. It then follows from \cref{eq:alias_vector} that
\begin{align}
	|\valias| \leqslant |\vdftindex' + \vdftindex'' + \vdftindex|\leqslant 3 \dftindex[]^{\max}.
	\label{eq:spherical_truncation_condition}
\end{align}
The remaining aliasing vectors in \cref{eq:phase_shift_aliases} all satisfy $|\valias| = \sqrt{2}$. Then, solved modes cannot produce any residual aliases if
\begin{align}
	3 \dftindex[]^{\max} < \sqrt{2} \quad \Rightarrow \quad \dftindex[]^{\max} < \frac{\sqrt{2}}{3}.
	\label{eq:spherical_truncation_radius}
\end{align}
This truncation, introduced by \citet{patterson71}, produces a spherical region of solved modes [see \cref{fig:dealiasing_regions}(b, e)]. Because this cutoff depends only on $|\vdftindex|$, it treats all directions equally in (dimensionless) DFT-frequency space. It is thus well suited to systems whose smallest dynamically relevant length scales occupy comparable fractions of the grid spacing in each direction. When these grid spacings are equal, the solved region is also spherical in physical wavenumber space.

\subsubsection{Polyhedral truncation}
\label{sec:polyhedral_truncation}
In strongly anisotropic systems, such as those frequently encountered in the context of magnetised plasmas, such a spherical truncation may discard physically relevant modes that could be retained without introducing residual aliases. Consider one of the aliases \cref{eq:phase_shift_aliases} that remains after the phase-shift procedure and has nonzero components in directions $\rindex$ and $\rindexalt$, such that $|\alias_\rindex| = |\alias_\rindexalt| = 1$. Suppose that every solved mode satisfies the pairwise bound
\begin{align}
	|\dftindex[\rindex]| + |\dftindex[\rindexalt]| \leqslant \dftindex[]^{\max},
	\label{eq:polyhedral_truncation_condition}
\end{align}
for every pair of active directions. For an aliased contribution to a solved output mode, a re-arrangement of \cref{eq:alias_vector} gives:
\begin{align}
	|\dftindex + \alias_\rindex|+|\dftindex[\rindexalt] + \alias_\rindexalt| \geqslant |\alias_\rindex| + |\alias_\rindexalt| - |\dftindex| - |\dftindex[\rindexalt]| \geqslant 2 - \dftindex[]^{\max}.
	\label{eq:polyhedral_inequality_output}
\end{align}
Simultaneously, two solved input modes satisfy:
\begin{align}
	|\dftindex' + \dftindex''|+|\dftindex[\rindexalt]' + \dftindex[\rindexalt]''| \leqslant 2\dftindex[]^{\max}.
	\label{eq:polyhedral_inequality_input}
\end{align}
The definition of the aliasing vector \cref{eq:alias_vector} requires that the quantities bounded in both \cref{eq:polyhedral_inequality_output} and \cref{eq:polyhedral_inequality_input} be equal. Such aliases are therefore impossible when 
\begin{align}
	2\dftindex[]^{\max} < 2 - \dftindex[]^{\max} \quad \Rightarrow \quad \dftindex[]^{\max} < \frac{2}{3}.
	\label{eq:polyhedral_truncation_radius}
\end{align}
Applying this condition to each pair of directions gives the solved region:
\begin{align}
	|\dftindex[x]| + |\dftindex[y]| < \frac{2}{3}, \quad |\dftindex[x]| + |\dftindex[z]| < \frac{2}{3},\quad |\dftindex[y]| + |\dftindex[z]| <  \frac{2}{3}.
	\label{eq:polyhedral_truncation_region}
\end{align}
The intersection of these pairwise bounds forms the polyhedral truncation \citep{rogallo81,canuto88} [see \cref{fig:dealiasing_regions}(c, f)]. It contains the spherical region defined by \cref{eq:spherical_truncation_radius} and is thus in some sense the `optimal' truncation, in that it retains modes up to the boundary permitted by each possible aliasing vector. That said, the additional modes retained are useful only when the dynamically relevant spectral support in a given system of equations lies within the resulting solved region. As anticipated above, this can occur naturally in strongly anisotropic systems, for which modes may extend to large DFT frequencies in one direction while remaining at relatively smaller frequencies in the other regions, occupying the additional regions retained near the coordinate axes when compared with the spherical truncation. This means that polyhedral truncation can then contain all dynamically relevant modes that would be retained on the full grid while the spherical truncation would discard some of them (see, e.g., \cref{fig:dealiasing_krehm_spectra_total_grid}), though this is of course not guaranteed for all anisotropic systems.

\section{Numerical results}
\label{sec:numerics}
With the algorithm and dealiasing schemes of \fouriersolver established, we now present a series of numerical studies to test its implementation. These calculations use the three systems currently available within the project organisation: isothermal KREHM (\texttt{KREHMFourier}), cold-ion ITG (\texttt{ColdITG2DFourier}), and collisional ETG (\texttt{CollisionalETGFourier}). Details about these systems --- including their governing equations, normalisation, and validation against known results --- are provided in \cref{app:systems}. They are otherwise unimportant for the present studies, which test either asymptotic scalings or the relative performance of different \fouriersolver methods, rather than the details of any particular physical setup. Simulation parameters are therefore given only where relevant. Applying these tests to systems with distinct governing equations and physical behaviour helps to establish that the results reflect the numerical methods of \fouriersolver rather than peculiarities of any individual model.

\subsection{Convergence of Pad\'e approximants}
\label{sec:numerics_pade}

\begin{figure}
	\centering
	%	\vspace*{0.5cm}
	\includegraphics[width=\figscale\textwidth]{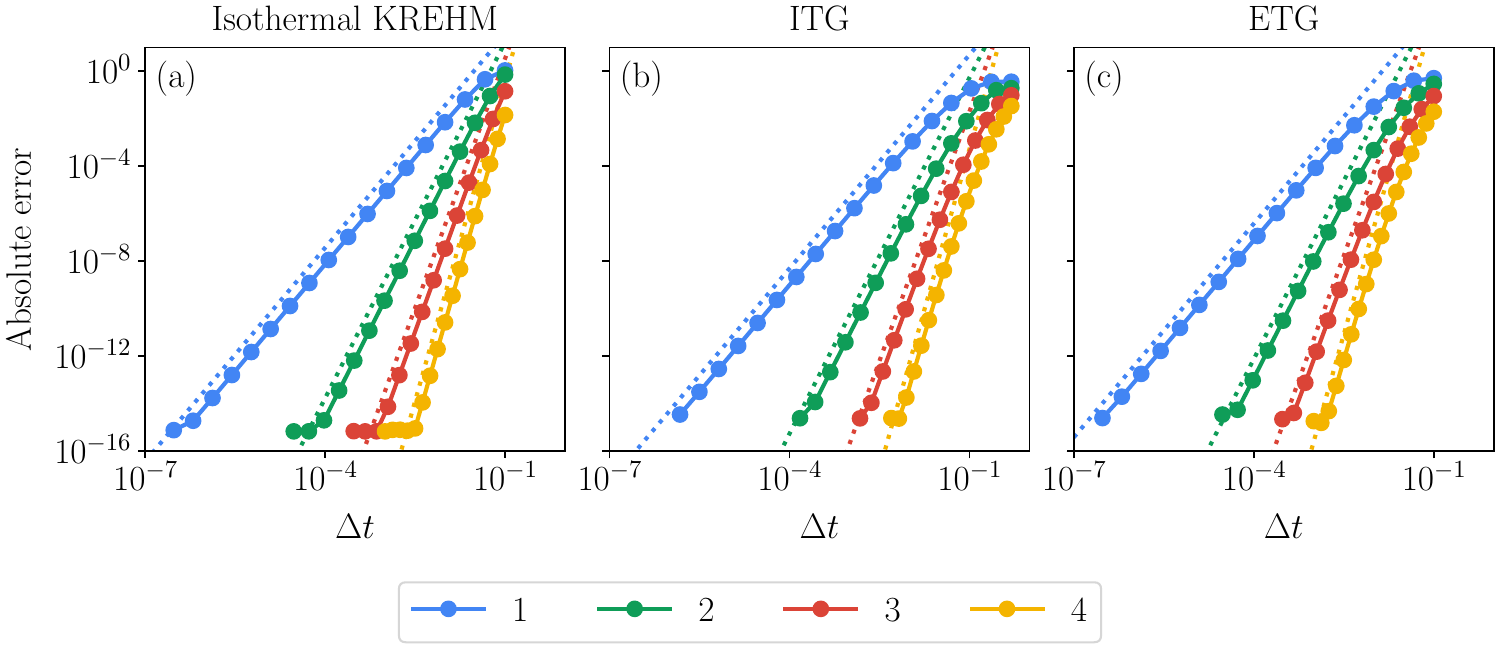}
	%	\vspace*{0.5cm}
	\caption{\normalfont Convergence of the linear propagator \cref{eq:pade_propagator} using double-precision floating-point numbers for the (a) isothermal KREHM, (b) ITG, and (c) ETG systems. The markers show the measured propagator error \cref{eq:propoagator_error} at a given time step size $\dt$. The colours indicate the order $\padeorder$ of the Pad\'e approximant, while the dotted lines show the expected $\dt^{2 \padeorder + 1}$ scaling. }
	\label{fig:propagator_convergence}
\end{figure}

The order $[\padeorder,\padeorder]$ Pad\'e approximation to the propagator \cref{eq:pade_propagator} agrees with the Taylor expansion of the exponential through order $2\padeorder$, meaning that the error in the propagator over an interval of length $\dt$ will be $\order{\dt^{2\padeorder+1}}$. To test this convergence, we construct the numerical linear propagator $\propagator^{(\padeorder)}$ used by the CUDA kernels and calculate its eigenvalues for each resolved Fourier wavenumber. We then calculate the eigenvalues of the exact propagator $\propagator$ via exponentiation of the eigenvalues of the linear matrix. The maximum over all wavenumbers of the absolute difference between these two sets of propagator eigenvalues then gives us an estimate of the error associated with the order $\padeorder$ diagonal Pad\'e approximant:
\begin{align}
	\max_{\vec{k}} \abs{\text{eig} \left[\propagator^{(\padeorder)} \right] - \text{eig} \left[\propagator\right]},
	\label{eq:propoagator_error}
\end{align}
where $|\dots |$ denotes the absolute value. While the scaling of \cref{eq:propoagator_error} with $\dt$ depends only on $\padeorder$, its absolute value can nevertheless depend on the spatial resolution due to the latter's effect on the largest resolved linear frequencies and damping rates. We thus perform a scan in the (normalised) time step size $\dt$ at fixed spatial resolution ($\totalmodes = 64^3 = 512^2$) for all systems. The results of this are shown in \cref{fig:propagator_convergence}. In all cases, the measured error \cref{eq:propoagator_error} follows the expected scaling at sufficiently small values of $\dt$, before eventually becoming limited by the double-precision numerical noise floor. It is worth noting that the fourth-order approximant encounters this floor at a time step of $\dt \lesssim 10^{-3}$. For lower values of $\dt$, the truncation error of the Pad\'e approximation itself is smaller than the floating-point errors involved in constructing and diagonalising the propagator, meaning that the numerical propagator is therefore effectively indistinguishable from an exact representation in this regime. 

\subsection{Convergence of time-stepping schemes}
\label{sec:numerics_timesteppers}
\begin{figure}
	\centering
	%	\vspace*{0.5cm}
	\includegraphics[width=\figscale\textwidth]{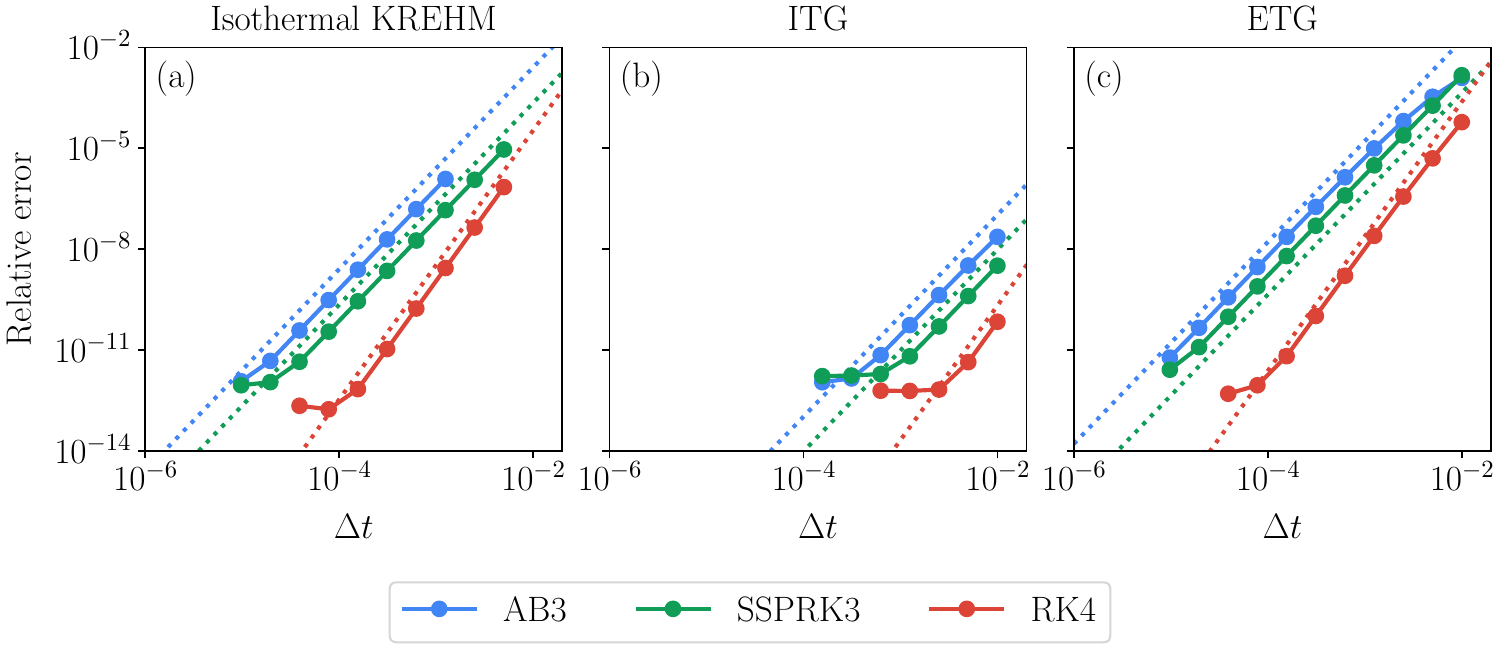}
	%	\vspace*{0.5cm}
	\caption{\normalfont Convergence of the time-stepping schemes using double-precision floating-point numbers for the (a) isothermal KREHM, (b) ITG, and (c) ETG systems. The markers show the measured relative error \cref{eq:timestepper_relative_error} at a given time step size $\dt$. The colours indicate the particular time-stepping scheme, while the dotted lines show the expected $\dt^{\timestepperorder}$ scaling, with $\timestepperorder$ the scheme order.}
	\label{fig:timestepper_convergence}
\end{figure}

The preceding section verifies the construction of the individual linear propagators. Their correct incorporation into the available time-stepping schemes is a separate requirement: since the integral on the right-hand side of \cref{eq:discrete_equation_exact} contains the propagator-weighted explicit terms, each scheme must apply the appropriate propagator over every intermediate-stage or full-step interval. Recovering the expected order of convergence is therefore a stringent test of the numerical implementation, as errors in the stage times and/or propagation intervals would generally spoil the observed convergence. This was tested as follows. Each of the systems was advanced using the chosen time-stepping scheme (AB3, SSPRK3, or RK4) until it reached a fully-developed nonlinear state, at which point the double-precision complex field data $\vfields$ was saved (two-thirds dealiasing was used in all cases). We then performed a series of restarts from this data, each evolving the system to some fixed time using successively smaller time steps.\footnote{Particular care must be taken in the case of AB3. While both SSPRK3 and RK4 are single-step methods, requiring no time history from previous steps, AB3 requires time history from the two previous steps (see \cref{app:ab3}), and therefore cannot be trivially restarted from field data saved at a single time. For the purposes of this convergence study, the first two steps were taken using RK4, and the resulting data was used to self-consistently calculate the correct AB3 time history that must be used for its third step, after which the time-stepping scheme proceeds as usual. This `bootstrap' procedure ensures the third-order convergence of AB3, as is clear from \cref{fig:timestepper_convergence}.} The solution obtained using the smallest time step was then taken to be the `reference' solution for a given time-stepping scheme. The resulting relative error was then defined as:
\begin{align}
	\left(\frac{ \sum_\vk\left|\vfields(\dt) - \vfields(\min[\dt]) \right|^2}{ \sum_\vk\left|\vfields(\min[\dt]) \right|^2 } \right)^{1/2},
	\label{eq:timestepper_relative_error}
\end{align}
where the fields $\vfields$ are evaluated at the final time. For a time-stepping scheme of order $\timestepperorder$, this error \cref{eq:timestepper_relative_error} will be $\order{\dt^{\timestepperorder}}$; this scaling is recovered by our implementation, as can be seen in \cref{fig:timestepper_convergence}. At sufficiently small time steps, the convergence error \cref{eq:timestepper_relative_error} becomes comparable to the accumulated floating-point differences between a given simulation and its numerical reference solution, causing the measured error to approach some numerical floor rather than continuing to decrease with the expected scaling. Both the value of this floor and the time step at which it is encountered may vary between systems due to the differences in their nonlinear trajectories, as can be seen in \cref{fig:timestepper_convergence}.

\begin{figure}
	\centering
	%	\vspace*{0.5cm}
	\includegraphics[width=\figscale\textwidth]{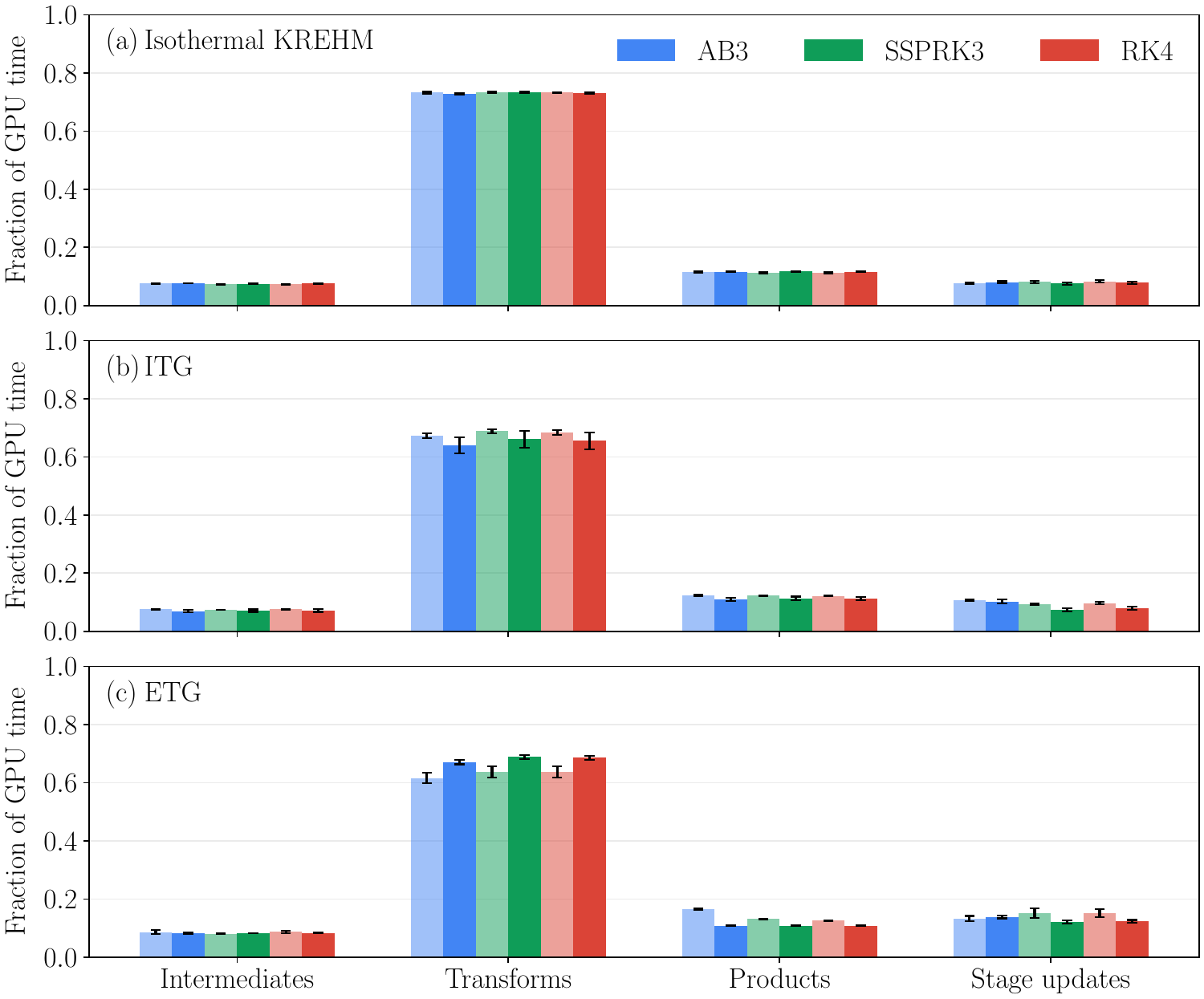}
	%	\vspace*{0.5cm}
	\caption{\normalfont The average fraction of GPU kernel execution time spent in various operations for the (a) isothermal KREHM, (b) ITG, and (c) ETG systems. In terms of the time step loop shown in \cref{fig:timestep_loop}: intermediates corresponds to step 2; transforms and products, to step 3; and stage updates, to steps 4 and 5. The colours indicate the time-stepping scheme (these are the same as in \cref{fig:timestepper_convergence}), with lighter and darker opacities denoting single and double precision, respectively. The error bars indicate the standard deviation in the average over different total numbers of modes $\totalmodes$. All other operations, including calculation of the linear propagators \cref{eq:pade_propagator}, take less than $2\%$ of the kernel execution time.}
	\label{fig:timestepper_gpu_kernel_fractions}
\end{figure}

\begin{figure}
	\centering
	%	\vspace*{0.5cm}
	\includegraphics[width=\figscale\textwidth]{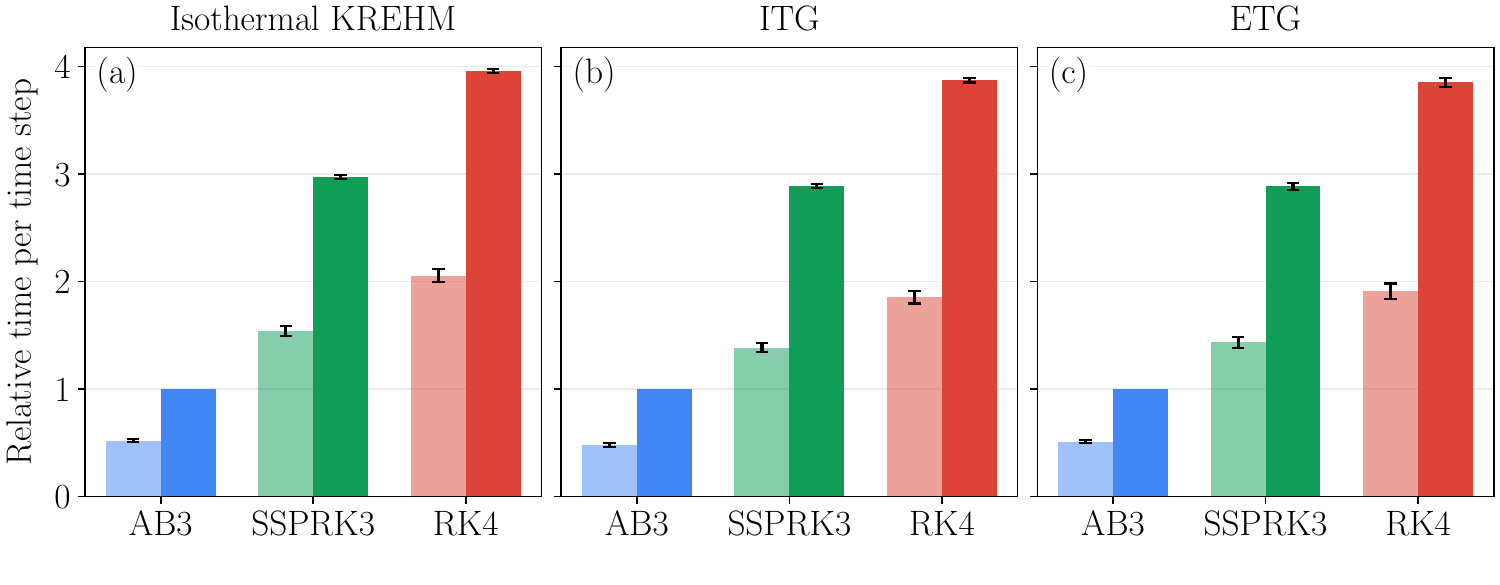}
	%	\vspace*{0.5cm}
	\caption{\normalfont The average wall time per time step relative to that of the double-precision AB3 time-stepping scheme for (a) isothermal KREHM, (b) ITG, and (c) ETG systems. The colours indicate the time-stepping scheme (these are the same as in \cref{fig:timestepper_convergence}), with the lighter and darker opacities denoting single and double precision, respectively. The error bars indicate the standard deviation in the average over different total numbers of modes $\totalmodes$.}
	\label{fig:timestepper_time_per_time_step}
\end{figure}

Another consistency check of the time-stepping implementation can be obtained by measuring the time taken per time step for each scheme. The performance of \fouriersolver is dominated by the evaluation of the explicit terms appearing on the right-hand side of \cref{eq:discrete_equation_exact}, assuming that the total number of modes $\totalmodes$ is sufficiently large. To demonstrate this, we measured the wall time associated with $10^4$ steps of each of the time-stepping schemes across a range of $\totalmodes > 2 \times 10^6$, and, using NVIDIA Nsight Systems, computed the fraction of the GPU kernel execution time taken up by the various operations comprising the time step loop (see \cref{fig:timestep_loop}). The average results of this are shown in \cref{fig:timestepper_gpu_kernel_fractions}, from which it is evident that the majority of performing a time step is spent evaluating the right-hand side explicit terms (and, within each such evaluation, by the Fourier transforms). AB3 requires a single evaluation of these terms, while SSPRK3 and RK4, being multi-stage methods, require three and four evaluations, respectively. Thus, we expect to find that the measured time per time step for SSPRK3 and RK4 is approximately three and four times larger than that for AB3, respectively. This relative time per time step is shown in \cref{fig:timestepper_time_per_time_step}, which behaves consistently with the order of the schemes. We also note that single precision is nearly twice as fast as double precision on the NVIDIA A100-SXM4-40GB GPUs that were used to obtain the results of this section.  

\begin{figure}
	\centering
	%	\vspace*{0.5cm}
	\includegraphics[width=0.337\textwidth]{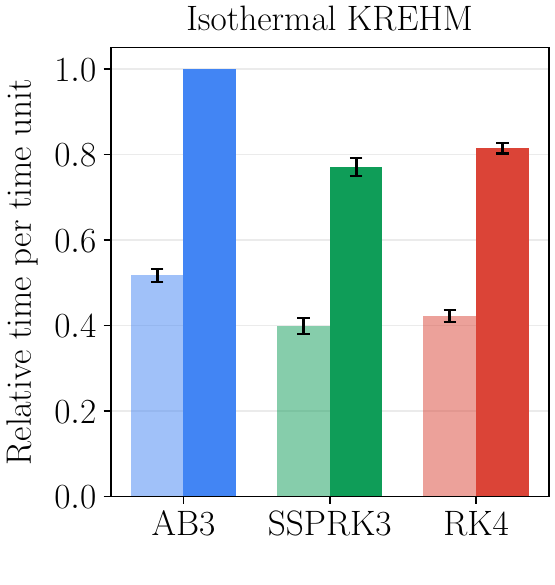}
	%	\vspace*{0.5cm}
	\caption{\normalfont The average wall time per normalised (physical) time unit relative to that of the AB3 time-stepping scheme for the isothermal KREHM system. The details of the figure are otherwise the same as in \cref{fig:timestepper_time_per_time_step}.}
	\label{fig:timestepper_time_per_time_unit}
\end{figure}

The time per time step does not, however, provide a complete measure of computational efficiency. Since the propagators alleviate much of the stability restriction imposed by the linear terms, the time-step size is generally limited by the explicitly evaluated nonlinear terms, imposed here through a system-specific Courant-Friedrichs-Lewy (CFL)  condition \citep{courant67}; see \cref{app:timestepping_algorithms} for further details. The larger stability regions of SSPRK3 and RK4 may therefore permit significantly larger time-step sizes than AB3, potentially offsetting their greater cost per step. The relevant measure of efficiency is thus the wall time required to advance through a fixed interval of simulated time, which depends strongly on the nonlinear dynamics of the system. As an example, we measured the wall time required to evolve a particular setup of the isothermal KREHM system to a fixed final time using each time-stepping scheme across a range of $\totalmodes > 2 \times 10^6$. Both SSPRK3 and RK4 required at least 20\% less wall time than AB3 for this problem and particular choice of CFL conditions (1.0 and 1.3, respectively), as shown in \cref{fig:timestepper_time_per_time_unit}. This ordering is not universal, however, and the optimal time-stepping scheme must therefore be determined for each application of \fouriersolver.
	
\subsection{Comparison of dealiasing methods}
\label{sec:numerics_dealiasing}
\begin{figure}
	\centering
	%	\vspace*{0.5cm}
	\includegraphics[width=0.592\textwidth]{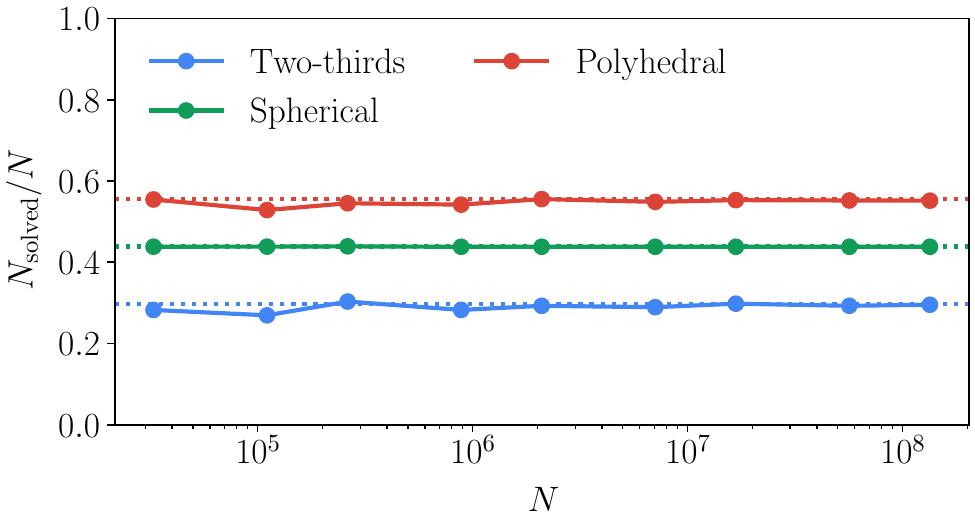}
	%	\vspace*{0.5cm}
	\caption{\normalfont The fraction of solved modes $\solvedmodes/\totalmodes$ as a function of the total number of modes $\totalmodes$ for the three-dimensional isothermal KREHM system. The colours distinguish the dealiasing methods, while the horizontal dotted lines show the corresponding theoretical limits from \cref{eq:dealiasing_solved_modes} for $\dimension=3$: two-thirds $\approx 0.296$; spherical $\approx 0.439$; and polyhedral $\approx 0.556$.}
	\label{fig:dealiasing_solved_modes_fraction}
\end{figure}

The choice of dealiasing method --- either the standard two-thirds (\cref{sec:twothirds_padding}) method or phase-shift dealiasing with a spherical (\cref{sec:spherical_truncation}) or polyhedral (\cref{sec:polyhedral_truncation}) truncation --- determines the number of solved modes, which we henceforth denote by $\solvedmodes$. For a system of dimension $\dimension \in \{2, 3\}$, the large-$\totalmodes$ solved-mode fraction is
\begin{align}
	\frac{\solvedmodes}{\totalmodes} \approx
	\begin{cases}
		\displaystyle \left(\frac{2}{3}\right)^\dimension ,
		& \text{two-thirds},\\[1ex]
		\displaystyle  \frac{\pi^{\dimension/2}}{\Gamma(1+\dimension/2)} \left(\frac{\sqrt{2}}{3}\right)^\dimension,
		& \text{spherical},\\[1ex]
		\displaystyle 2 \left(\frac{2}{3}\right)^\dimension - \left(\frac{1}{3}\right)^\dimension,
		& \text{polyhedral},
	\end{cases}
	\label{eq:dealiasing_solved_modes}
\end{align}
where $\Gamma$ is the standard gamma function \citep{abramowitz72}. By way of example, \cref{fig:dealiasing_solved_modes_fraction} shows the measured fraction for the three-dimensional isothermal KREHM system; this fraction depends only on the grid and choice of dealiasing scheme, rather than on the specific system. The measurements approach the limits in \cref{eq:dealiasing_solved_modes} as $\totalmodes$ increases; the deviations at smaller $\totalmodes$ arise because the number of discrete Fourier modes satisfying each truncation condition differs from the corresponding continuous-volume estimate. Consistent with \cref{fig:dealiasing_regions}, polyhedral dealiasing retains the greatest number of solved modes at fixed $\totalmodes$. Mode count alone, however, does not determine the effective resolution: the additional modes are useful only if they occupy dynamically relevant regions of wavenumber (or DFT-frequency) space.

\begin{figure}
	\centering
	%	\vspace*{0.5cm}
	\includegraphics[width=\figscale\textwidth]{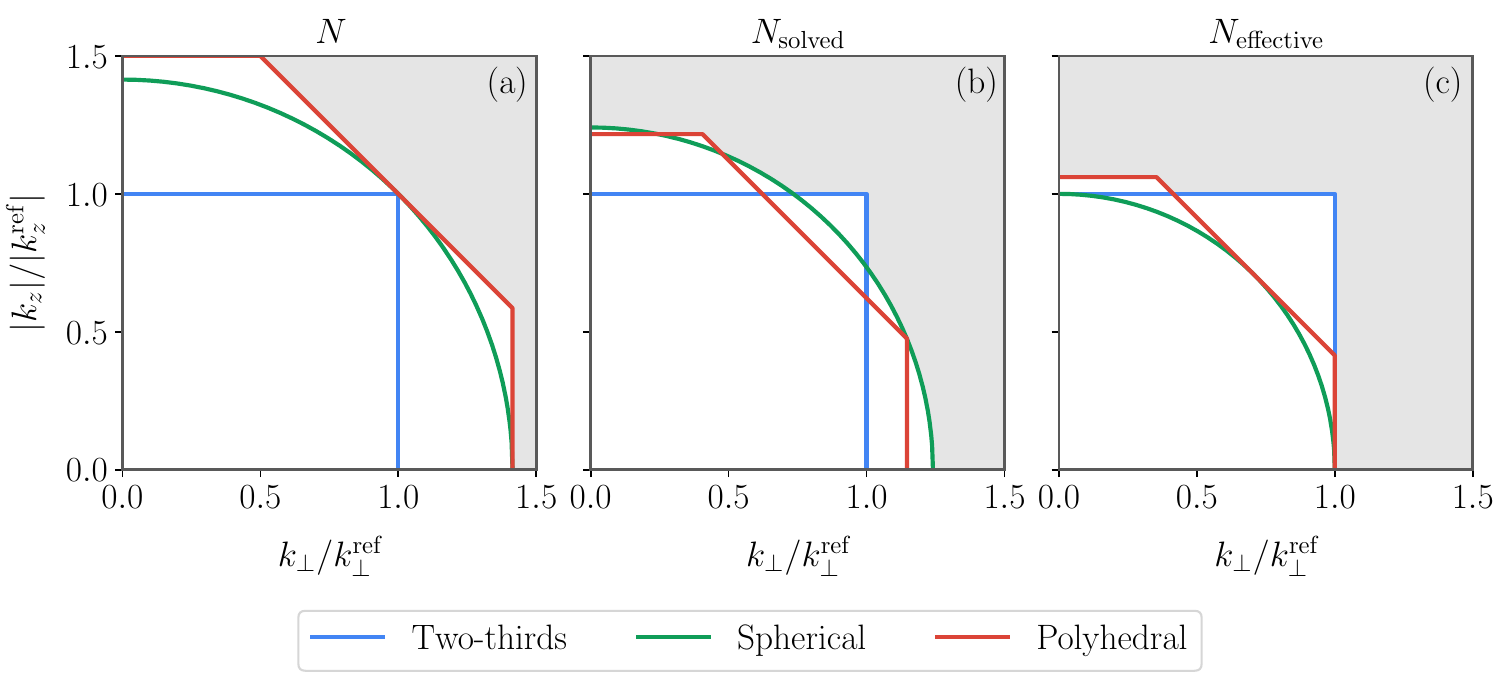}
	%	\vspace*{0.5cm}
	\caption{\normalfont A comparison of the solved regions in wavenumber space resulting from the various dealiasing methods at fixed (a) total number of modes $\totalmodes$, (b) number of solved modes $\solvedmodes$, and (c) effective number of modes $\effectivemodes$. Like in \cref{fig:dealiasing_regions}, the horizontal axis indicates fully retained axisymmetric subsets with $k_x^2 + k_y^2 = \kperp^2$. The coloured lines indicate the truncation boundaries, while the grey region contains modes excluded by all three methods. The wavenumbers are normalised by $k_z^\text{ref}$ and $\kperp^\text{ref}$, the maximum parallel and perpendicular wavenumbers, respectively retained by the two-thirds truncation.}
	\label{fig:dealiasing_regions_comparison}
\end{figure}

To quantify isotropic spectral coverage, we define an effective number of modes $\effectivemodes$ through an equivalent rectangular grid whose largest inscribed sphere has the same radius as that of the actual solved region, where distances are here measured in normalised DFT-frequency coordinates. In the limit of large resolution in every active dimension,
\begin{align}
	\frac{\effectivemodes}{\totalmodes} \approx
	\begin{cases}
		\displaystyle \left(\frac{2}{3}\right)^\dimension,
		& \text{two-thirds},\\[1ex]
		\displaystyle  \left(\frac{2 \sqrt{2}}{3}\right)^\dimension,
		& \text{spherical or polyhedral}.
	\end{cases}
	\label{eq:dealiasing_effective_modes}
\end{align}
For two-thirds dealiasing, this equivalent grid clearly coincides with the actual solved grid, viz., $\effectivemodes = \solvedmodes$ [cf. \cref{eq:dealiasing_solved_modes}]. For the phase-shift methods, it instead provides a common measure of isotropic spectral coverage, without implying that all modes of the equivalent rectangular grid are retained (they are not). \Cref{fig:dealiasing_regions_comparison} compares the solved wavenumber regions in each of the dealiasing methods at fixed $\totalmodes$, $\solvedmodes$, and $\effectivemodes$. At fixed $\effectivemodes$, all three methods contain the same largest isotropic region, although the phase-shift methods retain fewer modes near the diagonals of wavenumber space. They may therefore provide similar effective resolution for sufficiently isotropic dynamics, while using different total and solved mode counts. Whether the additional modes retained by a particular truncation improve the resolution depends on the spectral support of the dynamics, which we examine below. Throughout the remainder of this section, we show numerical results using the isothermal KREHM system, whose dynamics provide the clearest demonstration of the differences between the dealiasing methods. The performance benchmarks use the SSPRK3 time-stepping scheme, with all simulations performed on NVIDIA A100-SXM4-80GB GPUs. The measured statistics are specific to this configuration and should therefore be regarded as illustrative rather than universal: they depend on the system, choice of time-stepping scheme, precision, and hardware.

\begin{figure}
	\centering
	%	\vspace*{0.5cm}
	\includegraphics[width=0.6825\textwidth]{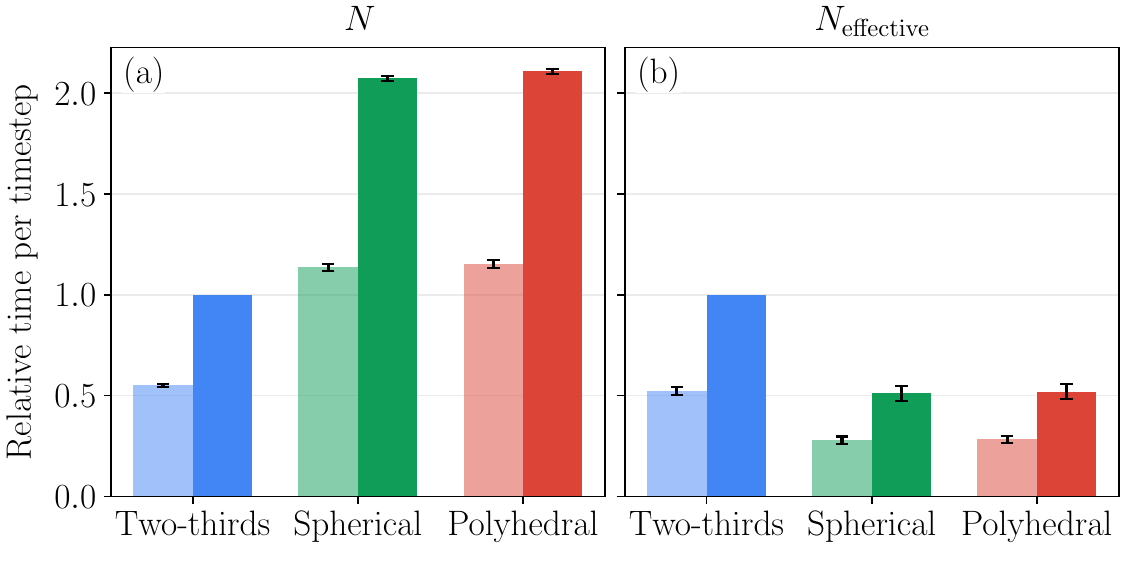}
	%	\vspace*{0.5cm}
	\caption{\normalfont The average wall time per time step relative to that of the two-thirds dealiasing method for fixed (a) total number of modes $\totalmodes$ and (b) effective number of modes $\effectivemodes$. The colours indicate the particular method, with the lighter and darker opacities denoting single and double precision, respectively. The error bars indicate the standard deviation in the average over different total numbers of modes $\totalmodes$.}
	\label{fig:dealiasing_time_per_timestep}
\end{figure}

\begin{figure}
	\centering
	%	\vspace*{0.5cm}
	\includegraphics[width=0.6825\textwidth]{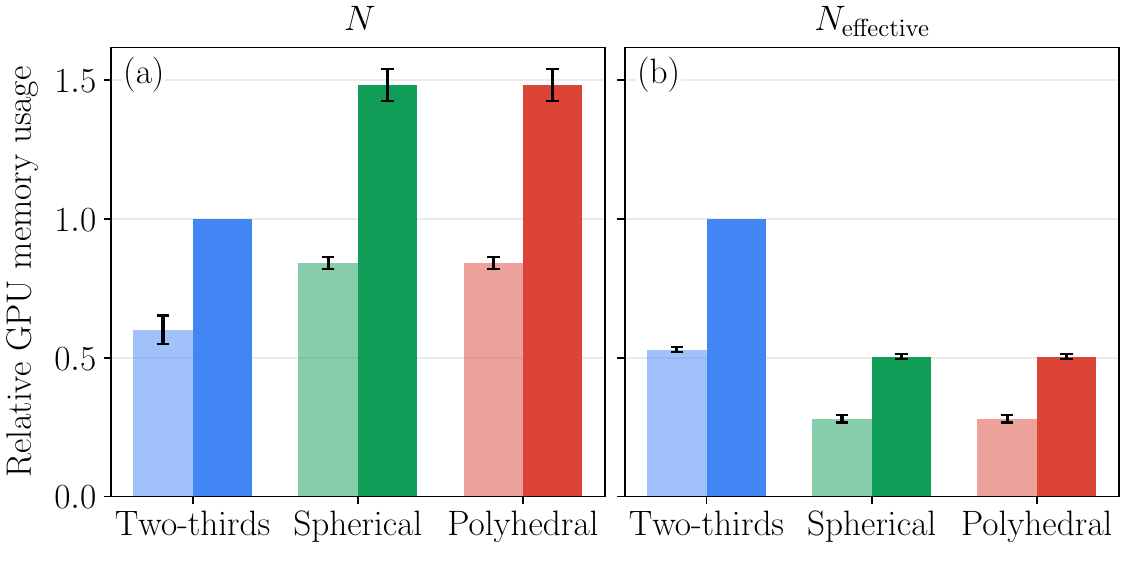}
	%	\vspace*{0.5cm}
	\caption{\normalfont The average GPU memory usage relative to that of the two-third dealiasing method for fixed (a) total number of modes $\totalmodes$ and (b) effective number of modes $\effectivemodes$. The details of the figure are otherwise the same as in \cref{fig:dealiasing_time_per_timestep}.}
	\label{fig:dealiasing_memory}
\end{figure}

\begin{figure}
	\centering
	%	\vspace*{0.5cm}
	\includegraphics[width=\figscale\textwidth]{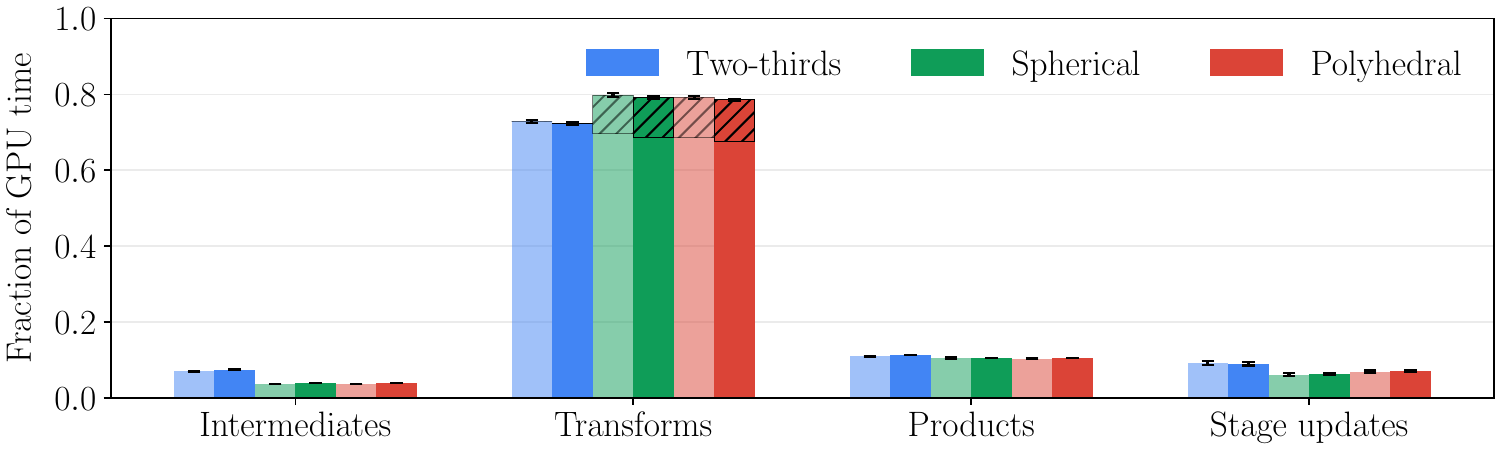}
	%	\vspace*{0.5cm}
	\caption{\normalfont The average fraction of GPU kernel execution time spent in various operations for the three dealiasing methods: two-thirds (blue), spherical (green), and polyhedral (red). In terms of the time step loop \cref{fig:timestep_loop}: intermediates corresponds to step 2; transforms and products, to step 3; and stage updates to steps 4 and 5. The lighter and darker opacities denote single and double precision, respectively. The error bars indicate the standard deviation in the average over different total numbers of modes $\totalmodes$, while the hatched regions indicate the portion of the time spent applying and undoing the phase shifts. All other operations, including calculation of the linear propagators \cref{eq:pade_propagator}, take less than $2\%$ of the kernel execution time.}
	\label{fig:dealiasing_gpu_kernel_fractions}
\end{figure}

Having established the different spectral regions retained by each dealiasing method, we now compare their computational costs. The relative time per time step and GPU memory usage are shown in figures \ref{fig:dealiasing_time_per_timestep} and \ref{fig:dealiasing_memory}, respectively. As in \cref{sec:numerics_timesteppers}, these results were obtained by measuring the wall time associated with $10^4$ steps of each dealiasing method across a range of $\totalmodes > 2 \times 10^6$ and computing the fractions of the GPU kernel execution time using NVIDIA Nsight Systems. As noted above, these measurements apply to the particular system and implementation considered here, rather than representing universal properties of the dealiasing methods across all such configurations. At fixed $\totalmodes$, the phase-shift methods require approximately twice the wall time per time step and between 35\% and 50\% more GPU memory than two-thirds dealiasing [see panel (a) in figures \ref{fig:dealiasing_time_per_timestep} and \ref{fig:dealiasing_memory}]. This additional cost arises because phase-shift dealiasing evaluates the nonlinear terms on both the original and shifted real-space grids [see \cref{eq:convolution_phase_shift_equal}], requiring a second set of Fourier transforms together with the operations and storage needed to shift and combine their contributions. Consequently, Fourier transforms account for a larger fraction of the GPU execution time for the phase-shift methods, as shown in \cref{fig:dealiasing_gpu_kernel_fractions}. Spherical and polyhedral truncation have almost identical costs because they employ the same phase-shift procedure and differ only in the Fourier modes retained during the stage updates. Fixed $\totalmodes$, however, does not provide a like-for-like comparison of the effective resolution of each dealiasing method. Indeed, \cref{eq:dealiasing_effective_modes} implies that the total grid size required by either phase-shift method at fixed $\effectivemodes$ satisfies
\begin{align}
	\frac{\totalmodes_{\text{phase-shift}}}
	{\totalmodes_{\text{two-thirds}}}
	= \frac{1}{2^{\dimension/2}}
	\label{eq:dealiasing_resolution_ratio}
\end{align}
The phase-shift methods therefore operate on substantially smaller arrays when the isotropic spectral coverage is held fixed. This reduction more than compensates for the additional shifted-grid evaluation: at fixed $\effectivemodes$, both spherical and polyhedral dealiasing require approximately half the wall time per time step and GPU memory of two-thirds dealiasing; see panel (b) in figures \ref{fig:dealiasing_time_per_timestep} and \ref{fig:dealiasing_memory}. The precise ratios remain dependent on the implementation and hardware, but their reversal between fixed $\totalmodes$ and fixed $\effectivemodes$ follows directly from the different spectral coverage of the methods.

\begin{figure}
	\centering
	%	\vspace*{0.5cm}
	\includegraphics[width=0.637\textwidth]{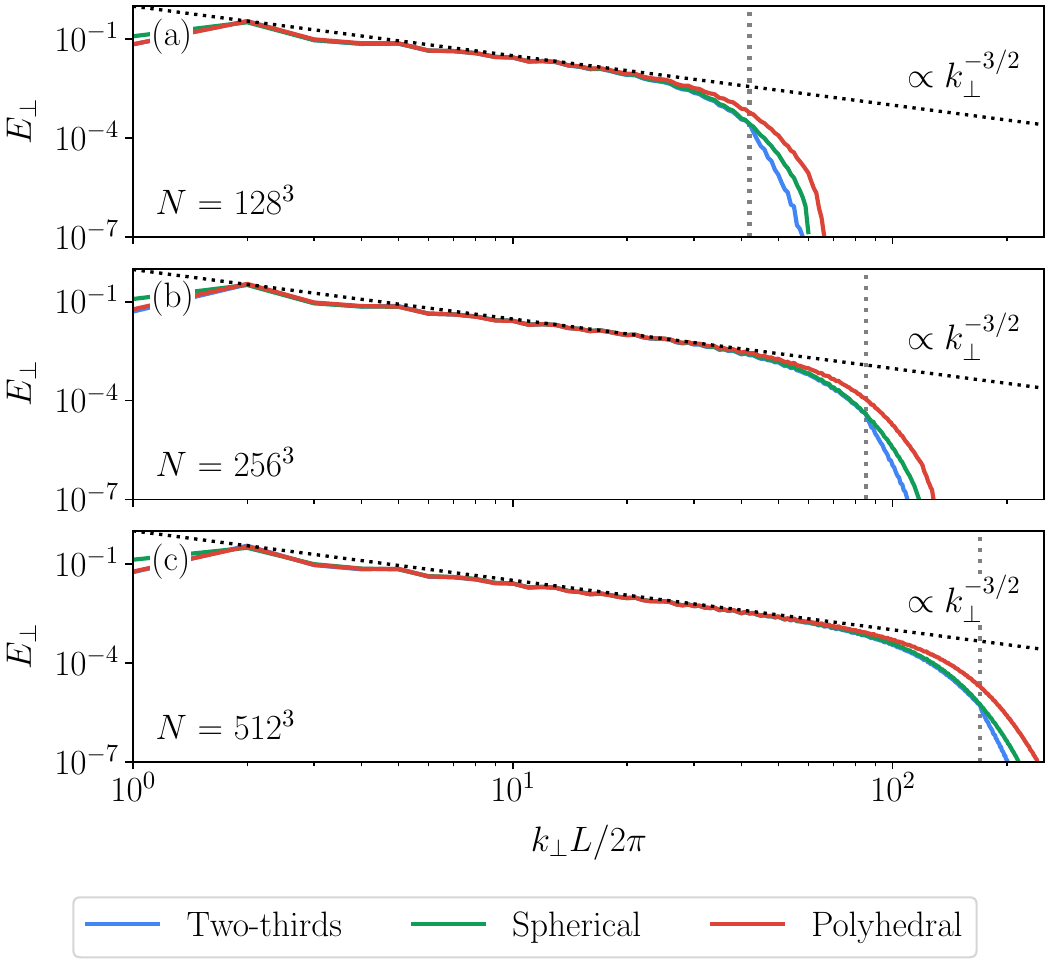}
	%	\vspace*{0.5cm}
	\caption{\normalfont Time-averaged perpendicular energy spectra from RMHD simulations of the isothermal KREHM system for different total number of modes $\totalmodes$: (a) $128^3$, (b) $256^3$, and (c) $512^3$. The colours correspond to the choice of dealiasing method, while the vertical dotted line shows the largest complete perpendicular wavenumber for the two-thirds case. The sloped dotted line shows the expected scaling, and $L$ is the perpendicular domain size.}
	\label{fig:dealiasing_krehm_spectra_total_grid}
\end{figure}

\begin{figure}
	\centering
	%	\vspace*{0.5cm}
	\includegraphics[width=0.637\textwidth]{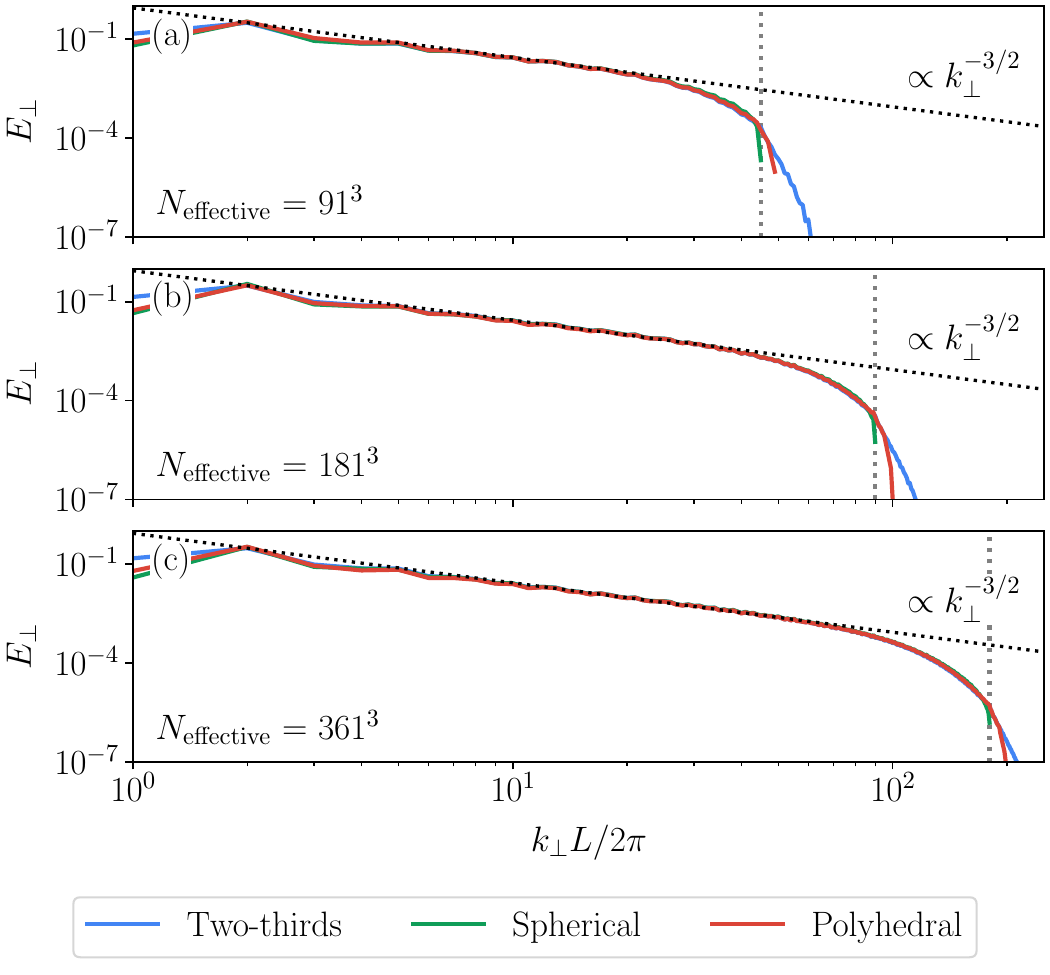}
	%	\vspace*{0.5cm}
	\caption{\normalfont Time-averaged perpendicular energy spectra from RMHD simulations of the isothermal KREHM system for different effective number of modes $\effectivemodes$: (a) $91^3$, (b) $181^3$, and (c) $361^3$. The details of the figure are otherwise the same as in \cref{fig:dealiasing_krehm_spectra_total_grid}.}
	\label{fig:dealiasing_krehm_spectra_eff_grid}
\end{figure}

It remains to verify that $\effectivemodes$ provides a meaningful measure of the dynamically resolved scales. For this purpose, we performed simulations of isothermal KREHM in the reduced magnetohydrodynamic limit (RMHD, \citealt{kadomtsev74,strauss76}); see \cref{app:krehm_validation} for further details of the underlying physics and simulation setup. The dissipation rate normalised to the maximum parallel and perpendicular wavenumbers was held fixed across all simulations to ensure the same rate of numerical damping at the grid scale. In \cref{fig:dealiasing_krehm_spectra_total_grid}, we plot the time-averaged perpendicular energy spectra [defined from \cref{eq:krehm_free_energy} analogously to \cref{eq:krehm_perp_spectra}] from simulations at fixed $\totalmodes$. As expected, the spectra show good agreement throughout their common well-resolved range, with the phase-shift methods retaining more spectral power at larger perpendicular wavenumbers, consistent with the solved regions shown in \cref{fig:dealiasing_regions_comparison}(a). It is clear that equal values of $\totalmodes$ therefore correspond to different spectral coverage. A more direct comparison is obtained at fixed $\effectivemodes$, as shown in \cref{fig:dealiasing_krehm_spectra_eff_grid}. Identical physical dissipation was used for all three methods in these simulations, ensuring that any difference arises from their spectral truncations rather than from different dissipation scales. The spectra once again agree closely throughout their common resolved range, including the onset of the dissipative cutoff, demonstrating that the phase-shift methods reproduce the effective resolution of two-thirds dealiasing on their smaller computational grids. The residual spectral tail produced by two-thirds dealiasing beyond the common isotropic cutoff arises from the additional modes retained near the diagonals of its rectangular solved region, but has no appreciable effect on the spectrum at lower wavenumbers. Similarly, the additional modes retained by polyhedral relative to spherical truncation do not materially alter the spectrum for this configuration. Increasing $\effectivemodes$ extends the range over which the spectra remain unchanged, providing an independent indication of numerical convergence.

Taken together, these results show that, for the RMHD dynamics considered here, phase-shift dealiasing achieves the same effective spectral resolution as two-thirds dealiasing while requiring approximately half of both the wall time per time step and GPU memory. This advantage is not universal: modes outside the common isotropic region may be dynamically important in systems with different spectral anisotropies, in which case the distinction between spherical and polyhedral truncation would become relevant, as discussed in \cref{sec:polyhedral_truncation}.

\subsection{Performance scaling}
\label{sec:numerics_performance}
Figures \ref{fig:timestepper_gpu_kernel_fractions} and \ref{fig:dealiasing_gpu_kernel_fractions} demonstrate that the computational cost of a time step is dominated by the evaluation of the explicitly treated nonlinear terms and, within each such evaluation, by the Fourier transforms. Since the number of arrays transformed during each nonlinear evaluation is independent of the grid size, we expect the time per time step to exhibit the $\order{\totalmodes\log\totalmodes}$ scaling of the underlying fast Fourier transforms at sufficiently large $\totalmodes$. Similarly, because the evolved fields and intermediate quantities are stored in arrays whose sizes are proportional to $\totalmodes$, the GPU memory usage should scale linearly with $\totalmodes$. The proportionality constants in both cases depend on the system, time-stepping scheme, and numerical precision.

\begin{figure}
	\centering
	%	\vspace*{0.5cm}
	\includegraphics[width=\figscale\textwidth]{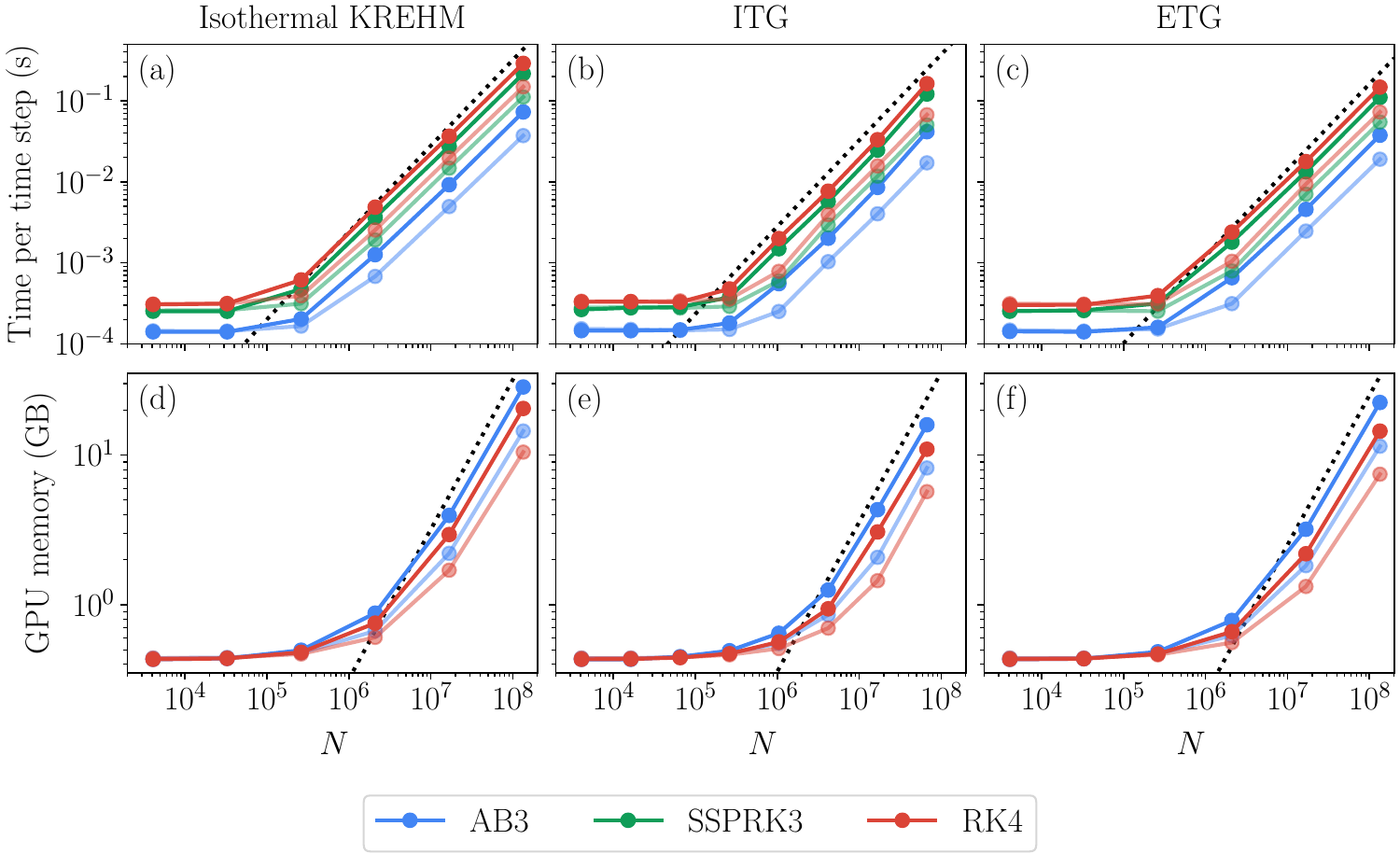}
	%	\vspace*{0.5cm}
	\caption{\normalfont Performance scalings as a function of the total number of modes $\totalmodes$ for the (a, d) isothermal KREHM, (b, e) ITG, and (c, f) ETG systems. The top row shows the wall time per time step in seconds, while the bottom row shows the GPU memory usage in gigabytes. The colours indicate the time-stepping scheme, while the lighter and darker opacities denote single and double precision, respectively. The dotted lines show the expected scalings. Note that the SSPRK3 curves in the bottom row lie behind the RK4 ones.}
	\label{fig:performance_timings}
\end{figure}

To test these expectations, we benchmarked the performance of the isothermal KREHM, ITG, and ETG systems using each of the available time-stepping schemes in both single and double precision. As in \cref{sec:numerics_timesteppers} and \cref{sec:numerics_dealiasing}, these results were obtained by measuring the wall time associated with $10^4$ steps of each method for each system and across a range of $\totalmodes$, with the fractions of the GPU kernel execution time computed using NVIDIA Nsight Systems. All measurements were performed on NVIDIA A100-SXM4-40GB GPUs. The resulting time per time step and GPU memory usage are shown in \cref{fig:performance_timings}. At small $\totalmodes$, both quantities approach approximately constant values because fixed allocations and per-step overheads remain comparable to the cost of operating on the grid. As $\totalmodes$ increases, the time per time step approaches the expected $\order{\totalmodes\log\totalmodes}$ scaling in each system, while the GPU memory usage approaches linear scaling. The separation between the timing curves is also consistent with the number of nonlinear evaluations required by each time-stepping scheme: one for AB3, three for SSPRK3, and four for RK4. AB3 is therefore the least expensive per step, followed by SSPRK3 and then RK4. At sufficiently large $\totalmodes$, the double-precision calculations require approximately twice the memory of their single-precision counterparts (see also \cref{fig:dealiasing_memory}), with a comparable separation also evident in the timing results (see also figures \ref{fig:timestepper_time_per_time_step} and \ref{fig:dealiasing_time_per_timestep}). Note that absolute memory requirements differ between both systems and time-stepping schemes because each stores a different number of evolved fields and intermediate arrays. In particular, AB3 requires more memory than either of the Runge-Kutta schemes --- despite requiring only one nonlinear evaluation per time step --- because it retains the explicitly treated right-hand-side histories from the two preceding steps (see \cref{app:ab3}). By contrast, SSPRK3 and RK4 have the same memory footprint in the present implementation, with intermediate stage arrays reused during each time step (sometimes referred to as a `low storage' scheme).

\begin{figure}
	\centering
	%	\vspace*{0.5cm}
	\includegraphics[width=0.501\textwidth]{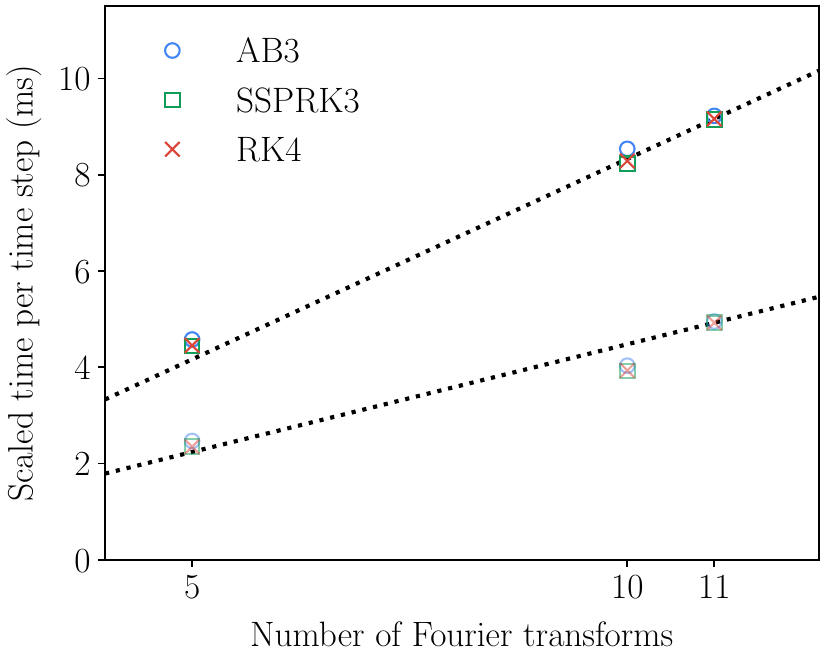}
	%	\vspace*{0.5cm}
	\caption{\normalfont The scaled wall time per time step as a function of the total number of individual Fourier transforms performed during each nonlinear evaluation. To compare the different timestepping schemes, we have divided the measured wall time by the number of nonlinear evaluations performed in each scheme (one, three, and four for AB3, SSPRK3, and RK4, respectively).  All measurements use the same total grid size, $\totalmodes=16\,777\,216=64 \times 512^2= 4096^2 = 256^3$. The colours and markers indicate the time-stepping scheme, while the lighter and darker opacities denote single and double precision, respectively. The points at $5$, $10$, and $11$ transforms correspond to the ETG, ITG, and isothermal KREHM systems, respectively. The dotted lines show proportional scaling with the number of transforms.}
	\label{fig:performance_fft_scaling}
\end{figure}

The remaining variation in the time per nonlinear evaluation between systems should arise primarily from the number of Fourier transforms required to construct their nonlinear terms. The ETG, ITG, and isothermal KREHM systems perform $5$, $10$, and $11$ individual transforms per nonlinear evaluation, respectively, counting complex-to-real and real-to-complex transforms separately. To test this dependence while removing the leading effect of the different numbers of time-stepping stages, \cref{fig:performance_fft_scaling} shows the time per time step divided by the number of nonlinear evaluations performed by the corresponding time-stepping scheme. All measurements use the same total grid size, $\totalmodes=16\,777\,216$, corresponding to $256^3$, $4096^2$, and $64 \times 512^2$ for the three-dimensional isothermal KREHM, two-dimensional ITG, and three-dimensional ETG systems, respectively. The resulting measurements for AB3, SSPRK3, and RK4 lie close to one another for each system and precision, confirming that their leading difference in cost arises from the number of nonlinear evaluations. The scaled wall time also increases approximately linearly with the number of Fourier transforms, as expected when the transforms dominate the cost of each nonlinear evaluation. Although the quantity that we have measured includes the complete time-step cost, rather than the Fourier transforms alone, the remaining stage updates, propagator operations, and system-specific kernels introduce only comparatively small departures from this proportional scaling, as could have been anticipated from figures \ref{fig:timestepper_gpu_kernel_fractions} and \ref{fig:dealiasing_gpu_kernel_fractions}. Together, these results confirm that the asymptotic large-$\totalmodes$ performance of \fouriersolver is governed primarily by the number and size of the Fourier transforms required by the chosen system and time-stepping scheme.

\section{Summary and future work}
\label{sec:conclusion}
We have presented \flucs, a framework for implementing and solving simple systems of partial differential equations efficiently on GPU hardware. Its present release provides a Fourier pseudospectral solver, \fouriersolver, for one-, two-, and three-dimensional periodic domains solved on a single NVIDIA GPU (\cref{sec:fourier_solver_algorithm}). The numerical machinery includes integrating-factor time integration using Pad\'e approximations to the linear propagators (\cref{sec:time_integration}), several explicit time-stepping schemes, and a choice of dealiasing methods for polynomial-type nonlinear terms (\cref{sec:dealiasing}). Implementations of a set of example systems --- isothermal KREHM, cold-ion ITG, and collisional ETG --- demonstrate how distinct physical models can use this shared infrastructure, with application benchmarks recovering known features of their dynamics (\cref{app:systems}). Our numerical investigations recover the expected convergence of the linear propagators and time-stepping schemes, and illustrate the timing and memory costs of the available numerical options using one of the example systems (\cref{sec:numerics_pade,sec:numerics_timesteppers}). At sufficiently large grid sizes, the computational cost is dominated by the Fourier transforms required to evaluate the nonlinear terms, with the time per step approaching the expected $\order{\totalmodes\log\totalmodes}$ scaling. The dealiasing comparisons in \cref{sec:numerics_dealiasing} also illustrate the distinction between computational grid size and dynamically useful spectral resolution: for the particular case considered here, phase-shift dealiasing reproduces the resolved perpendicular spectra on smaller computational grids, reducing both memory requirements and wall time per time step. The measured performance ratios depend on the particular system and numerical configuration, but demonstrate the practical value of retaining control over these choices.

A central objective of \flucs is to occupy a useful boundary between a code dedicated to one physical model and a general-purpose framework for arbitrary equations. Its clear distinction between solvers and systems allows the numerical methods and data representation to be specialised for a well-defined class of problems without restricting their use to a single set of governing equations. Individual systems retain control over their equation-specific operations, allowing these to be combined and optimised according to their computational requirements, while reusing the solver's time integration, dealiasing, and other numerical infrastructure. Python provides an accessible interface for running simulations and developing new systems, while compiled CUDA C++ kernels and GPU numerical libraries perform the computationally intensive work. This approach requires more explicit implementation work when adding a new system compared to frameworks in which equations can be entered symbolically, such as Dedalus \citep{burns20}. The benefit is that \flucs systems can be highly specialised and optimised for the particular form and structure of the considered equations, thus taking better advantage of the computational performance of the hardware. The increasing availability of large-language-model-based coding tools may, however, reduce this practical barrier by assisting with the translation of governing equations into the prescribed system interface and the construction of equation-specific kernels. Existing systems provide templates for this process, while the shared solver limits the amount of numerical machinery that must be implemented anew. Such assistance does not remove the need for physical understanding, numerical verification, or performance testing. It does, however, offer a route towards making specialised implementations easier to construct without relinquishing the opportunities for optimisation that motivate the explicit system interface.

Future developments will extend the range of numerical problems supported by the framework. In particular, we intend to implement a solver that retains Fourier representations and periodic boundary conditions in two (perpendicular) coordinates while allowing the linear and nonlinear operators to vary along a third (parallel) coordinate. This would extend the framework beyond the homogeneous periodic formulation considered here, enabling investigations of, e.g., magnetised-plasma turbulence in systems with spatially varying equilibria and magnetic geometry. Execution on multiple GPUs is another potential extension, enabling calculations whose memory requirements exceed what is practical on a single device. The present emphasis on single-GPU execution is nevertheless deliberate: avoiding inter-device communication is particularly valuable for smaller problems, for which communication overhead can outweigh the benefit of distributing the calculation. Increasing GPU memory capacities may also allow a substantial range of reduced-model investigations to remain on a single device. The aim of future parallelisation would therefore be to extend the accessible problem sizes while preserving the efficient single-GPU execution relevant to many of the calculations that motivate \flucs.

\section*{Acknowledgements}
The authors would like to thank 
W. Clarke, 
P. Hill,
F.~J. Escoto,
R. Madan,
L. Pattinson,
C. Skene,
L.-P. Turica,
and
Y. Zhang
for helpful discussions and suggestions at various stages of this project.

\section*{Funding}
The work of T.A. was supported in part by the Laboratory Directed Research and Development (LDRD) Program at the Princeton Plasma Physics Laboratory for the U.S. Department of Energy under Contract No. DE-AC02-09CH11466. The United States Government retains a non-exclusive, paid-up, irrevocable, world-wide license to publish or reproduce the published form of this manuscript, or allow others to do so, for United States Government purposes. The simulations presented in this article were performed on computational resources managed and supported by Princeton University's Research Computing, and on the Hydra HPC system managed by the Rudolf Peierls Centre for Theoretical Physics at the University of Oxford.  

\section*{Declaration of interests}
The authors declare that they have no known competing financial interests or personal relationships that could have appeared to influence the work reported in this paper.

\section*{Data availability}
Data for this publication will be made available on request. The code discussed is open-source, available at \url{https://github.com/flucs-code} under an MIT licence.

\section*{Declaration of generative AI and AI-assisted technologies in the manuscript preparation process}
Generative AI and AI-assisted technologies were used in the process of preparing this manuscript for the purposes of checking for typographical errors in prose and equations where appropriate.

% End credits
\printcredits

\appendix

\section{Time-stepping algorithms}
\label{app:timestepping_algorithms}
This appendix details the numerical implementation of the various explicit time-stepping schemes that are available in \fouriersolver for time-integrating the right-hand side of \cref{eq:integrating_factor_equation}. Although these methods are standard within the literature, we include them here for the sake of completeness, and to highlight details of the implementation that are specific to \flucs. All three methods permit the time-step size to vary during a simulation. This adaptivity is controlled by a Courant-Friedrichs-Lewy (CFL) condition \citep{courant67}, rather than an estimate of the local truncation error as used by embedded schemes. Before advancing each nonlinear time step, each system evaluates a system-specific rate $\cflomega$ (with dimensions of inverse time), from which we can define
\begin{align}
	\cfl_\tindex = \dtn \cflomega.
	\label{eq:cfl_number}
\end{align}
The time-step size is then adjusted to enforce
\begin{align}
	\cfl_\tindex \leqslant \cflmax,
	\label{eq:cfl_condition}
\end{align}
where \cflmax is a quantity specified by the user. The exact definition of $\cflomega$ is system-specific because the processes determining the maximum stable time-step size depend intrinsically on the equations being solved. Each implemented system must therefore calculate the appropriate local inverse time scale during the evaluation of the explicit terms, with the maximum over the computational domain used to find $\cflomega$. The time-step controller itself is common to all systems: the time-step size is bounded above by a user-specified value, and its increase between successive steps is limited to prevent abrupt changes. The user may choose either continuous adjustment, for which the time-step size is recalculated after every evaluation of $\cflomega$, or discrete adjustment, for which it is retained for multiple time steps unless the condition \cref{eq:cfl_condition} is violated or when an increase is permitted after a prescribed number of steps.

Finally, to allow the expressions that follow to remain compact, we introduce the step-local notation
\begin{align}
	\vfields[\tindex]\equiv\vfields(t_\tindex),\quad \rhsexplicit[\tindex]\equiv\rhsexplicit\!\left[\vfields[\tindex]\right], \quad \propagator[\tindex]({\Delta}t, {\Delta}t')\equiv\propagator(t_\tindex+{\Delta}t,t_\tindex+{\Delta}t').
	\label{eq:timestepping_shorthand}
\end{align}

\subsection{Third-order Adams-Bashforth}
\label{app:ab3}
The third-order adaptive Adams-Bashforth scheme \citep{marciniak20,tavakkol21} is obtained by approximating the explicitly treated right-hand side in the integrating-factor variable by a quadratic polynomial through its values at $t_\tindex$, $t_{\tindex-1}$, and $t_{\tindex-2}$. The resulting time step update is
\begin{align}
	\vfields[\tindex+1] = \propagator[\tindex](0, \dtn)\vfields[\tindex] + \dtn &\left[\abcoeffs_{\tindex,0}\propagator[\tindex](0, \dtn)\rhsexplicit[\tindex] +\abcoeffs_{\tindex,1}\propagator[\tindex](0, \dtn)\propagator[\tindex-1](0, \dtn[-1])\rhsexplicit[\tindex-1] \right.	\label{eq:ab3_update}\\
	&\hspace{2.95cm}
	\left.+ \:\abcoeffs_{\tindex,2}\propagator[\tindex](0, \dtn)\propagator[\tindex-1](0, \dtn[-1])\propagator[\tindex-2](0,\dtn[-2])\rhsexplicit[\tindex-2]
	\right],
	\nonumber
\end{align}
where the variable-step coefficients are
\begin{align}
	\abcoeffs_{\tindex,0} &= 1+\frac{1}{6}\frac{\dtn}{\dtn[-1]} \frac{2\dtn+6\dtn[-1]+3\dtn[-2]}{\dtn[-1]+\dtn[-2]}, 	\label{eq:ab3_coefficients_0}\\
	\abcoeffs_{\tindex,1} &= \quad - \frac{1}{6}\frac{\dtn}{\dtn[-1]} \frac{2\dtn+3\dtn[-1]+3\dtn[-2]}{\dtn[-2]},	\label{eq:ab3_coefficients_1}\\
	\abcoeffs_{\tindex,2} &=\quad + \frac{1}{6} \frac{\dtn}{\dtn[-2]} \frac{2\dtn+3\dtn[-1]}{\dtn[-1]+\dtn[-2]}. 	\label{eq:ab3_coefficients_2}
\end{align}
For a constant time-step, these reduce to $\abcoeffs_{\tindex,0} = 23/12$, $\abcoeffs_{\tindex,1}= -4/3$, and $\abcoeffs_{\tindex,2}=5/12$. Within \fouriersolver, the propagated right-hand-side histories appearing in \cref{eq:ab3_update} are stored in a circular buffer in global memory. At each time step, the current full-step propagator is applied to the newly evaluated right-hand side and to the two stored histories. The latter are thereby advanced to the new reference time before the circular buffer is updated, avoiding the need to directly reconstruct the products of propagators in \cref{eq:ab3_update}. When this time history is absent (e.g., when starting a simulation from initial conditions, or restarting from data at a given time), the time-step sizes appearing in \cref{eq:ab3_coefficients_0}-\cref{eq:ab3_coefficients_2} are set to ensure that the first and second time steps use forward Euler and second-order Adams-Bashforth, respectively \citep{hairer93}. This startup procedure generally introduces an \(\order{\dt^2}\) initial error. Third-order convergence from the initial condition requires starting values of sufficient accuracy, as provided by the RK4 bootstrap used in the convergence study \cref{sec:numerics_timesteppers}.

\subsection{Third-order strong-stability-preserving Runge-Kutta}
\label{app:ssprk3}
The third-order strong-stability-preserving Runge-Kutta scheme of \citet{shu88} is a three-stage, single-step method and therefore requires no time history from previous steps. Let $\vfields[\tindex]^{(1)}$ and $\vfields[\tindex]^{(2)}$ denote the fields at the two intermediate stages, and define their corresponding explicitly treated right-hand sides by
\begin{align} 
	\rhsexplicit[\tindex]^{(1)}\equiv\rhsexplicit\!\left[\vfields[\tindex]^{(1)}\right],\quad \rhsexplicit[\tindex]^{(2)}\equiv\rhsexplicit\!\left[\vfields[\tindex]^{(2)}\right]. \label{eq:ssprk3_rhs_shorthand} 
\end{align}
The first stage is a forward-Euler update of the integrating-factor field over the complete time step, which can be written in terms of the original fields as [cf. \cref{eq:integrating_factor_equation}]:
\begin{align} 
	\vfields[\tindex]^{(1)}=\propagator[\tindex](0, \dtn)\left[\vfields[\tindex]+\dtn\rhsexplicit[\tindex]\right]. 
	\label{eq:ssprk3_stage_1} 
\end{align}
The right-hand side $\rhsexplicit[\tindex]^{(1)}$ is then evaluated using this stage field at time $t_{\tindex+1}$. The second stage combines the initial field and a forward-Euler update from the first stage after propagating both contributions to the intermediate time $t_\tindex + \dtn/2$, viz., 
\begin{align} 		 		
	\vfields[\tindex]^{(2)}=\frac{3}{4}\propagator[\tindex]\left(0, \frac{1}{2}\dtn\right)\vfields[\tindex]+\frac{1}{4}\propagator[\tindex]\left(\dtn,\frac{1}{2} \dtn \right)\left[\vfields[\tindex]^{(1)}+\dtn\rhsexplicit[\tindex]^{(1)}\right]. 
	\label{eq:ssprk3_stage_2} 
\end{align}
After evaluating $\rhsexplicit[\tindex]^{(2)}$ using the second stage field at this intermediate time, the final field is obtained as
\begin{align} 
	\vfields[\tindex+1]=\frac{1}{3}\propagator[\tindex](0, \dtn)\vfields[\tindex]+\frac{2}{3}\propagator[\tindex]\left(\frac{1}{2}\dtn,\dtn\right)\left[\vfields[\tindex]^{(2)}+\dtn\rhsexplicit[\tindex]^{(2)}\right]. 
	\label{eq:ssprk3_stage_3} 
\end{align}
The three evaluations of the explicitly treated right-hand side therefore occur at $t_\tindex$, $t_{\tindex+1}$, and $t_\tindex+\dtn/2$, respectively. This requires propagators over a complete forward step, a forward half-step, and a backward half-step. Note that the latter propagator $\propagator[\tindex](\dtn,\dtn/2)$, appearing in \cref{eq:ssprk3_stage_2}, is evaluated directly as a backwards propagation, rather than explicitly inverting the forward half-step propagator according to \cref{eq:propagator_identities}. This means that for dissipative linear operators [e.g., hyperdissipation; see, e.g., \cref{eq:krehm_hyperdissipation}], this backwards propagator `reverses' the damping over this intermediate stage, amplifying the corresponding modes. Although this amplification is compensated by the subsequent forward propagation in \cref{eq:ssprk3_stage_3}, the scheme can easily become numerically unstable if the chosen time step is not sufficiently small.

\subsection{Fourth-order Runge-Kutta}
\label{app:rk4}
The classical fourth-order Runge--Kutta scheme \citep{runge95,kutta01,hairer93} is a four-stage, single-step method and therefore requires no time history from previous steps. As with SSPRK3, let $\vfields[\tindex]^{(1)}$, $\vfields[\tindex]^{(2)}$, and $\vfields[\tindex]^{(3)}$ denote the three intermediate stage fields, with their corresponding explicit right-hand sides defined analogously to \cref{eq:ssprk3_rhs_shorthand}. The first stage is a forward-Euler update of the integrating-factor field over half a time step, which can be written in terms of the original fields as
\begin{align} 
	\vfields[\tindex]^{(1)}=\propagator[\tindex]\left(0,\frac{1}{2}\dtn\right)\left[\vfields[\tindex]+\frac{1}{2}\dtn\rhsexplicit[\tindex]\right]. 
	\label{eq:rk4_stage_1} 
\end{align}
The right-hand side $\rhsexplicit[\tindex]^{(1)}$ is then evaluated using this stage field at time $t_\tindex+\dtn/2$. A second approximation to the fields at the same intermediate time is constructed using this right-hand side:
\begin{align} 
	\vfields[\tindex]^{(2)}=\propagator[\tindex]\left(0,\frac{1}{2}\dtn\right)\vfields[\tindex]+\frac{1}{2}\dtn\rhsexplicit[\tindex]^{(1)}. 
	\label{eq:rk4_stage_2} 
\end{align}
After evaluating $\rhsexplicit[\tindex]^{(2)}$ using this second intermediate field, the third stage advances from $t_\tindex$ to $t_{\tindex+1}$ using the second-stage right-hand side:
\begin{align} 
	\vfields[\tindex]^{(3)}=\propagator[\tindex](0,\dtn)\vfields[\tindex]+\dtn\propagator[\tindex]\left(\frac{1}{2}\dtn,\dtn\right)\rhsexplicit[\tindex]^{(2)}. 
	\label{eq:rk4_stage_3} 
\end{align}
The right-hand side $\rhsexplicit[\tindex]^{(3)}$ is then evaluated using this third stage field at $t_{\tindex+1}$. The final accumulated update is then:
\begin{align}
	 \vfields[\tindex+1]=\propagator[\tindex](0,\dtn)\vfields[\tindex]+\frac{1}{6}\dtn\left[\propagator[\tindex](0,\dtn)\rhsexplicit[\tindex]+2\propagator[\tindex]\left(\frac{1}{2}\dtn,\dtn\right)\rhsexplicit[\tindex]^{(1)}+2\propagator[\tindex]\left(\frac{1}{2}\dtn,\dtn\right)\rhsexplicit[\tindex]^{(2)}+\rhsexplicit[\tindex]^{(3)}\right]. 
	 \label{eq:rk4_update} 
 \end{align}
The four evaluations of the explicit right-hand side therefore occur at $t_\tindex$, $t_\tindex+\dtn/2$, $t_\tindex+\dtn/2$, and $t_{\tindex+1}$, respectively, requiring a complete forward-step propagator and propagators over the two forward half-step intervals. For a time-independent linear operator, the two half-step propagators are identical and only the complete-step and half-step propagators need be calculated. Unlike SSPRK3, the RK4 stage times are non-decreasing, so no backward propagator is required. To reduce the memory footprint of the scheme, two auxiliary field arrays are used alternately to store the three intermediate stage fields, while the weighted contributions appearing on the right-hand side of \cref{eq:rk4_update} are accumulated as the stages are evaluated.

\section{Currently implemented systems}
\label{app:systems}
This appendix details the various systems that are currently implemented within \flucs: isothermal KREHM (\cref{app:krehm}), two-dimensional cold-ion ITG (\cref{app:itg}), and collisional ETG (\cref{app:etg}). For each, we briefly outline the equations and their dimensionless variables before validating their implementation by reproducing known results from prior publications.

\subsection{Isothermal KREHM}
\label{app:krehm}
The \texttt{KREHMFourier} system implements the isothermal kinetic-reduced-electron-heating-model \citep[isothermal KREHM; see, e.g.,][]{adkins24}. The model describes the local evolution of low-frequency, anisotropic fluctuations in a strongly magnetised, low-beta plasma, retaining finite-ion-Larmor-radius effects and electron inertia while neglecting electron Landau damping. The latter simplification follows from assuming that the electrons remain isothermal along the exact magnetic-field lines, thereby closing the electron kinetic equation at the parallel-velocity moment. The resultant two-field system of equations is:
\begin{align}
	&\frac{\rmd}{\rmd t}\left(\taubar^{-1}\dphipot\right) - \frac{v_A^2 \rhos^2}{c} \gpar\gperp^2\dApar =0,\label{eq:krehm_phi_equation}\\
	&\frac{\rmd}{\rmd t}\left(\dApar-d_e^2\gperp^2\dApar\right)+c\left(\frac{\partial\dphipot}{\partial z}+\gpar\taubar^{-1}\dphipot\right) =0,\label{eq:krehm_apar_equation}
\end{align}
Here, $c$ is the speed of light, $v_A$ is the Alfv\'en speed associated with an equilibrium magnetic field $\vB$ with magnitude $B = |\vB|$ and unit vector $\ub = \vB/B$, $\rhos$ is the ion sound radius, related to the ion Larmor radius $\rhoi$ by $\rhos = \sqrt{Z/2\tau}\rhoi$, $\tau = \Ti/\Te$ is the ratio of the equilibrium ion temperature to that of the electrons, $Z$ is the charge number of the ions, and $d_e = c/\omega_{\text{pe}} = (m_e c^2/4\pi n_e e^2)^{1/2}$ is the electron-inertial length, in which $-e$, $m_e$, and $\dens[e]$ are the charge, mass, and equilibrium density of the electrons, respectively. Note that $d_e = \rhoe/\sqrt{\beta_e}$, where $\rhoe$ is the electron Larmor radius and $\beta_e = 8\pi\ne \Te/B^2$ the electron beta. The convective derivative encodes advection by the $\exb$ flow due to the perturbed electrostatic potential $\dphipot$:
\begin{align}
	\frac{\rmd}{\rmd t} = \frac{\partial}{\partial t} + \duperp \cdot \gperp, \quad \duperp = \frac{c}{B} \ub \times \gperp \dphipot,
	\label{eq:convective_derivative}
\end{align}
while the parallel derivative is along the exact magnetic field, including the perturbation of the magnetic-field direction arising from the parallel component of the magnetic-vector potential $\dApar$:
\begin{align}
	\gpar = \ub \cdot \grad  = \frac{\partial}{\partial z} + \frac{\dBperp}{B} \cdot \gperp, \quad \dBperp = - \ub \times \gperp \dApar.
	\label{eq:parallel_derivative}
\end{align}
Note that the advection terms appearing in both \cref{eq:convective_derivative} and \cref{eq:parallel_derivative} can be written in terms of the Poisson bracket, viz., for two spatially-varying functions $f$ and $g$:
\begin{align}
	\{f, g\} = \ub \cdot (\grad f \times \grad g) = \frac{\partial f}{\partial x} \frac{\partial g}{\partial y} - \frac{\partial f}{\partial y} \frac{\partial g}{\partial x}.
	\label{eq:poisson bracket}
\end{align}
Finally, finite-ion-Larmor-radius effects enter through
\begin{align}
	\taubar^{-1}=\frac{Z}{\tau}\left(1-\hat{\Gamma}_0\right),
	\label{eq:taubar}
\end{align} 
where the operator $\hat{\Gamma}_0$ can be expressed, in Fourier space, in terms of the modified Bessel function of the first kind: $\Gamma_0 = \rmI_0(\besselargint[i]) e^{-\besselargint[i]}$, where $\besselargint[i] = (\kperp \rhoi)^2/2$ and $\kperp^2 = k_x^2 + k_y^2$. This becomes $1 - \hat{\Gamma}_0 \approx - \rhoi^2\gperp^2/2$ at large scales $\kperp \rhoi \ll 1$ and $1 - \hat{\Gamma}_0 \approx 1$ at small scales $\kperp \rhoi \gg 1$.

\subsubsection{Dimensionless variables}
\label{app:krehm_variables}
Within \texttt{KREHMFourier}, the equilibrium magnetic field is taken to lie along the $z$ direction, $\ub=\hat{\vec{z}}$, without loss of generality. The parallel coordinate is normalised to an arbitrary reference scale $\Lpar$, while the perpendicular coordinates are normalised to a second arbitrary scale $\rhoperp$. Defining the Alfv\'en crossing time $t_A=\Lpar/v_A$, the dimensionless variables are thus
\begin{align}
	\frac{t}{t_A} \equiv \texttt{time}, \quad \frac{z}{\Lpar} \equiv \texttt{z}, \quad \frac{x}{\rhoperp} \equiv \texttt{x}, \quad \frac{y}{\rhoperp} \equiv \texttt{y},  \quad \frac{c t_A}{B \rhoperp^2} \dphipot \equiv \texttt{phi} , \quad \frac{\Lpar}{B \rhoperp^2} \dApar \equiv \texttt{apar}.
	\label{eq:krehm_dimensionless}
\end{align}
The resulting physical parameters entering the dimensionless equations are consequently 
\begin{align}
   \frac{\rhoi}{\rhoperp} \equiv \texttt{rhoi},  \quad \frac{d_e}{\rhoperp} \equiv \texttt{de}, \quad \frac{Z}{\tau} \equiv \texttt{ZTe\_over\_Ti},
	\label{eq:krehm_parameters}
\end{align}
Note that rather than evaluating \cref{eq:taubar} directly, the implementation works with $(1 - \Gamma_0)/\besselargint[i]$ where possible, whose limit as $\rhoi/\rhoperp \rightarrow 0$ is evaluated analytically. This allows one to obtain the equations of reduced magnetohydrodynamics \citep[RMHD,][]{kadomtsev74,strauss76} exactly by setting $\rhoi = d_e = 0$, rather than introducing a numerically small ion Larmor radius. Similarly, the finite-Larmor-radius-MHD (FLR-MHD) system studied in \cite{meyrand21} can be obtained directly by setting $d_e = 0$, which reduces to both RMHD and electron reduced magnetohydrodynamics \citep[ERMHD,][]{sch09,boldyrev13} at large ($\kperp \rhoi \ll 1$) and small ($\kperp \rhoi \gg 1$) scales, respectively.

\subsubsection{Validation against known results}
\label{app:krehm_validation}
Although \cref{eq:krehm_phi_equation} and \cref{eq:krehm_apar_equation} provide the most straightforward formulation of the isothermal KREHM system for numerical implementation, it is often useful to recast the same dynamics in terms of the so-called `generalised Els\"asser potentials'. These can be expressed, in Fourier space, as\footnote{Note that we have adopted the sign convention that the `plus' field is the one propagating in the $+\ub$ direction. This is the opposite to that used in both \cite{meyrand21} and \cite{adkins24} [see, e.g., equation (2.12) in the latter].}
\begin{align}
	\thetapmk = \sqrt{1 + \kperp^2 d_e^2 } \left[\frac{c}{B} \frac{\vphase (\kperp)/v_A}{(\kperp \rhos)^2} \taubar^{-1} \dphipotk \pm \frac{\dApark}{\sqrt{4 \pi \dens[i] m_i}}\right],
	\label{eq:krehm_elsasser_potentials}
\end{align}
in which
\begin{align}
	\vphase(\kperp) = \kperp \rhos \left(\frac{1+\taubar}{1 + \kperp^2 d_e^2}\right)^{1/2} v_A
	\label{eq:krehm_vphase}
\end{align}
is the perpendicular-wavenumber-dependent phase velocity of the linear modes. These satisfy the dispersion relation $\omega=\pm\kpar\vphase$, with the generalised Els\"asser potentials \cref{eq:krehm_elsasser_potentials} providing the corresponding linear eigenfunctions. The latter have the property that on the largest scales $\kperp \ll \rhoi^{-1}, d_e^{-1}$, they reduce to the standard Els\"asser potentials of RMHD \citep[see, e.g., ][]{sch09}.

The isothermal KREHM system conserves two nonlinear invariants: the free energy $\fenergy$ and the (generalised) helicity $\helicity$, which can be written in terms of the Els\"asser potentials \cref{eq:krehm_elsasser_potentials} as
\begin{align}
	\fenergy = \frac{\dens[i] m_i}{4} \sum_\vk \left[ | \kperp \thetapk |^2 + |\kperp \thetamk |^2 \right]
	\label{eq:krehm_free_energy}
\end{align}
and
\begin{align}
	\helicity = \frac{\dens[i] m_i}{4} \sum_\vk  \frac{| \kperp \thetapk|^2 - |\kperp \thetamk|^2}{\vphase(\kperp)/v_A},
	\label{eq:krehm_helicity}
\end{align}
respectively \citep[see][for the corresponding expressions in terms of the potentials $\dphipot$ and $\dApar$]{adkins24}. In the remainder of this section, we will consider cases where free energy \cref{eq:krehm_free_energy} and helicity \cref{eq:krehm_helicity} are injected into our isothermal KREHM system at constant rates $\fluxw$ and $\fluxh$, respectively, by some large-scale stirring on turbulent fluctuations, as is typically the case in many astrophysical environments such as the solar wind. The presence of a non-zero helicity-injection rate $\fluxh$ causes the turbulence to be \textit{imbalanced}, in that it is energetically dominated by fluctuations propagating in one direction along the equilibrium magnetic field. The Els\"asser potentials \cref{eq:krehm_elsasser_potentials} provide a natural basis for studying the resulting dynamics: the rates of energy injection into the forward- and backward-propagating fluctuations are given by
\begin{align}
	\fluxpm = \frac{\fluxw \pm \fluxh}{2},
	\label{eq:krehm_fluxes}
\end{align}
which are clearly unequal for nonzero $\fluxh$.  

Both balanced and imbalanced RMHD turbulence have been studied extensively --- although important theoretical questions remain unresolved in the latter case \citep[see][and references therein]{sch22} --- making it a natural starting point for validating the numerical implementation of \cref{eq:krehm_phi_equation} and \cref{eq:krehm_apar_equation}. Since the system is scale-invariant in the perpendicular direction for $\rhoi = d_e = 0$, the associated reference scale is chosen to be $\rhoperp = L/2\pi$, with $L = L_x = L_y$ the perpendicular domain size, and similarly for $\Lpar = L_z/2\pi$. All three directions are resolved using $n_x=n_y=n_z=512$ grid points, with phase-shift dealiasing using the polyhedral truncation \cref{eq:polyhedral_truncation_condition} and single precision. The SSPRK3 algorithm (\cref{app:ssprk3}) is used for the time-stepping, with a [4, 4] Pad\'e approximation to the linear propagators. Hyperdissipation in both the perpendicular and parallel directions is introduced by replacing the Fourier-space time-derivatives on the left-hand sides of \cref{eq:krehm_phi_equation} and \cref{eq:krehm_apar_equation} by
\begin{align}
	\frac{\rmd}{\rmd t} + \nu_\perp \left(\frac{\kperp}{\kperp^\text{max}}\right)^6 + \nu_z \left(\frac{k_z}{k_z^\text{max}}\right)^6,
	\label{eq:krehm_hyperdissipation}
\end{align}
where $\kperp^\text{max}$ and $k_z^\text{max}$ are the maximum perpendicular and parallel wavenumbers in the simulation domain. Such hyperdissipation is provided directly by \texttt{FourierSolver}, and is thus available to all systems. In this instance, the rates $\nu_\perp$ and $\nu_z$ are allowed to be adaptive, viz., they are re-evaluated at each time step to ensure that dissipation occurs near the grid scale, taking initial values of $\nu_\perp = \nu_z = 1.5 t_A^{-1}$. Fluctuations are forced at large scales at $\kperp \rhoperp = 2$, $|k_z| \Lpar = 1$ through the form of negative damping \citep{meyrand21,adkins24,adkins25}, allowing the rates of free-energy and helicity injection to be controlled directly while producing sufficiently random motions to generate turbulence.  

\begin{figure}
	\centering
	%	\vspace*{0.5cm}
	\includegraphics[width=\figscale\textwidth]{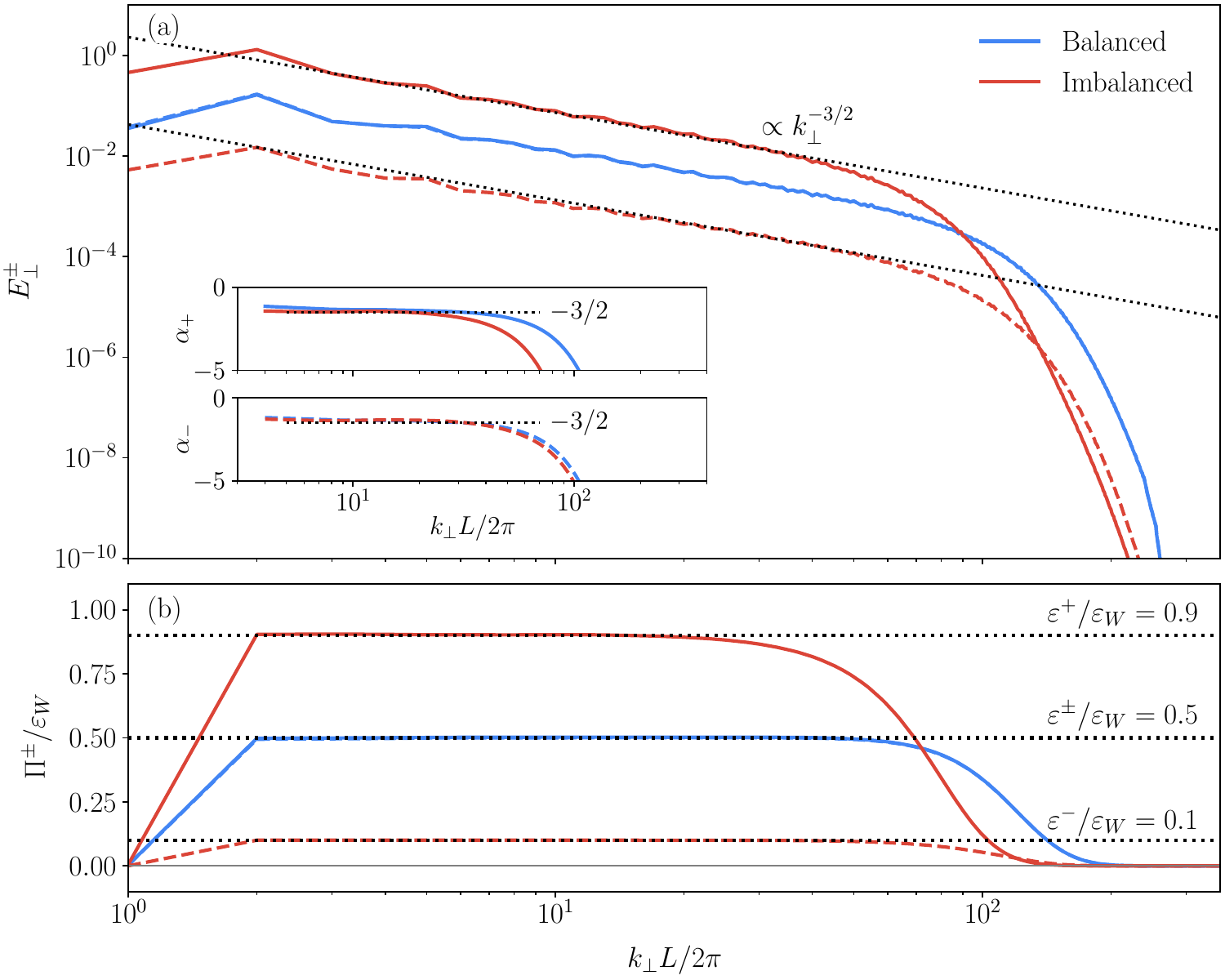}
	%	\vspace*{0.5cm}
	\caption{\normalfont Time-averaged results from simulations of balanced ($\fluxh/\fluxw=0$, $\fluxp/\fluxm=1$, blue) and imbalanced ($\fluxh/\fluxw=0.8$, $\fluxp/\fluxm=9$, red) RMHD turbulence. The solid and dashed lines denote quantities corresponding to $\thetap$ and $\thetam$, respectively, which almost exactly overlay in the balanced case. (a) Perpendicular spectra of the Els\"asser energies \cref{eq:krehm_perp_spectra} normalised to $(\rhoperp/\Lpar)^2 (B^2/8\pi)$. The dotted lines show the expected inertial-range scaling, while the inset panels display the measured local scaling exponents $\alpha_\pm = \rmd \log E_\perp^\pm/\rmd \log \kperp$ for each spectrum. (b) Perpendicular spectral energy fluxes $\Pi^\pm$ normalised to the energy-injection rate $\fluxw$. The horizontal dotted lines indicate the injected values of the fluxes \cref{eq:krehm_fluxes}.}
	\label{fig:krehm_rmhd_benchmark}
\end{figure}

To test the implementation across both zero and substantial cross-helicity injection, we compare the results from a balanced simulation with $\fluxh/\fluxw=0$ to an imbalanced one with $\fluxh/\fluxw=0.8$. In \cref{fig:krehm_rmhd_benchmark}(a), we plot the perpendicular energy spectra of the Els\"asser energies
\begin{align}
	E_\perp^\pm(\kperp) = 2 \pi \kperp \int_{-\infty}^\infty \rmd \kpar \: \frac{\dens[i] m_i}{4} \left< | \kperp \thetapmk |^2\right>,
	\label{eq:krehm_perp_spectra}
\end{align}
where the brackets denote an ensemble average, approximated here as a time average. In all cases, the spectra are well described by a power law with exponent $-3/2$ --- differing by only their outer-scale amplitudes --- consistent with previous simulations of forced RMHD turbulence. Note that this is not the same as the $-5/3$ spectrum predicted by the standard Kolmogorov-style phenomenology \citep[see, e.g.,][]{GS95,sch09}; the shallower $-3/2$ scaling has been attributed to scale-dependent dynamic alignment \citep{boldyrev06}, with subsequent theories incorporating the effects of intermittency \citep{chandran15,mallet17a}, and is supported by observations of imbalanced turbulence \citep{chen20}. \Cref{fig:krehm_rmhd_benchmark}(b) shows the time-averaged perpendicular spectral energy fluxes $\Pi^\pm(\kperp)$ computed directly from the nonlinear terms appearing in \cref{eq:krehm_phi_equation} and \cref{eq:krehm_apar_equation}. The fluxes of both fields are approximately constant throughout the inertial range and equal to their injected values \cref{eq:krehm_fluxes}, confirming the expectation that the RMHD system supports a constant-flux cascade of energy and helicity to small perpendicular spatial scales. 

\begin{figure}
	\centering
	%	\vspace*{0.5cm}
	\includegraphics[width=\figscale\textwidth]{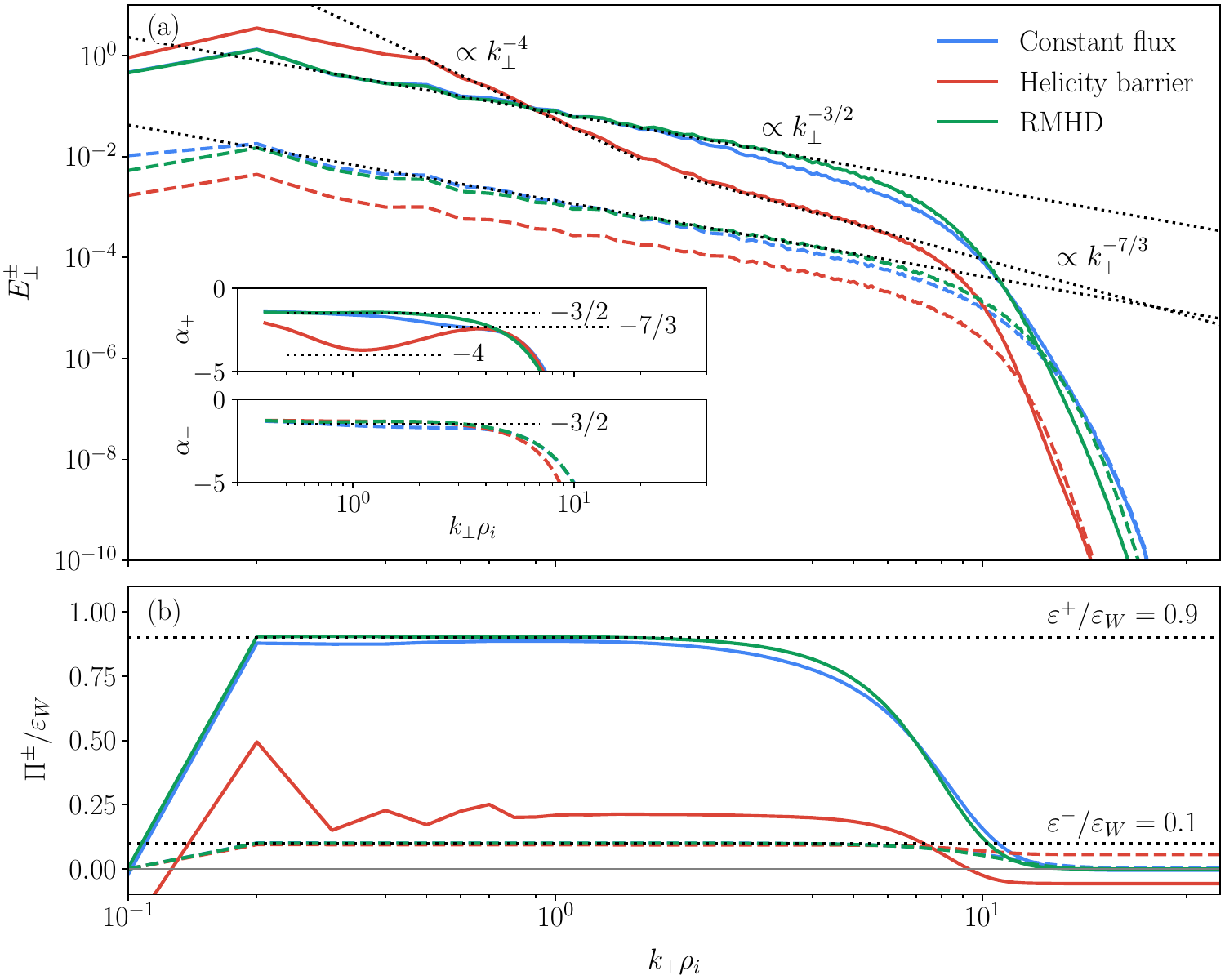}
	%	\vspace*{0.5cm}
	\caption{\normalfont Time-averaged results from simulations of imbalanced isothermal KREHM turbulence with $\fluxh/\fluxw = 0.8$: constant flux ($\beta/\betacrit = 0.64$, blue), helicity barrier ($\beta/\betacrit = 2.56$, red), and RMHD (same as the simulation from \cref{fig:krehm_rmhd_benchmark}, green). Note that the axis of the RMHD simulation has been rescaled for comparison with the other two cases. The solid and dashed lines denote quantities corresponding to $\thetap$ and $\thetam$, respectively. (a) Perpendicular spectra of the Els\"asser energies \cref{eq:krehm_perp_spectra} normalised to $(\rhoi/\Lpar)^2 (B^2/8\pi)$. The dotted lines show the expected scalings, while the inset panels display the measured local scaling exponents $\alpha_\pm = \rmd \log E_\perp^\pm/\rmd \log \kperp$ for each spectrum. (b) Perpendicular spectral energy fluxes $\Pi^\pm$ normalised to the energy-injection rate $\fluxw$. The horizontal dotted lines indicate the injected values of the fluxes \cref{eq:krehm_fluxes}.}
	\label{fig:krehm_helicity_barrier_benchmark}
\end{figure}

Such a cascade is not, however, universally supported by the isothermal KREHM system \cref{eq:krehm_phi_equation} and \cref{eq:krehm_apar_equation}. Recent work has shown that the constant-flux assumption must break down above a critical value of the electron beta \citep{adkins24,adkins25}
\begin{align}
	\betacrit = \frac{2Z}{1+\tau/Z} \frac{m_e}{m_i} \left(\frac{\fluxw}{\fluxh}\right)^2,
	\label{eq:krehm_betacrit}
\end{align}
due to the formation of the so-called `helicity barrier' \citep{meyrand21,squire22}, which prevents the imbalanced portion of the cascade from reaching small perpendicular scales. The resulting secular accumulation of energy in the $\thetap$ field would, in a physical system, eventually be arrested by the excitation of ion-cyclotron waves that resonantly heat the ion distribution function \citep{squire22,johnston25,zhang25}. Although this saturation channel lies outside the low-frequency gyrokinetic ordering underpinning the isothermal KREHM system, its absence does not invalidate the helicity-barrier dynamics captured by the model. Rather, once ion-cyclotron physics becomes important, isothermal KREHM no longer provides a complete description of the subsequent evolution. Nevertheless, the existence of the critical beta \cref{eq:krehm_betacrit}, and the corresponding switch in turbulent behaviour that occurs across it, provides a useful test of our numerical implementation. 

\begin{figure}
	\centering
	%	\vspace*{0.5cm}
	\includegraphics[width=\figscale\textwidth]{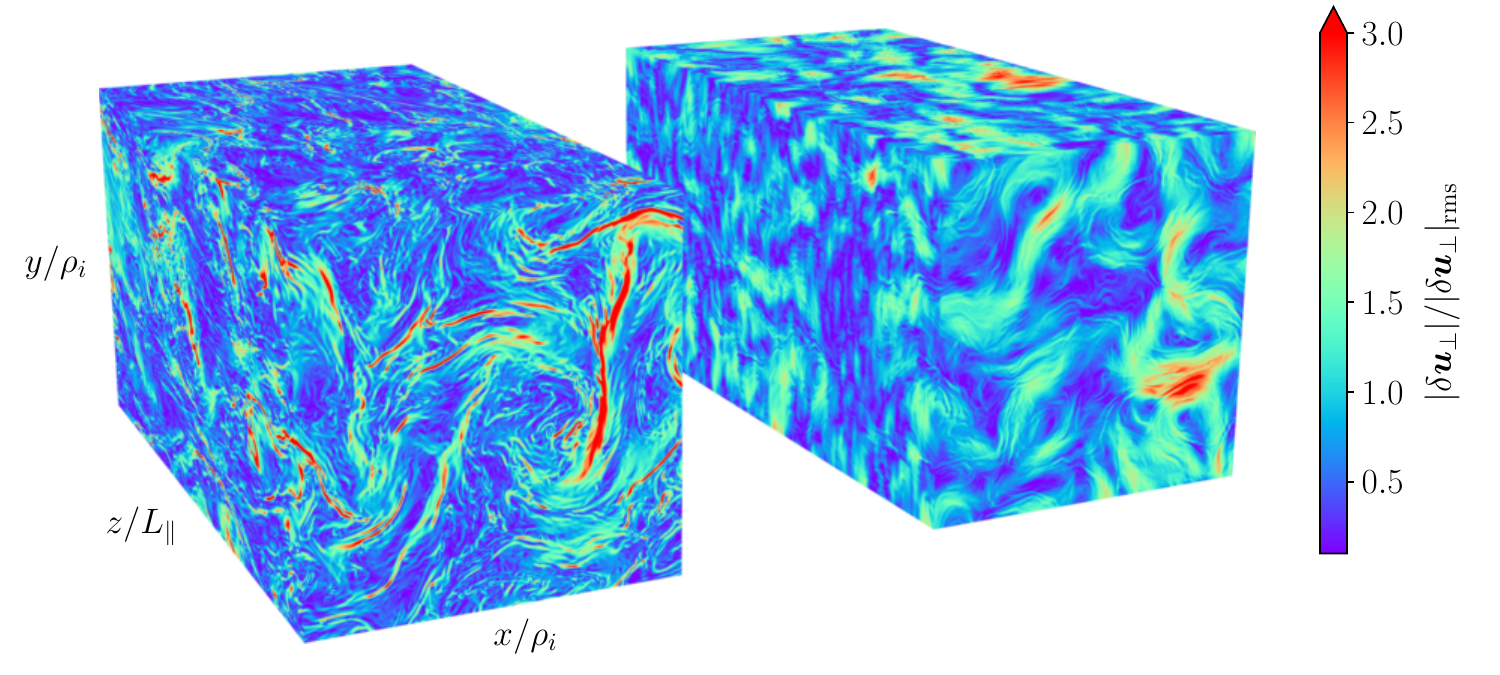}
	%	\vspace*{0.5cm}
	\caption{\normalfont Real-space snapshots of the $\exb$ flow $\duperp$ [see \cref{eq:convective_derivative}] for the `constant flux' (left) and `helicity barrier' (right) isothermal KREHM simulations. The colours indicate the magnitude of $\duperp$ relative to its (spatial) root-mean-square value, while the coordinate directions are as shown.}
	\label{fig:krehm_helicity_barrier_realspace}
\end{figure}

For this, we consider two imbalanced simulations that lie on either side of the critical beta \cref{eq:krehm_betacrit}, with values of $d_e$ that differ by a factor of two but have parameters that are otherwise identical. The helicity injection rate $\fluxh/\fluxw = 0.8$ used for both simulations corresponds to a critical beta $\betacrit(m_e/m_i)^{-1} = 1.56$: the first, with $\beta/\betacrit = 0.64$ ($d_e = \rhoi$), is expected to saturate via a constant flux cascade, while the other, with $\beta/\betacrit = 2.56$ ($d_e = \rhoi/2$), is expected to form a helicity barrier. The numerical setup is otherwise the same as for the RMHD simulations, except we now choose our reference lengthscale to be $\rhoperp = \rhoi$ and $L = L_x = L_y = 0.1 \rhoi$. As with \cref{fig:krehm_rmhd_benchmark}, we plot the perpendicular energy spectra and energy fluxes measured in these simulations in panels (a) and (b) of \cref{fig:krehm_helicity_barrier_benchmark}, respectively, including the imbalanced RMHD simulation from \cref{fig:krehm_rmhd_benchmark} for reference. Note that up to details of the dissipation range, \cref{fig:krehm_helicity_barrier_benchmark}(a) is almost identical to figure 2(a) of \cite{adkins24}. It is clear that both the `constant flux' and RMHD simulations indeed undergo a constant-flux cascade to small scales, with the former showing the transition to the $-7/3$ scaling expected below the ion Larmor scale (see \citealt{cho04,sch09}, though note the steeper scalings predicted by \citealt{boldyrev12,meyrand13,zhou23MNRAS}). The `helicity barrier' simulation, on the other hand, displays a sharp $\sim \kperp^{-4}$ break straddling the ion Larmor scale before flattening at smaller scales $\kperp \rhoi \gtrsim 1$, accompanied by a significant reduction in the $\thetap$ flux when compared with the corresponding constant-flux value. The stark contrast in turbulent behaviour between the `constant flux' and `helicity barrier' simulations, despite their small difference in physical parameters, is reflected in the real-space snapshots of the turbulence shown in \cref{fig:krehm_helicity_barrier_realspace}: the majority of the energy resides in large-scale structures in the latter case because it is prevented from cascading to small scales by the helicity barrier \citep[cf. figure 6 of][]{adkins24}.

\subsection{Two-dimensional cold-ion ITG}
\label{app:itg}
The \texttt{ColdITG2DFourier} system implements the two-dimensional long-wavelength fluid model of ion-temperature-gradient (ITG) driven turbulence due to \cite{ivanov20}. It is obtained from the electrostatic ion gyrokinetic equation \citep[see, e.g.,][]{abel13} in the highly-collisional, cold-ion limit, with the resulting equations evolving the perturbations to the electrostatic potential and ion temperature $\dTi$: 
\begin{align}
	&\frac{\rmd}{\rmd t} \left(\frac{e \dphipot'}{\Te} - \rhos^2 \gperp^2 \frac{e \dphipot}{\Te}\right) - \frac{2\rhos c_s}{L_B} \frac{\partial}{\partial y} \left(\frac{e \dphipot}{\Te} + \frac{\tau}{Z}\frac{\dTi}{\Ti}\right)+ \frac{\tau}{Z}\frac{\rhos c_s}{L_T} \frac{\partial}{\partial y} \left(\rhos^2 \gperp^2 \frac{e \dphipot}{\Te} \right)  \nonumber  \\
	& \hspace*{1.5cm} + \rhos c_s \left(\rhos^2 \gperp \cdot \pbra{\gperp \frac{e \dphipot}{\Te}}{\frac{\tau}{Z}\frac{\dTi}{\Ti}}\right) = - \chi \rhos^2 \gperp^4 \left(a \frac{e \dphipot}{\Te} - b \frac{\tau}{Z}\frac{\dTi}{\Ti} \right), \label{eq:itg_phi_equation} \\
	&\frac{\rmd}{\rmd t} \left(\frac{\dTi}{\Ti}\right) +  \frac{\rhos c_s}{L_T} \frac{\partial}{\partial y} \left(\frac{e \dphipot}{\Te}\right) = \chi \gperp^2 \left(\frac{\dTi}{\Ti}\right).
	\label{eq:itg_temp_equation}
\end{align}
Here, $\vthi = \sqrt{2 \Ti/\mi}$ is the thermal velocity of the ions, with $m_i$ their mass, $c_s = \sqrt{Z\Te/m_i}$ is the ion sound speed, $\rhos$ is the sound radius as in \cref{eq:krehm_apar_equation}, $\chi = (8/9)\sqrt{2/\pi} \nu_{ii} \rhoi^2$ is the thermal diffusivity, and $\nu_{ii}$ is the ion-ion collision frequency, while $a = 9/40$ and $b = 67/160$ are constants whose values are specific to the Landau collision operator [see appendix A of \cite{ivanov20}]. The coordinates $x$ and $y$ are taken to be the radial and poloidal directions, respectively, and all perturbed quantities are independent of the field-parallel coordinate. The convective derivative is as defined in \cref{eq:convective_derivative}. While the equilibrium magnetic field and ion temperature vary in the radial direction, the local nature of the model means that the corresponding scale lengths
\begin{align}
	\frac{1}{L_B} = -\frac{\rmd \log B}{\rmd x},  \quad \frac{1}{L_T} = - \frac{\rmd \log \Ti}{\rmd x},
	\label{eq:itg_gradients}
\end{align}
are assumed constant across the simulation domain. Finally, $\phi'$ denotes the nonzonal component of $\phi$, viz., $\phi' = \phi - \avgzonal{\phi}$, where the overline denotes an average over $y$.

\subsubsection{Dimensionless variables}
\label{app:itg_variables}
Within \texttt{ColdITG2DFourier}, the equilibrium scale lengths are normalised to an arbitrary reference scale $\Lperp$, meaning that the dimensionless variables are
\begin{align}
	\frac{2c_s t}{L_\perp}\equiv\texttt{time}, \quad \frac{x}{\rhos}\equiv\texttt{x}, \quad \frac{y}{\rhos}\equiv\texttt{y}, \quad \frac{ L_\perp}{2\rhos}\frac{e \dphipot}{\Te}\equiv\texttt{phi}, \quad \frac{L_\perp}{2\rhos} \frac{\tau}{Z}\frac{\dTi}{\Ti}\equiv\texttt{T}.
	\label{eq:itg_dimensionless}
\end{align}
The resulting dimensionless parameters entering the equations are
\begin{align}
	\frac{L_\perp}{L_B}\equiv\texttt{kappaB}, \quad \frac{\tau L_\perp}{2ZL_T}\equiv\texttt{kappaT},\quad \frac{L_\perp\chi}{2\rhos^2c_s}\equiv\texttt{chi}.
	\label{eq:itg_parameters}
\end{align}
The model considered by \citet{ivanov20} is recovered by choosing $\Lperp=L_B$, such that $\texttt{kappaB}=1$. Note that while the coefficients $a$ and $b$ can be calculated directly from the Landau collision operator, both are exposed as the inputs \texttt{coeffa} and \texttt{coeffb}, respectively, to permit their variation but taking their default (physical) values unless specified.

\subsubsection{Validation against known results}
\label{app:itg_validation}
The main source of energy for the fluctuations described by \cref{eq:itg_phi_equation,eq:itg_temp_equation} is the radial ion heat flux, which can be written in terms of the Fourier fields as
\begin{equation}
	Q_i = \frac{3}{2}\ni\Ti c_s
    \Re \sum_{\vec{k}}
	i k_y\rhos
	\frac{e\dphipot_{\vec{k}}^*}{\Te}
	\frac{{\dTi}_{\vec{k}}}{\Ti}
	,
	\label{eq:Qi_def}
\end{equation}
where `$\Re$' denotes the real part and `$^*$' the complex conjugate. \citet{ivanov20} demonstrated that the cold-ion ITG system is a simple, qualitative model of the so-called Dimits transition \citep{dimits00} between a low-transport state regulated by quasi-static long-wavelength zonal \(\exb\) flows and a high-transport, `strongly driven' state that arises once the system no longer supports the zonal flows.

\begin{figure}
	\centering
	%	\vspace*{0.5cm}
	\includegraphics[width=\figscale\textwidth]{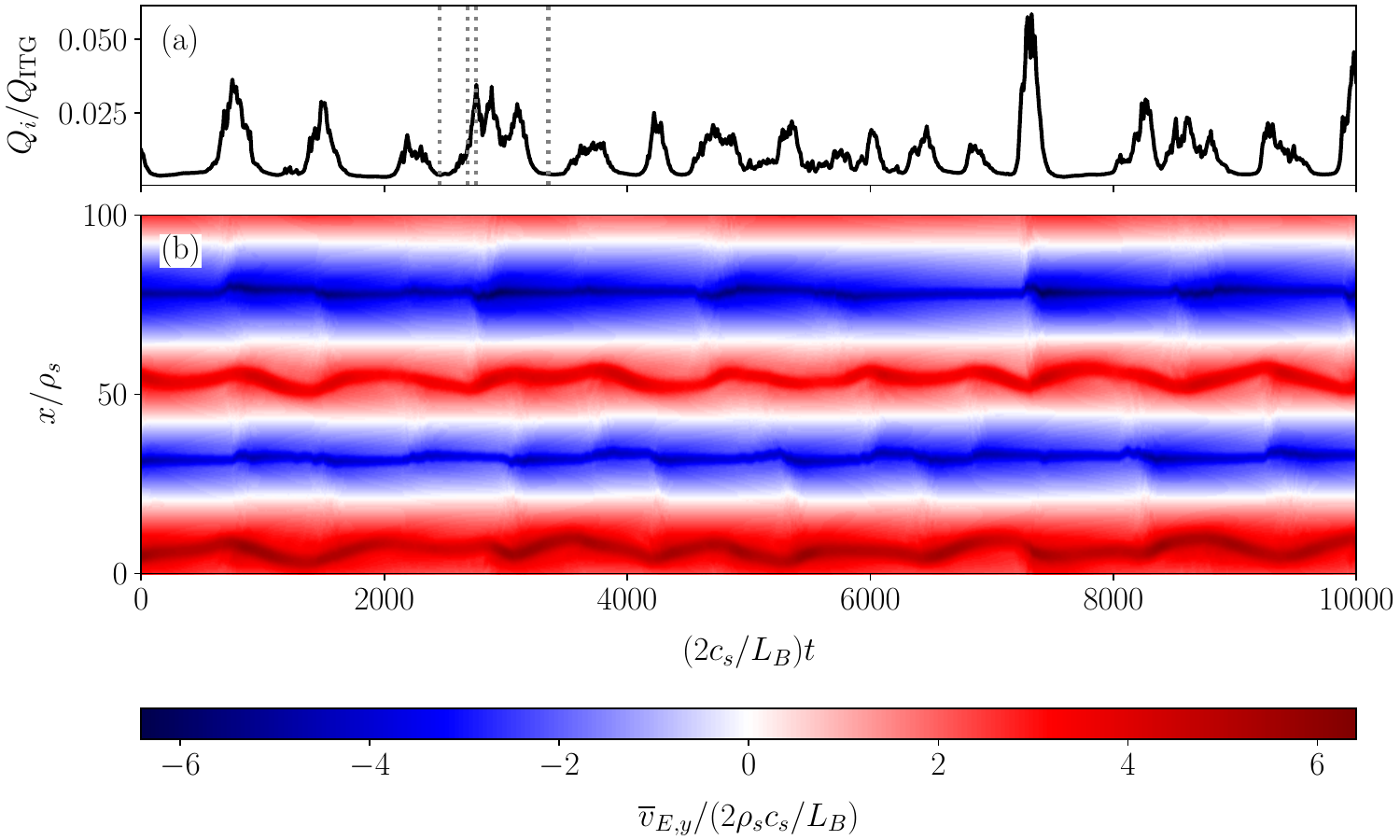}
	%	\vspace*{0.5cm}
	\caption{\normalfont Heat flux and zonal flows from a cold-ion ITG simulation with ${\tau L_B/2ZL_T} = 0.36$ and ${L_B\chi/2\rhos^2c_s}=0.1$. (a) Time trace of the ion heat flux \cref{eq:Qi_def} normalised to \(Q_\text{ITG} = 3Q_{\text{gB}i}/\sqrt{2}(\tau/Z)^{5/2}\), where \(Q_{\text{gB}i} = \ni\Ti\vthi (\rhoi/\LB)^2\) is the ion gyro-Bohm heat flux. The vertical dotted lines correspond to the times shown in \cref{fig:itg_turbulent_burst_realspace}. (b) Contours of the zonal flow \(\overline{v}_{E, y} = (c/B) \px\overline{\phi}\).}
	\label{fig:itg_heatflux_and_zonal_flow}
\end{figure}

\begin{figure}
	\centering
	%	\vspace*{0.5cm}
	\includegraphics[width=\figscale\textwidth]{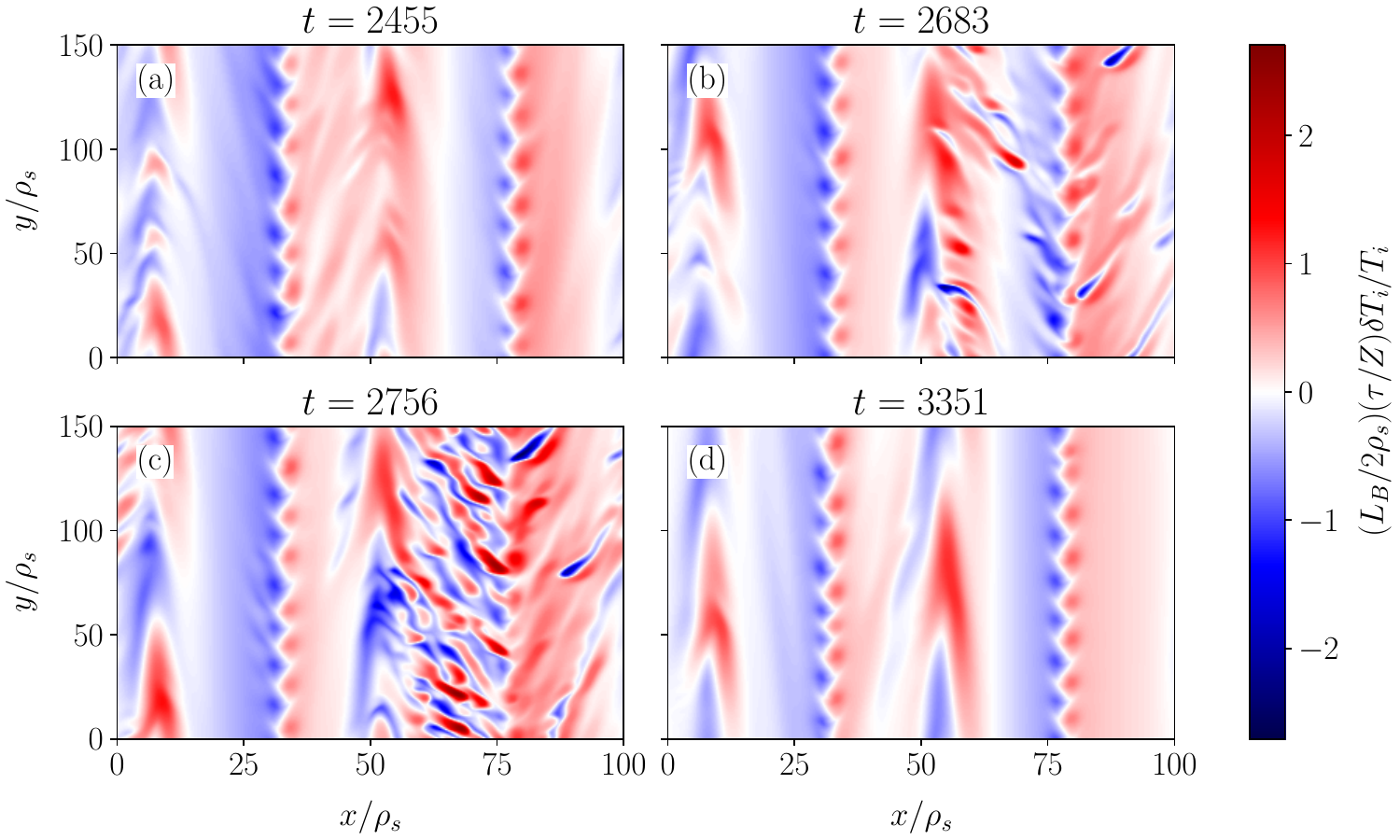}
	%	\vspace*{0.5cm}
	\caption{\normalfont Snapshots of the temperature fluctuations of \cref{eq:itg_phi_equation,eq:itg_temp_equation} in (a) the pre-burst quiescent state, (b) at the beginning of a burst, (c) at the peak of the burst, (d) the post-burst quiescent state. The times at which the snapshots are taken are indicated in \cref{fig:itg_heatflux_and_zonal_flow}(a) by vertical dotted lines.}
	\label{fig:itg_turbulent_burst_realspace}
\end{figure}

\begin{figure}
	\centering
	%	\vspace*{0.5cm}
	\includegraphics[width=\figscale\textwidth]{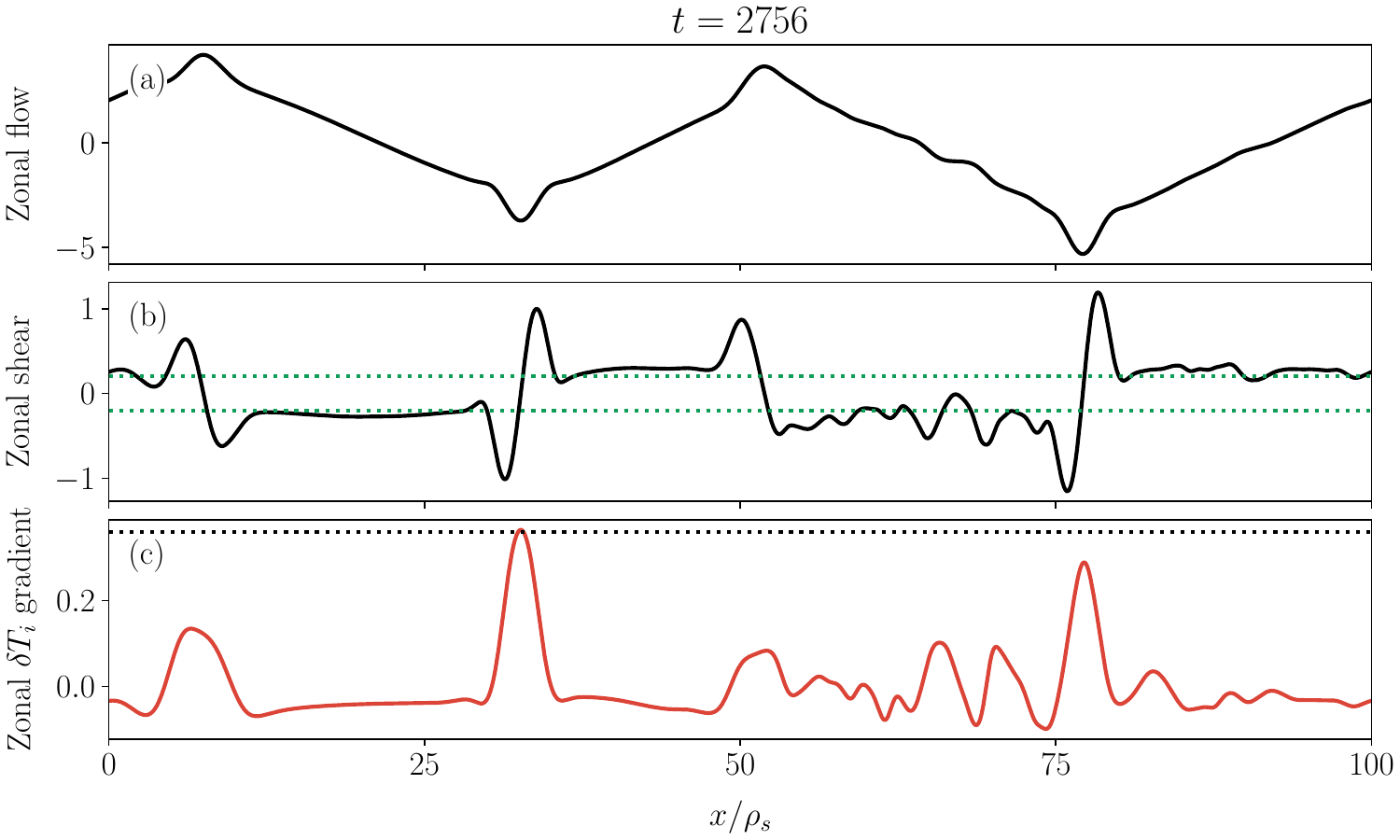}
	%	\vspace*{0.5cm}
	\caption{\normalfont Radial profiles of (a) the zonal flow \(\zf\), in units \(2\rhos c_s/\LB\); (b) the zonal shear \(\px \zf\), in units \(2 c_s/\LB\); (c) the zonal temperature gradient \(\px \zonaldTi\), in units \(2Z\Ti/\tau\LB\). The horizontal green dotted lines in panel (b) denote the maximum linear growth rate \(\pm\gamma_\text{max}\) in the same units as the zonal shear. The horizontal black dotted line in panel (c) denotes the negative of the imposed equilibrium temperature gradient in the same units as the zonal temperature gradient.}
	\label{fig:itg_turbulent_burst_zonal_profiles}
\end{figure}

\begin{figure}
	\centering
	%	\vspace*{0.5cm}
	\includegraphics[width=\figscale\textwidth]{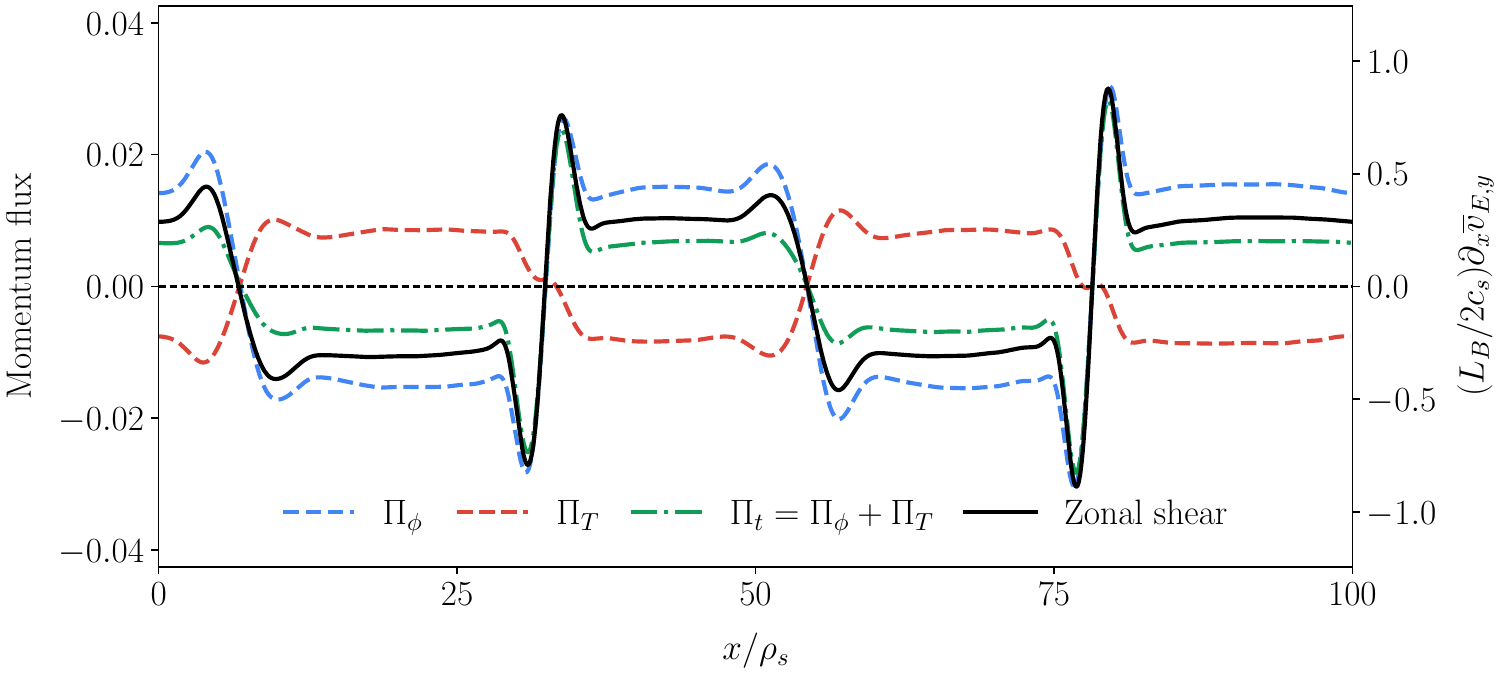}
	%	\vspace*{0.5cm}
	\caption{\normalfont Radial profiles of the turbulent stresses \cref{eq:Pi_phi,eq:Pi_T}, in units of \((2\rhos c_s /\LB)^2\), and averaged over the time interval shown in \cref{fig:itg_heatflux_and_zonal_flow}.}
	\label{fig:itg_momentum_flux}
\end{figure}

To validate the \flucs implementation of the cold-ion ITG system, we reproduce the predator-prey dynamics and bursty heat flux reported by \citet{ivanov20} for their `SimH' parameters ${\tau L_B/2ZL_T} = 0.36$ and ${L_B\chi/2\rhos^2c_s}=0.1$. We used the RK4 timestepper with \([4,4]\) Pad\'e integrating factors and two-thirds padding on a grid \((n_x, n_y) = (768, 512)\), yielding \((511, 341)\) solved modes. This is in contrast to \citet{ivanov20}, who used an AB3 method without integrating factors, Crank-Nicolson (equivalent to \([1,1]\) Pad\'e) linear propagator and retained only \((507, 337)\) solved modes. Here, we focus on the last interval of 10000 normalised time units in a simulation covering 60000 total normalised time units. \Cref{fig:itg_heatflux_and_zonal_flow} shows the bursty heat-flux dynamics and quasi-static zonal flow behaviour observed in \cite{ivanov20}, with the real-space dynamics of these turbulent bursts and the coherent structures (`ferdinons') that dominate the transport shown in \cref{fig:itg_turbulent_burst_realspace}. These figures are similar to figures 11 and 9 of \cite{ivanov20}, respectively, though we note that the average heat flux in \cref{fig:itg_heatflux_and_zonal_flow}(a) is \(\approx0.011\), a factor of 2 larger than that reported \(\approx0.006\) by \citet{ivanov20} due to a missing factor of two in their code. The \flucs implementation of \cref{eq:itg_phi_equation,eq:itg_temp_equation} also correctly recovers the structure of the zonal flows and fluctuating temperature. As \cref{fig:itg_turbulent_burst_zonal_profiles} demonstrates, the saturated-state zonal flows have well-defined `shear zones' wherein the shear in the zonal flows is approximately constant and comparable to the linear growth rate, which allows it to suppress the growing fluctuations \citep{waltz94}. Between the shear zones, where the zonal shear changes sign, a strong localised temperature perturbation flattens the driving background gradient, particularly around the minima of the zonal flow, where, as shown in \cref{fig:itg_turbulent_burst_zonal_profiles}(c), the zonal temperature gradient is approximately the negative of the equilibrium one.

The evolution of the zonal flow \(\zf = (c/B)\px\overline{\dphipot}\) is determined by the momentum equation
\begin{equation}
	\partd{\zf}{t} + \partd{}{x}\left(\Pi_\phi + \Pi_T + \Pi_d\right) = 0,
	\label{eq:zonal_momentum_equation}
\end{equation}
where
\begin{align}
	\Piphi &= -\frac{c^2}{B^2} \avgzonal{\partd{\dphipot}{x}\partd{\dphipot}{y}}, \label{eq:Pi_phi} \\
	\PiT &= -\frac{\rhoi\vthi}{2}\frac{c}{B} \avgzonal{\partd{\dphipot}{x}\partd{}{y}\frac{\dTi}{\Ti}}, \label{eq:Pi_T} \\
	\Pid &= - \chi \gperp^2 \left(a \frac{c}{B}\overline{\dphipot} - b \frac{\rhoi\vthi}{2}\frac{\zonaldTi}{\Ti} \right) \label{eq:Pi_d}
\end{align}
are the Reynolds, diamagnetic, and diffusive stresses, respectively. The two turbulent stresses, viz., \(\Piphi\) and \(\PiT\), are in competition: the former drives the zonal flow by providing an effective negative turbulent diffusivity for large-scale zonal flows, while the diamagnetic stress opposes it. Whether a large-scale zonal flow is formed or not depends on the outcome of this competition. In the low-transport regime, illustrated in \crefrange{fig:itg_heatflux_and_zonal_flow}{fig:itg_turbulent_burst_zonal_profiles}, the Reynolds stress dominates and thus strong, long-wavelength zonal flows can be generated and maintained. This is evident in \cref{fig:itg_momentum_flux} where we plot the time-averaged turbulent contributions to \cref{eq:zonal_momentum_equation} from the same regime. As detailed in \citet{ivanov20}, the anti-diffusive nature of the Reynolds stress is exemplified by the positive correlation between local zonal shear and \(\Piphi\). When the diamagnetic stress wins over the Reynolds one, the zonal-flow-mediated saturated state cannot be established. In this case, \cref{eq:itg_phi_equation,eq:itg_temp_equation} fail to achieve finite-amplitude saturation; instead, a growing box-scale `streamer' (i.e., mode with \(k_x \approx 0\)) dominates. This behaviour is reproduced by the current implementation at identical parameters, though the details thereof are not shown here given that the final state is not stationary, and therefore cannot be meaningfully compared with previous cases.

\subsection{Collisional ETG}
\label{app:etg}
The \texttt{CollisionalETGFourier} system implements the three-dimensional collisional fluid model of electron-temperature-gradient (ETG) driven turbulence due to \cite{adkins23}. The model describes electrostatic fluctuations on scales much larger than the electron Larmor radius but much smaller than the ion Larmor radius, allowing the ions to be treated as adiabatic. The resulting equations evolve the perturbed electrostatic potential $\dphipot$ and electron temperature $\dTe$: 
\begin{align}
	&\frac{\partial}{\partial t}\left(\frac{e\dphipot}{\Te}\right) - \frac{c_1 \vthe^2}{2 \nu_{ei}} \frac{\partial^2}{\partial z^2} \left[\left(1 + \frac{\tau}{Z}\right) \frac{e \dphipot}{\Te} - \frac{\tau}{Z}\left(1 + \frac{c_2}{c_1}\right) \frac{\dTe}{\Te}\right] + \frac{\rhoe \vthe}{L_B} \frac{\partial}{\partial y} \left[\left(1 + \frac{\tau}{Z}\right)\frac{e \dphipot}{\Te} - \frac{\tau}{Z} \frac{\dTe}{\Te}\right] \nonumber \\
	& \hspace*{1.65cm} - \frac{\tau}{Z} \frac{\rhoe \vthe}{2 L_n} \frac{\partial}{\partial y} \left(\frac{e \dphipot}{\Te}\right) = 0.  \label{eq:etg_phi_equation} \\
	& \frac{\rmd}{\rmd t} \left(\frac{\dTe}{\Te}\right) + \frac{2}{3} \frac{c_1 \vthe^2}{2 \nu_{ei}} \frac{\partial^2}{\partial z^2} \left[\left(1 + \frac{Z}{\tau}\right)\left(1 + \frac{c_2}{c_1}\right) \frac{e \dphipot}{\Te} - \left[\frac{c_3}{c_1} + \left(1 + \frac{c_2}{c_1}\right)^2\right]\frac{\dTe}{\Te}\right] \nonumber \\
	& \hspace*{1.65cm} - \frac{2}{3} \frac{\rhoe \vthe}{L_B} \frac{\partial}{\partial y}\left[\left(1 + \frac{Z}{\tau}\right) \frac{e \dphipot}{\Te} - \frac{7}{2} \frac{\dTe}{\Te}\right] + \frac{\rhoe\vthe}{2 L_T} \frac{\partial}{\partial y} \left(\frac{e \dphipot}{\Te}\right) = 0. \label{eq:etg_temp_equation}
\end{align}
Here, $\vthe = \sqrt{2\Te/\me}$ and $\rhoe$ are the thermal velocity and Larmor radius of the electrons, respectively, and $\nu_{ei}$ is the electron-ion collision frequency. Similar to \cref{app:itg}, the order-unity coefficients $c_1$, $c_2$, and $c_3$ are constants whose values are specific to the Landau collision operator and depend on the ion charge number $Z$ [see appendix B of \cite{adkins23}]. The equilibrium magnetic-field direction is taken to be $\ub=\hat{\vec{z}}$, while $x$ and $y$ denote the radial and binormal directions, respectively. The scale lengths of the equilibrium magnetic-field strength, electron density, and electron temperature are defined analogously to \cref{eq:itg_gradients}. Note that the presence of the second derivatives with respect to $z$ in both \cref{eq:etg_phi_equation} and \cref{eq:etg_temp_equation} is a consequence of the perturbations to the parallel electron velocity and parallel electron heat flux being determined by a static balance between parallel streaming and collisions within the asymptotic limit used to derive this system. 

\subsubsection{Dimensionless variables}
\label{app:etg_variables}
As in \cref{app:itg_variables}, \texttt{CollisionalETGFourier} normalises the equilibrium scale lengths to an arbitrary reference scale $\Lperp$, while perpendicular coordinates are normalised to some reference scale $\rhoperp  = \rhoe \Lperp/\sigma \lambdae$, where $\lambdae = \vthe/\nu_{ei}$ is the electron-ion mean free path and $\sigma$ is an arbitrary constant satisfying $\beta_e \ll \sigma \ll 1$. The dimensionless variables are then
\begin{align}
	\frac{\rhoe \vthe}{2 \rhoperp \Lperp} t \equiv \texttt{time}, \quad \sqrt{\sigma}\frac{z}{\Lperp} \equiv \texttt{z}, \quad \frac{x}{\rhoperp} \equiv \texttt{x}, \quad \frac{y}{\rhoperp} \equiv \texttt{y}, \quad \frac{\Lperp}{\rhoperp} \frac{e\dphipot}{\Te} \equiv \texttt{phi}, \quad \frac{\Lperp}{\rhoperp} \frac{\dTe}{\Te} \equiv \texttt{T}.
	\label{eq:etg_variables}
\end{align}
The resulting dimensionless parameters entering the equations are
\begin{align}
	\frac{\Lperp}{L_B}\equiv\texttt{kappaB},\quad \frac{\Lperp}{L_n}\equiv\texttt{kappaN},\quad \frac{\Lperp}{L_T}\equiv\texttt{kappaT},\quad \tau\equiv\texttt{tratio}, \quad Z\equiv\texttt{charge}.
	\label{eq:etg_parameters}
\end{align}
The code evaluates $c_1$, $c_2$, and $c_3$ from \texttt{charge} and supplies them internally as \texttt{coeffa}, \texttt{coeffb}, and \texttt{coeffc}, respectively. These values may also be specified directly, allowing these collisional coefficients to be varied independently of the ion charge number. A particularity of this system is that the only nonlinear term is the $\exb$ advection of $\dTe$ contained in the convective derivative in \cref{eq:etg_temp_equation}; the corresponding advection of $\dphipot$ vanishes due to the properties of the Poisson bracket \cref{eq:poisson bracket}. This makes \cref{eq:etg_phi_equation} and \cref{eq:etg_temp_equation} particularly inexpensive to solve when compared to other three-dimensional fluid systems due to the low number of Fourier-transform pairs required.

\subsubsection{Validation against known results}
\label{app:etg_validation}
Due to the presence of the equilibrium temperature gradient, the collisional ETG system of equations \cref{eq:etg_phi_equation} and \cref{eq:etg_temp_equation} supports a radial (electron) heat flux that can be written in terms of the Fourier fields as
\begin{align}
	Q_e = \frac{3}{4} \dens[e] \Te \vthe \Re \sum_{\vec{k}} i k_y \rhoe \frac{e \dphipot_{\vk}^*}{T_e} \frac{{\dTe}_{\vk}}{\Te}.
	\label{eq:etg_heatflux}
\end{align}
\cite{adkins23} showed that the scale invariance of \cref{eq:etg_phi_equation} and \cref{eq:etg_temp_equation}, encoded in the arbitrary nature of $\sigma$, implies a particular scaling of the heat flux with the square of the normalised parallel system size, viz., 
\begin{align}
	Q_e \propto \left(\frac{L_z \sqrt{\sigma}}{\LT}\right)^2,
	\label{eq:etg_scaling}
\end{align}
assuming that the perpendicular domain sizes (i.e., $L_x$ and $L_y$) are sufficiently large relative to the characteristic scales of the turbulence. \cite{adkins23} also recovered the scaling in \cref{eq:etg_scaling} using constant-flux, critically-balanced cascade arguments, related versions of which have been applied more broadly to gradient-driven turbulence in a number of settings \citep{barnes11,sch16,adkins22,ivanov25,nies26,adkins26}. A detailed account of these arguments lies beyond the scope of the present paper.

To validate our implementation of the collisional ETG system, we repeat the same scan in the parallel domain size that was performed in \cite{adkins23} to verify the prediction \cref{eq:etg_scaling}. Terms involving the density and magnetic-field gradients in \cref{eq:etg_phi_equation} and \cref{eq:etg_temp_equation} were set to zero, setting the equilibrium reference lengthscale as $L_\perp = \LT$ and $\tau = Z = 1$. The perpendicular domain size and number of grid points were held fixed at $L_x = L_y = 40 \rhoperp$ and $n_x = n_y = 288$, respectively, while the parallel domain $L_z\sqrt{\sigma}/\LT$ is varied between 15 and 55 while maintaining a fixed ratio of $L_z\sqrt{\sigma}/\LT$ to the number of grid points $n_z$, with $n_z = 48$ at $L_z\sqrt{\sigma}/\LT = 20$. This ensures that the maximum parallel wavenumber $k_z^\text{max}$, and therefore the maximum parallel-damping rate arising from the parallel thermal conduction terms in \cref{eq:etg_phi_equation} and \cref{eq:etg_temp_equation}, is also held fixed. We use standard two-thirds padding (\cref{sec:twothirds_padding}) for dealiasing, and single precision. To provide a sink of energy at small perpendicular scales, hyperdissipation is introduced by replacing the Fourier-space time-derivatives of the left-hand sides of \cref{eq:etg_phi_equation} and \cref{eq:etg_temp_equation} by [cf. \cref{eq:krehm_hyperdissipation}]
\begin{align}
    \frac{\rmd}{\rmd t} + \nu_\perp \left(\frac{\kperp}{\kperp^\text{max}}\right)^6,
	\label{eq:etg_hyperdissipation}
\end{align}
with the (constant) coefficient equal to $\nu_\perp/\omega_\perp = 100$, while $\omega_\perp = \rhoe \vthe /2 \rhoperp \LT$ is the value of the electron drift frequency at $k_y \rhoperp = 1$ [cf. the first expression in \cref{eq:etg_variables}]. This numerical setup is the same as in \cite{adkins23}, with the primary difference coming from the difference in time-stepping algorithm: here, we use SSPRK3 with a [4, 4] Pad\'e approximation to the linear propagators (\cref{app:ssprk3}), while the implementation of \cite{adkins23} used an algorithm analogous to the present AB3 one (\cref{app:ab3}), except without integrating factors and with the linear terms implemented using an explicit Crank-Nicolson solve (equivalent to the [1, 1] Pad\'e approximation). 

\begin{figure}
	\centering
	%	\vspace*{0.5cm}
	\includegraphics[width=\figscale\textwidth]{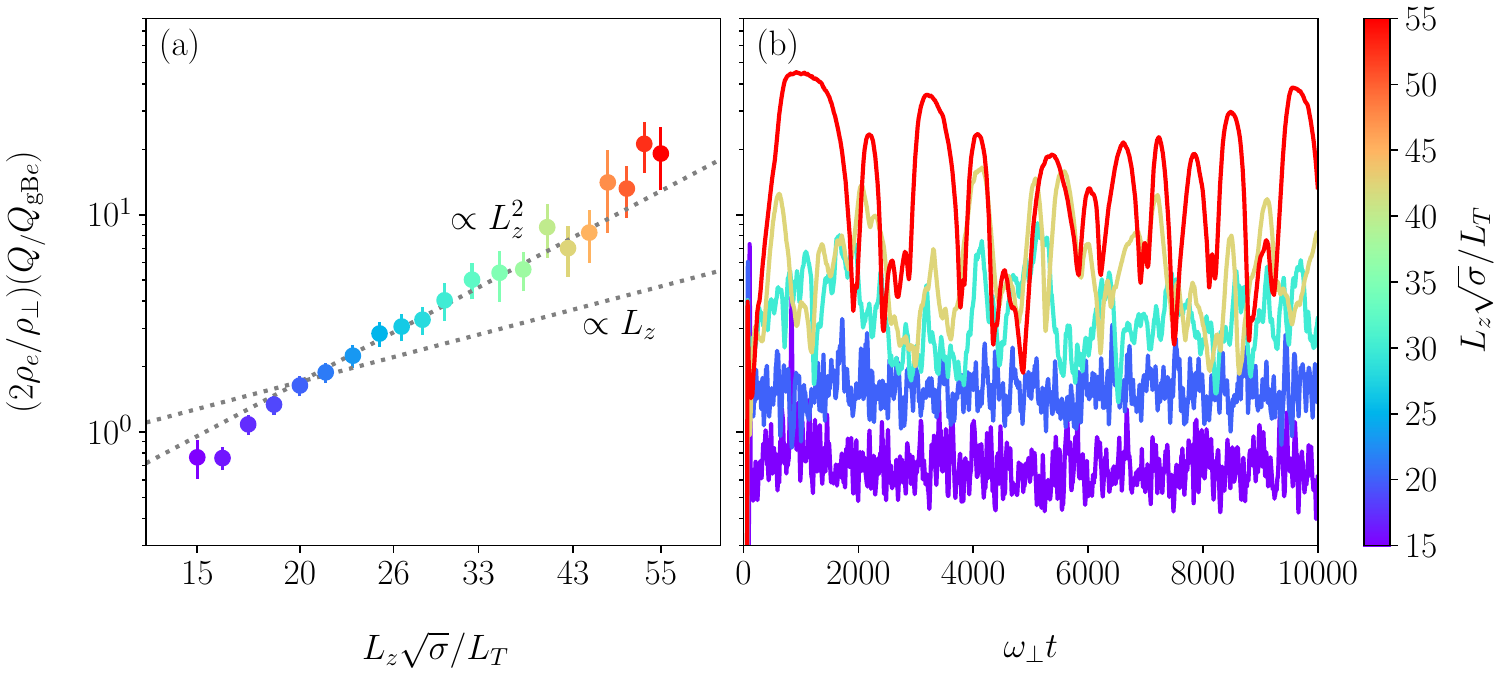}
	%	\vspace*{0.5cm}
	\caption{\normalfont Scaling results from the collisional ETG simulations, with the colour showing the value of the parallel domain size $L_z \sqrt{\sigma}/\LT$ for each case. (a) Time-averaged heat fluxes as a function of the parallel domain size (circles), with the error bars showing the standard deviation over time. The dotted lines show the predicted scaling \cref{eq:etg_scaling} alongside a linear $\propto L_z$ scaling for reference. (b) Heat fluxes as a function of time for the simulations with $L_z \sqrt{\sigma}/\LT = 15.0, 20.0, 30.0, 42.5, 55.0$.}
	\label{fig:etg_lpar_scan}
\end{figure}

\begin{figure}
	\centering
	%	\vspace*{0.5cm}
	\includegraphics[width=\figscale\textwidth]{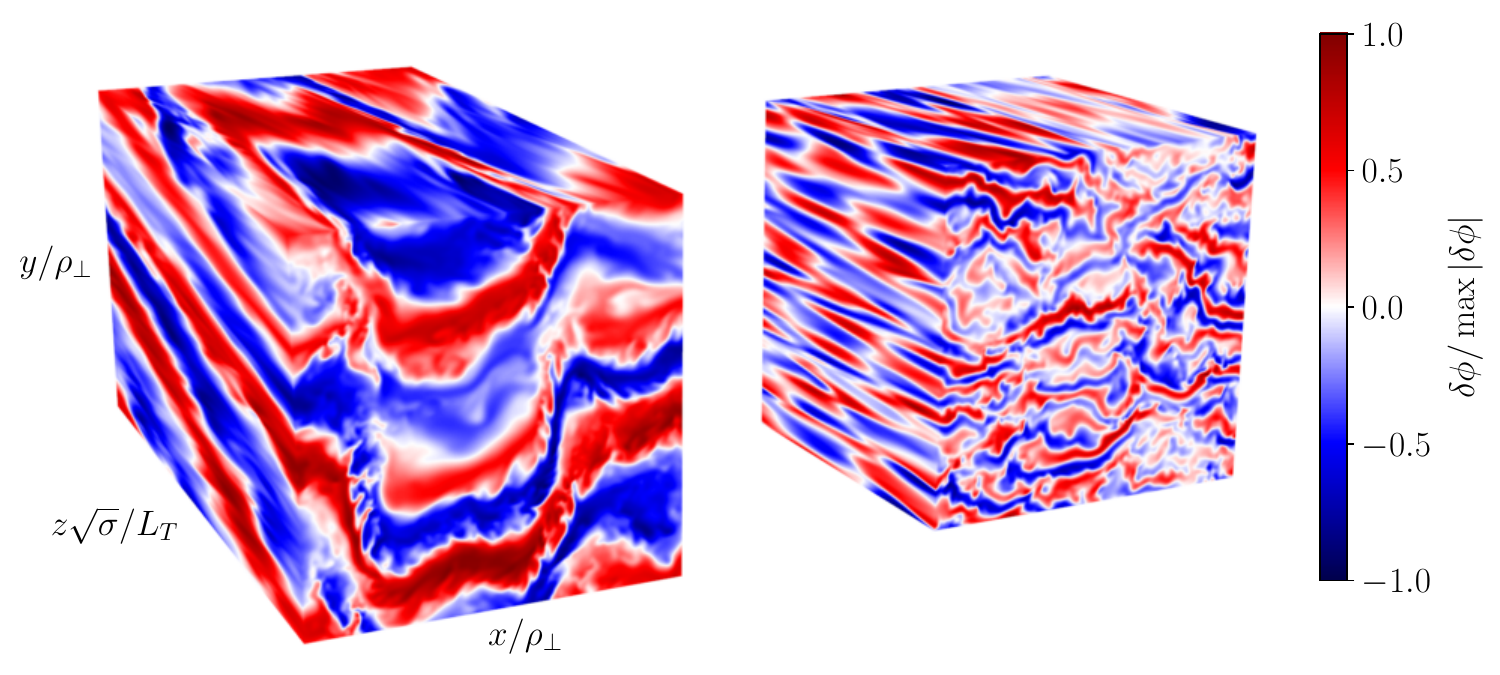}
	%	\vspace*{0.5cm}
	\caption{\normalfont Real-space snapshots of the electrostatic potential $\dphipot$ for collisional ETG simulations with $L_z \sqrt{\sigma}/\LT = 40$ (left) and $L_z \sqrt{\sigma}/\LT = 20$ (right). The colours indicate the value of $\dphipot$ relative to its maximum absolute value. The domains are scaled so as to preserve their relative aspect ratio, i.e., the parallel extent of the left-hand domain is twice that of the right.}
	\label{fig:etg_lpar_scan_realspace}
\end{figure}

In \cref{fig:etg_lpar_scan}(a), we plot the time-averaged heat fluxes from these simulations as a function of their parallel domain size. The data clearly reproduces the predicted scaling \cref{eq:etg_scaling}. Deviations occur at small parallel domain sizes, where the turbulence is weakly driven and therefore strongly influenced by (artificial) dissipative effects, and at large parallel domain sizes, where the finite perpendicular extent of the computational domain begins to constrain the allowed size of the largest turbulent eddies. \Cref{fig:etg_lpar_scan}(b) shows the corresponding heat fluxes as a function of time, from which it is clear that increasing the parallel domain size produces larger-amplitude oscillations on progressively longer timescales, consistent with the underlying scale invariance of the equations. Except for the largest parallel domain sizes, the time-averaged heat fluxes agree to within 10\% of those obtained by \cite{adkins23}. At the largest sizes, the longer turnover time means that the available interval samples fewer statistically independent turnover times; extending the simulation to achieve comparable statistical sampling would be prohibitively expensive, leaving the resulting average and comparison with \cite{adkins23} subject to a greater finite-time uncertainty. Finally, the scaling \cref{eq:etg_scaling} can also be seen in the behaviour of the turbulence in real space. \Cref{fig:etg_lpar_scan_realspace} compares real-space snapshots from two simulations, one of which has half the parallel domain size as the other. It is straightforward to see that the parallel extent of the eddies always remains comparable to the parallel domain size, with the perpendicular extent of the eddies, and therefore the turbulent heat flux, growing accordingly, a visual manifestation of the critically-balanced dynamics underpinning \cref{eq:etg_scaling}.

\bibliographystyle{cas-model2-names}

\bibliography{bibliography}

\end{document}